\pdfoutput=1
\documentclass[10pt,a4paper,twocolumn,english,prd,showpacs,showkeys,floatfix,longbibliography]{revtex4-1}
\usepackage[T1]{fontenc}
\usepackage[utf8]{inputenc}
\usepackage{color}
\usepackage{babel}
\usepackage{array}
\usepackage{float}
\usepackage{booktabs}
\usepackage{units}
\usepackage{amsmath}
\usepackage{amssymb}
\usepackage{mathdots}
\usepackage{graphicx}
\usepackage{esint}
\usepackage[unicode=true,
 bookmarks=true,bookmarksnumbered=false,bookmarksopen=true,bookmarksopenlevel=2,
 breaklinks=true,pdfborder={0 0 0},pdfborderstyle={},backref=false,colorlinks=true]
 {hyperref}
\hypersetup{pdftitle={Spectral properties and chiral symmetry violations of (staggered) domain wall fermions in the Schwinger model},
 pdfauthor={Christian Hoelbling, Christian Zielinski},
 linkcolor=BlueViolet,anchorcolor=BlueViolet,citecolor=BlueViolet,filecolor=BlueViolet,urlcolor=BlueViolet}

\makeatletter

\pdfpageheight\paperheight
\pdfpagewidth\paperwidth

\providecommand{\tabularnewline}{\\}

 \@ifundefined{textcolor}{}
 {%
   \definecolor{BLACK}{gray}{0}
   \definecolor{WHITE}{gray}{1}
   \definecolor{RED}{rgb}{1,0,0}
   \definecolor{GREEN}{rgb}{0,1,0}
   \definecolor{BLUE}{rgb}{0,0,1}
   \definecolor{CYAN}{cmyk}{1,0,0,0}
   \definecolor{MAGENTA}{cmyk}{0,1,0,0}
   \definecolor{YELLOW}{cmyk}{0,0,1,0}
 }

\usepackage[usenames,dvipsnames]{xcolor}

\allowdisplaybreaks
\usepackage{bbm}
\usepackage{morefloats}

\@ifundefined{showcaptionsetup}{}{%
 \PassOptionsToPackage{caption=false}{subfig}}
\usepackage{subfig}
\makeatother

\begin{document}
\global\long\def\One{\mathbbm{1}}
\global\long\def\ii{\mathrm{i}}
\global\long\def\spec{\operatorname{spec}}
\global\long\def\sn{\operatorname{sn}}
\global\long\def\sign{\operatorname{sign}}
\global\long\def\erf{\operatorname{erf}}
\global\long\def\Tr{\operatorname{Tr}}
\global\long\def\diag{\operatorname{diag}}
\global\long\def\Uone{\mathrm{U}\!\left(1\right)}
\global\long\def\SUthree{\mathrm{SU}\!\left(3\right)}

\title{Spectral properties and chiral symmetry violations\\
of (staggered) domain wall fermions in the Schwinger model}

\author{Christian Hoelbling}

\address{Department of Physics, University of Wuppertal, D-42119 Wuppertal,
Germany}

\author{Christian Zielinski}

\address{Division of Mathematical Sciences, Nanyang Technological University,
Singapore 637371}

\address{Department of Physics, University of Wuppertal, D-42119 Wuppertal,
Germany}
\begin{abstract}
We follow up on a suggestion by Adams and construct explicit domain
wall fermion operators with staggered kernels. We compare different
domain wall formulations, namely the standard construction as well
as Boriçi's modified and Chiu's optimal construction, utilizing both
Wilson and staggered kernels. In the process, we generalize the staggered
kernels to arbitrary even dimensions and introduce both truncated
and optimal staggered domain wall fermions. Some numerical investigations
are carried out in the $\left(1+1\right)$-dimensional setting of
the Schwinger model, where we explore spectral properties of the bulk,
effective and overlap Dirac operators in the free-field case, on quenched
thermalized gauge configurations and on smooth topological configurations.
We compare different formulations using the effective mass, deviations
from normality and violations of the Ginsparg-Wilson relation as measures
of chirality.
\end{abstract}



\maketitle
{\footnotesize{}\tableofcontents{}}{\footnotesize \par}

\section{Introduction}

Chiral symmetry plays a crucial role in the understanding of hadron
phenomenology and the low-energy dynamics of quantum chromodynamics
(QCD). On the lattice, the overlap construction \cite{Narayanan:1992wx,Narayanan:1993ss,Narayanan:1993sk,Narayanan:1994gw,Neuberger:1997fp,Neuberger:1998my,Neuberger:1998wv}
allows one to implement a fermion operator with exact chiral symmetry
\cite{Ginsparg:1981bj,Hasenfratz:1998ri,Luscher:1998pqa}, thus evading
the Nielsen-Ninomiya theorem \cite{Nielsen:1980rz,Nielsen:1981hk,Nielsen:1981xu,Friedan:1982nk}.
In practice, the use of overlap fermions is limited by the fact that
they generically require a factor of $\mathcal{O}\left(10\text{--}100\right)$
more computational resources than Wilson fermions and tunneling between
topological sectors is severely suppressed even at moderate lattice
spacings \cite{Egri:2005cx,Fukaya:2006vs,Cundy:2008zc,Cundy:2011fz}.

Domain wall fermions \cite{Kaplan:1992bt,Shamir:1993zy,Furman:1994ky}
offer an alternative by formulating fermions with approximate chiral
symmetry in $d$ dimensions by means of massive interacting fermions
in $d+1$ dimensions ($d=2,4$). The limit of an infinite extension
of the extra dimension can again be expressed as an overlap operator
with exact chiral symmetry. For a finite extent, domain wall fermions
can then be seen as a truncation of overlap fermions. They offer the
possibility of reducing computational cost and are well suited for
parallel implementations. This comes at the price of replacing the
exact chiral symmetry by an approximate one. It is expected that chiral
symmetry violations are exponentially suppressed \cite{Shamir:1993zy,Vranas:1997da,Aoki:1997xg,Kikukawa:1997tf},
although in practice this suppression can still require large extents
of the extra dimension \cite{Chen:1998xw,Blum:1998ud,Fleming:1999eq,Hernandez:2000iw,Aoki:2000pc}.
However, these violations also facilitate the tunneling between topological
sectors.

Domain wall fermions are typically formulated with a Wilson kernel
\cite{Wilson:1975id}. Only recently has it been clarified by Adams
\cite{Adams:2009eb,Adams:2010gx} how to utilize the computationally
more efficient staggered fermions \cite{Kogut:1974ag,Banks:1975gq,Banks:1976ia,Susskind:1976jm}
in its place by giving staggered fermions a flavor-dependent mass;
see also Refs.~\cite{Golterman:1984cy,Hoelbling:2010jw,SharpeWorkshop}.
Subsequent numerical work \cite{deForcrand:2011ak,deForcrand:2012bm,Durr:2013gp,Adams:2013tya}
focused on the properties of these staggered Wilson fermions and their
use as a kernel for an overlap construction \cite{Adams:2010gx}.
The possibility of staggered domain wall fermions, which was also
suggested in Ref.~\cite{Adams:2010gx}, has however not been investigated
any further. The present work is meant as a first step in closing
this gap. We give explicit constructions of staggered domain wall
fermions and compare their spectral and chiral symmetry breaking properties
to those of traditional domain wall fermions with Wilson fermions
in the context of the Schwinger model \cite{Schwinger:1962tp}.

While we are eventually interested in QCD, the Schwinger model, i.e.~$\left(1+1\right)$-dimensional
quantum electrodynamics (QED), retains enough properties of QCD. In
particular, we find confinement and topological structure, making
it useful for conceptual investigations. On the other hand, it is
numerically simple enough to allow the computation of the complete
eigenvalue spectrum of fermion operators on nontrivial background
configurations. Moreover, the study of fermions in $1+1$ dimensions
naturally arises e.g.~in the low-energy description of conducting
electrons in metals, see Ref.~\cite{tsvelik2005quantum}.

This paper is organized as follows. In Sec.~\ref{sec:Kernel-operators},
we discuss the kernel operators, among them generalizations of staggered
Wilson fermions in an arbitrary even number of dimensions. In Sec.~\ref{sec:Domain-wall-fermions},
the construction of (staggered) domain wall fermions and their variations
are given, in Sec.~\ref{sec:Effective-Dirac-operator} we introduce
the effective Dirac operators and discuss the limiting overlap operators,
in Sec.~\ref{sec:Setting} we explain our approach of carrying out
the numerical calculations and in Sec.~\ref{sec:Numerical-results}
we discuss the numerical results. In Sec.~\ref{sec:Conclusions},
we conclude our work and give an outlook.

\section{Kernel operators \label{sec:Kernel-operators}}

We begin by giving a quick review of the kernel operators we are considering,
namely Wilson and staggered Wilson fermions. Here and in the following,
we are mostly interested in the $\left(1+1\right)$-dimensional case
($d=2$), but where convenient we write down the general $d$-dimensional
expressions.

\subsection{Wilson kernel}

For Wilson fermions \cite{Wilson:1974sk}, the Dirac operator reads
\begin{equation}
D_{\mathsf{w}}\left(m_{\mathsf{f}}\right)=\gamma_{\mu}\nabla_{\mu}+m_{\mathsf{f}}+W_{\mathsf{w}}.
\end{equation}
Here the $\gamma_{\mu}$ matrices refer to a representation of the
Dirac algebra $\left\{ \gamma_{\mu},\gamma_{\nu}\right\} =2\delta_{\mu,\nu}\One$
with $\mu\in\left\{ 1,\ldots,d\right\} $, $\delta_{\mu,\nu}$ to
the Kronecker delta, $\nabla_{\mu}$ to the covariant central finite
difference operator and $m_{\mathsf{f}}$ to the bare fermion mass.
The Wilson term reads 
\begin{equation}
W_{\mathsf{w}}=-\frac{ar}{2}\Delta
\end{equation}
with lattice spacing $a$, Wilson parameter $r\in\left(0,1\right]$
and the covariant lattice Laplacian $\Delta$. We note that $D_{\mathsf{w}}^{\dagger}=\gamma_{d+1}D_{\mathsf{w}}\gamma_{d+1}$,
where $\gamma_{d+1}$ is the chirality matrix, and that the $W_{\mathsf{w}}$
term breaks chiral symmetry explicitly. In terms of the parallel transport
\begin{align}
T_{\mu}\Psi\left(x\right) & =U_{\mu}\left(x\right)\Psi\left(x+a\hat{\mu}\right)
\end{align}
we have the following definitions: 
\begin{align}
\nabla_{\mu} & =\frac{1}{2a}\left(T_{\mu}-T_{\mu}^{\dagger}\right),\\
C_{\mu} & =\frac{1}{2}\left(T_{\mu}+T_{\mu}^{\dagger}\right),\\
\Delta & =\frac{2}{a^{2}}\sum_{\mu}\left(C_{\mu}-\One\right).
\end{align}
Through the Wilson term $W_{\mathsf{w}}$, the doublers acquire a
mass $\mathcal{O}\left(a^{-1}\right)$. In the continuum limit, the
number of flavors is then reduced from $2^{d}$ to one physical flavor.

\subsection{Staggered Wilson fermions}

Following Refs.~\cite{Golterman:1984cy,Adams:2009eb,Adams:2010gx,Hoelbling:2010jw,deForcrand:2012bm,Durr:2013gp},
in $d=4$ dimensions a staggered Wilson operator can be written as
\begin{equation}
D_{\mathsf{sw}}\left(m_{\mathsf{f}}\right)=D_{\mathsf{st}}+m_{\mathsf{f}}+W_{\mathsf{st}}
\end{equation}
with staggered Dirac operator
\begin{align}
D_{\mathsf{st}} & =\eta_{\mu}\nabla_{\mu},\\
\eta_{\mu}\chi\left(x\right) & =\left(-1\right)^{\sum_{\nu<\mu}x_{\nu}/a}\chi\left(x\right)
\end{align}
and bare fermion mass $m_{\mathsf{f}}$. The staggered Wilson term
is an operator that, up to discretization terms, is trivial in spin
but splits the different flavors. We also require the Dirac operator
to have a real determinant. The original suggestion by Adams \cite{Adams:2009eb,Adams:2010gx}
reads
\begin{equation}
W_{\mathsf{st}}=\frac{r}{a}\left(\One+\Gamma_{1234}C_{1234}\right)\label{eq:AdamsMassTerm}
\end{equation}
with a Wilson-like parameter $r>0$ and operators
\begin{align}
\Gamma_{1234}\chi\left(x\right) & =\left(-1\right)^{\sum_{\mu}x_{\mu}/a}\chi\left(x\right),\\
C_{1234} & =\eta_{1}\eta_{2}\eta_{3}\eta_{4}\left(C_{1}C_{2}C_{3}C_{4}\right)_{\mathsf{sym}}.
\end{align}
In the spin-flavor basis, this term has the form
\begin{equation}
W_{\mathsf{st}}\sim\frac{r}{a}\One\otimes\left(\One-\xi_{5}\right)+\mathcal{O}\left(a\right),
\end{equation}
which splits the four flavors into pairs of two according to their
``flavor chirality'', i.e.~the eigenbasis of $\xi_{5}$. The notation
$A\sim B$ means that $B$ corresponds to the respective spin$\,\otimes\,$flavor
interpretation \cite{Golterman:1984cy} of $A$ up to proportionality,
while the $\xi_{\mu}$ are a representation of the Dirac algebra in
flavor space. The determinant of $D_{\mathsf{sw}}$ is real due to
the $\epsilon$-Hermiticity
\begin{equation}
D_{\mathsf{sw}}^{\dagger}=\epsilon D_{\mathsf{sw}}\epsilon
\end{equation}
with
\begin{equation}
\epsilon\left(x\right)=\left(-1\right)^{\sum_{\mu}x_{\mu}/a}.
\end{equation}
In four dimensions, one may also split the flavors with respect to
the eigenbasis of different elements of the flavor Dirac algebra \cite{Golterman:1984cy,Hoelbling:2010jw}.
To retain $\epsilon$ Hermiticity and, thus, a real determinant, the
flavor structure of the mass term needs to be restricted to a sum
of products of an even number of $\xi_{\mu}$. A single flavor staggered
fermion in four dimensions can, thus, be obtained by e.g.
\begin{align}
W_{\mathsf{st}} & =\frac{r}{a}\left(2\cdot\One+W_{\mathsf{st}}^{12}+W_{\mathsf{st}}^{34}\right),\label{eq:DefD4MassTerm}\\
W_{\mathsf{st}}^{\mu\nu} & =\ii\,\Gamma_{\mu\nu}C_{\mu\nu},\label{eq:SingleMassTerm}
\end{align}
where the operators $\Gamma_{\mu\nu}$ and $C_{\mu\nu}$ are given
by
\begin{align}
\Gamma_{\mu\nu}\chi\left(x\right) & =\varepsilon_{\mu\nu}\left(-1\right)^{\left(x_{\mu}+x_{\nu}\right)/a}\chi\left(x\right),\\
C_{\mu\nu} & =\eta_{\mu}\eta_{\nu}\cdot\frac{1}{2}\left(C_{\mu}C_{\nu}+C_{\nu}C_{\mu}\right)\quad\left(\textrm{no sum}\right).
\end{align}
Here and in the following, $\varepsilon_{\mu_{1}\cdots\mu_{N}}$ refers
to the Levi-Civita symbol. To interpret the mass term in Eq.~\eqref{eq:SingleMassTerm},
note that
\begin{align}
\Gamma_{\mu\nu} & \sim\gamma_{\mu}\gamma_{\nu}\otimes\xi_{\mu}\xi_{\nu},\\
C_{\mu\nu} & \sim\gamma_{\mu}\gamma_{\nu}\otimes\One,\\
\epsilon & \sim\gamma_{5}\otimes\xi_{5},
\end{align}
up to discretization terms. As a result, we find for the staggered
Wilson term
\begin{equation}
W_{\mathsf{st}}^{\mu\nu}\sim\One\otimes\sigma_{\mu\nu}+\mathcal{O}\left(a\right)
\end{equation}
with $\sigma_{\mu\nu}=\ii\,\xi_{\mu}\xi_{\nu}$. This implies that
the number of physical fermion species of $W_{\mathsf{st}}$ is reduced
to one by giving all but a single flavor a mass $\mathcal{O}\left(a^{-1}\right)$.

These results can be formulated in an arbitrary even number of dimensions
$d$, where we can write a single flavor mass term as
\begin{equation}
W_{\mathsf{st}}=\frac{r}{a}\sum_{k=1}^{d/2}\left(\One+W_{\mathsf{st}}^{\left(2k-1\right)\left(2k\right)}\right)\label{eq:DefGeneralMassTerm}
\end{equation}
and Eq.~\eqref{eq:DefD4MassTerm} now follows as the special case
$d=4$. To construct a more general mass term, we define 
\begin{equation}
W_{\mathsf{st}}^{\mu_{1}\cdots\mu_{2n}}=\ii^{n}\,\Gamma_{\mu_{1}\cdots\mu_{2n}}C_{\mu_{1}\cdots\mu_{2n}},
\end{equation}
for an arbitrary $n\le d/2$, where
\begin{align}
\Gamma_{\mu_{1}\cdots\mu_{2n}}\chi\left(x\right) & =\varepsilon_{\mu_{1}\cdots\mu_{2n}}\left(-1\right)^{\sum_{i=1}^{2n}x_{\mu_{i}}/a}\chi\left(x\right),\\
C_{\mu_{1}\cdots\mu_{2n}} & =\eta_{\mu_{1}}\cdots\eta_{\mu_{2n}}\left(C_{\mu_{1}}\cdots C_{\mu_{2n}}\right)_{\mathsf{sym}}.
\end{align}
In the spin$\,\otimes\,$flavor interpretation, we find
\begin{align}
\Gamma_{\mu_{1}\cdots\mu_{2n}} & \sim\left(\gamma_{\mu_{1}}\ldots\gamma_{\mu_{2n}}\right)\otimes\left(\xi_{\mu_{1}}\ldots\xi_{\mu_{2n}}\right),\\
C_{\mu_{1}\cdots\mu_{2n}} & \sim\left(\gamma_{\mu_{1}}\ldots\gamma_{\mu_{2n}}\right)\otimes\One,\\
W_{\mathsf{st}}^{\mu_{1}\cdots\mu_{2n}} & \sim\One\otimes\left(\ii^{n}\,\xi_{\mu_{1}}\ldots\xi_{\mu_{2n}}\right),
\end{align}
up to discretization terms. In addition, the new mass terms fulfill
the $\epsilon$-Hermiticity relation
\begin{equation}
W^{\dagger}=\epsilon\,W\epsilon,\qquad W\equiv W_{\mathsf{st}}^{\mu_{1}\cdots\mu_{2n}}.
\end{equation}
We can, thus, replace Eq.~\eqref{eq:DefGeneralMassTerm} by a generic
\begin{equation}
W_{\mathsf{st}}=\sum_{n=1}^{d/2}\sum_{\boldsymbol{\mu}_{n}}\frac{r_{\boldsymbol{\mu}_{n}}}{a}\left(\One+W_{\mathsf{st}}^{\boldsymbol{\mu}_{n}}\right),\label{eq:DefMoreGeneralMassTerm}
\end{equation}
where $r_{\boldsymbol{\mu}_{n}}\geq0$ and the sum is over all multi-indices
$\boldsymbol{\mu}_{n}=\left(\mu_{1},\ldots,\mu_{2n}\right)$ with
$1\leq\mu_{i}\leq d$ for all $i$ with $1\leq i\leq2n$ \footnote{Note that not all of the possible combinations of mass terms in Eq.~\eqref{eq:DefMoreGeneralMassTerm}
are practically useful.}. Adams' original mass term in $d=4$ dimensions in Eq.~\eqref{eq:AdamsMassTerm}
then follows from setting $r_{1234}=r>0$ and $r_{\boldsymbol{\mu}_{n}}=0$
otherwise.

For the $d=2$ case that we will consider in the numerical part of
this paper, the definition is essentially unique and reads
\begin{equation}
W_{\mathsf{st}}=\frac{r}{a}\left(\One+W_{\mathsf{st}}^{12}\right).
\end{equation}
In this case the reduction is from two staggered flavors to a single
physical one.

Like in the Wilson case, all possible $W_{\mathsf{st}}$ terms break
chiral symmetry explicitly. Furthermore, there may be additional counterterms
if too many of the staggered symmetries are broken \cite{SharpeWorkshop}.

\section{Domain wall fermions \label{sec:Domain-wall-fermions}}

After having introduced the kernel operators, we now move on to the
domain wall fermion Dirac operators. Originally proposed by Kaplan
\cite{Kaplan:1992bt}, then refined by Shamir and Furman \cite{Shamir:1993zy,Furman:1994ky},
the domain wall construction implements approximately massless fermions
in $d$ dimensions by means of a $\left(d+1\right)$-dimensional theory.
Equivalently, domain wall fermions can be understood as a tower of
$N_{s}$ fermions in $d$ dimensions with a particular flavor structure.

We now give a quick summary of the well-known $\left(d+1\right)$-dimensional
formulations. For the remainder of the paper we fix the $d$-dimensional
lattice spacing to $a=1$ and the (staggered) Wilson parameter to
$r=1$.

\subsection{Standard construction}

We begin with the standard construction. First, let us define
\begin{equation}
D_{\mathsf{w}}^{\pm}=a_{d+1}D_{\mathsf{w}}\left(-M_{0}\right)\pm\One,
\end{equation}
where we explicitly write out the lattice spacing $a_{d+1}$ in the
extra dimension. The parameter $M_{0}$ is the so-called domain wall
height and must be suitably chosen for the description of a single
flavor. In the free-field case, we have $M_{0}\in\left(0,2r\right)$.

The Dirac operator reads
\begin{equation}
\overline{\Psi}D_{\mathsf{dw}}\Psi=\sum_{s=1}^{N_{s}}\overline{\Psi}_{s}\left[D_{\mathsf{w}}^{+}\Psi_{s}-P_{-}\Psi_{s+1}-P_{+}\Psi_{s-1}\right]
\end{equation}
with $\left(d+1\right)$-dimensional fermion fields $\overline{\Psi}$,
$\Psi$ and chiral projectors $P_{\pm}=\frac{1}{2}\left(\One\pm\gamma_{d+1}\right)$.
Here and in the following, the index $s$ refers to the additional
spatial (or equivalently flavor) coordinate. The gauge links are taken
to be the identity matrix along the additional coordinate. Furthermore,
we impose the following boundary conditions,
\begin{align}
P_{+}\left(\Psi_{0}+m\Psi_{N_{s}}\right) & =0,\\
P_{-}\left(\Psi_{N_{s}+1}+m\Psi_{1}\right) & =0,
\end{align}
where $m$ is related to the bare fermion mass, see Eq.~\eqref{eq:InducedMass}.
We note that in the special case of $m=0$ we find Dirichlet boundary
conditions in the extra dimension, while for $m=\pm1$ one recovers
(anti\nobreakdash-)periodic boundary conditions. If we write down
the Dirac operator in the extra dimension explicitly, we find
\begin{equation}
D_{\mathsf{dw}}=\left(\begin{array}{ccccc}
D_{\mathsf{w}}^{+} & -P_{-} &  &  & mP_{+}\\
-P_{+} & D_{\mathsf{w}}^{+} & -P_{-}\\
 & \ddots & \ddots & \ddots\\
 &  & -P_{+} & D_{\mathsf{w}}^{+} & -P_{-}\\
mP_{-} &  &  & -P_{+} & D_{\mathsf{w}}^{+}
\end{array}\right).\label{eq:DefDdw}
\end{equation}
One possibility of constructing the $d$-dimensional fermion fields
from the boundary is via
\begin{equation}
q=P_{+}\Psi_{N_{s}}+P_{-}\Psi_{1},\quad\overline{q}=\overline{\Psi}_{1}P_{+}+\overline{\Psi}_{N_{s}}P_{-}.\label{eq:DefLightFermions}
\end{equation}

Let us also define the reflection operator along the extra dimension
\begin{equation}
R=\left(\begin{array}{ccc}
 &  & \One\\
 & \iddots\\
\One
\end{array}\right).
\end{equation}
We find that $D_{\mathsf{dw}}$ is $R\gamma_{d+1}$-Hermitian
\begin{equation}
D_{\mathsf{dw}}^{\dagger}=R\gamma_{d+1}\cdot D_{\mathsf{dw}}\cdot R\gamma_{d+1},\label{eq:RGHermiticity}
\end{equation}
which ensures that $\det D_{\mathsf{dw}}\in\mathbb{R}$ and the applicability
of importance sampling techniques.

Besides this canonical formulation, several variations of domain wall
fermions have been proposed.

\subsection{Boriçi's construction}

One of them is the construction by Boriçi \cite{Borici:1999zw}, which
follows from the original proposal by the replacements
\begin{align}
P_{+}\Psi_{s-1} & \to-D_{\mathsf{w}}^{-}P_{+}\Psi_{s-1},\\
P_{-}\Psi_{s+1} & \to-D_{\mathsf{w}}^{-}P_{-}\Psi_{s+1}.
\end{align}
Note that this is an $\mathcal{O}\left(a_{d+1}\right)$ modification.
The Dirac operator in its full form reads
\begin{equation}
\overline{\Psi}D_{\mathsf{dw}}\Psi=\sum_{s=1}^{N_{s}}\overline{\Psi}_{s}\left[D_{\mathsf{w}}^{+}\Psi_{s}+D_{\mathsf{w}}^{-}P_{-}\Psi_{s+1}+D_{\mathsf{w}}^{-}P_{+}\Psi_{s-1}\right]
\end{equation}
or explicitly
\begin{multline}
D_{\mathsf{dw}}=\\
\left(\begin{array}{ccccc}
D_{\mathsf{w}}^{+} & D_{\mathsf{w}}^{-}P_{-} &  &  & -mD_{\mathsf{w}}^{-}P_{+}\\
D_{\mathsf{w}}^{-}P_{+} & D_{\mathsf{w}}^{+} & D_{\mathsf{w}}^{-}P_{-}\\
 & \ddots & \ddots & \ddots\\
 &  & D_{\mathsf{w}}^{-}P_{+} & D_{\mathsf{w}}^{+} & D_{\mathsf{w}}^{-}P_{-}\\
-mD_{\mathsf{w}}^{-}P_{-} &  &  & D_{\mathsf{w}}^{-}P_{+} & D_{\mathsf{w}}^{+}
\end{array}\right).\label{eq:DefDdwBorici}
\end{multline}
Furthermore, Eq.~\eqref{eq:DefLightFermions} generalizes to
\begin{align}
q & =P_{+}\Psi_{N_{s}}+P_{-}\Psi_{1},\\
\overline{q} & =-\overline{\Psi}_{1}D_{\mathsf{w}}^{-}P_{+}-\overline{\Psi}_{N_{s}}D_{\mathsf{w}}^{-}P_{-}
\end{align}
and Eq.~\eqref{eq:RGHermiticity} to
\begin{align}
\left(\mathcal{D}^{-1}D_{\mathsf{dw}}\right)^{\dagger} & =R\gamma_{d+1}\cdot\left(\mathcal{D}^{-1}D_{\mathsf{dw}}\right)\cdot R\gamma_{d+1},\label{eq:RGHermiticityBorici}\\
\mathcal{D} & =\One_{N_{s}}\otimes D_{\mathsf{w}}^{-}
\end{align}
(see Ref.~\cite{Brower:2004xi}). This formulation is also known
as ``truncated overlap fermions'' as the corresponding $d$-dimensional
effective operator equals the polar decomposition approximation \cite{Neuberger:1998my,higham1994matrix}
of order $N_{s}/2$ of Neuberger's overlap operator (for even $N_{s}$).

\subsection{Optimal construction \label{subsec:Optimal-construction}}

The last modification we consider are the optimal domain wall fermions
proposed by Chiu \cite{Chiu:2002ir,Chiu:2002kj}. The idea is to modify
$D_{\mathsf{dw}}$ in such a way, that the effective Dirac operator
is expressed through Zolotarev's optimal rational approximation of
the sign function \cite{zolotarev1877application,akhiezer1990elements,achieser2013theory}
(see also Refs.~\cite{vandenEshof:2002ms,Chiu:2002eh}). In the following,
we quote the central formulas of the construction given in Ref.~\cite{Chiu:2002ir}.

Starting from Boriçi's construction, the Dirac operator is modified
by introducing weight factors
\begin{multline}
\overline{\Psi}D_{\mathsf{dw}}\Psi=\sum_{s=1}^{N_{s}}\overline{\Psi}_{s}D_{\mathsf{w}}^{+}\left(s\right)\Psi_{s}\\
+\sum_{s=1}^{N_{s}}\overline{\Psi}_{s}\left[D_{\mathsf{w}}^{-}\left(s\right)P_{-}\Psi_{s+1}+D_{\mathsf{w}}^{-}\left(s\right)P_{+}\Psi_{s-1}\right],
\end{multline}
where
\begin{equation}
D_{\mathsf{w}}^{\pm}\left(s\right)=a_{d+1}\omega_{s}D_{\mathsf{w}}\left(-M_{0}\right)\pm\One.
\end{equation}
The weight factors $\omega_{s}$ are given by
\begin{equation}
\omega_{s}=\frac{1}{\lambda_{\mathsf{min}}}\sqrt{1-\kappa^{\prime2}\sn\left(v_{s},\kappa^{\prime}\right)}\label{eq:OptWeights}
\end{equation}
with $\sn\left(v_{s},\kappa^{\prime}\right)$ being the corresponding
Jacobi elliptic function with argument $v_{s}$ and modulus $\kappa^{\prime}$.
The modulus is defined by
\begin{equation}
\kappa^{\prime}=\sqrt{1-\lambda_{\mathsf{min}}^{2}/\lambda_{\mathsf{max}}^{2}}
\end{equation}
and $\lambda_{\mathsf{min}}^{2}$ ($\lambda_{\mathsf{max}}^{2}$)
is the respective smallest (largest) eigenvalue of $H_{\mathsf{w}}^{2}$
with
\begin{equation}
H_{\mathsf{w}}=\gamma_{d+1}D_{\mathsf{w}}\left(-M_{0}\right).
\end{equation}
The argument $v_{s}$ reads
\begin{equation}
v_{s}=\left(-1\right)^{s-1}M\sn^{-1}\left(\sqrt{\frac{1+3\lambda}{\left(1+\lambda\right)^{3}}},\sqrt{1-\lambda^{2}}\right)+\left\lfloor \frac{s}{2}\right\rfloor \frac{2K^{\prime}}{N_{s}},
\end{equation}
where
\begin{align}
\lambda & =\prod_{\ell=1}^{N_{s}}\frac{\Theta^{2}\left(2\ell K^{\prime}/N_{s},\kappa^{\prime}\right)}{\Theta^{2}\left(\left(2\ell-1\right)K^{\prime}/N_{s},\kappa^{\prime}\right)},\\
M & =\prod_{\ell=1}^{\left\lfloor N_{s}/2\right\rfloor }\frac{\sn^{2}\left(\left(2\ell-1\right)K^{\prime}/N_{s},\kappa^{\prime}\right)}{\sn^{2}\left(2\ell K^{\prime}/N_{s},\kappa^{\prime}\right)}.
\end{align}
Here $\left\lfloor \cdot\right\rfloor $ refers to the floor function,
$K^{\prime}=K\left(\kappa^{\prime}\right)$ to the complete elliptic
integral of the first kind with
\begin{equation}
K\left(k\right)=\intop_{0}^{\pi/2}\frac{\mathrm{d}\theta}{\sqrt{1-k^{2}\sin^{2}\theta}}.
\end{equation}
Furthermore we introduced the elliptic Theta function via
\begin{equation}
\Theta\left(w,k\right)=\vartheta_{4}\left(\frac{\pi w}{2K},k\right)
\end{equation}
with $K=K\left(k\right)$ and elliptic theta functions $\vartheta_{i}$
\cite{abramowitz2012handbook}. Some reference values for the weight
factors $\left\{ \omega_{s}\right\} $ can be found in App.~\ref{sec:Optimal-weights}.

Similarly to Boriçi's construction, Eq.~\eqref{eq:DefLightFermions}
now generalizes to
\begin{align}
q & =P_{+}\Psi_{N_{s}}+P_{-}\Psi_{1},\\
\overline{q} & =-\overline{\Psi}_{1}D_{\mathsf{w}}^{-}\left(1\right)P_{+}-\overline{\Psi}_{N_{s}}D_{\mathsf{w}}^{-}\left(N_{s}\right)P_{-}
\end{align}
and Eq.~\eqref{eq:RGHermiticity} again to Eq.~\eqref{eq:RGHermiticityBorici},
but now with
\begin{equation}
\mathcal{D}=\diag\left[D_{\mathsf{w}}^{-}\left(1\right),\ldots,D_{\mathsf{w}}^{-}\left(N_{s}\right)\right],
\end{equation}
as pointed out in Ref.~\cite{Brower:2004xi}. Optimal domain wall
fermions have been and are still extensively used. Some of the results
obtained can be found in Refs.~\cite{Chen:2011qy,Hsieh:2011qx,Chiu:2011dz,Chiu:2011bm,Chen:2012jya,Chiu:2012jm}.

We also note that there is a modified construction of optimal domain
wall fermions \cite{Chiu:2015sea}, which is reflection-symmetric
along the fifth dimension. For completeness we point out that all
the preceding domain wall fermion formulations can be seen as special
cases of Möbius domain wall fermions \cite{Brower:2004xi,Brower:2005qw,Brower:2009sb}.

\subsection{Staggered formulations}

As proposed in Ref.~\cite{Adams:2010gx}, we can use the staggered
Wilson kernel to formulate a staggered version of domain wall fermions.
We can write the Dirac operator in a general $d$-dimensional form
as
\begin{equation}
\overline{\Upsilon}D_{\mathsf{sdw}}\Upsilon=\sum_{s=1}^{N_{s}}\overline{\Upsilon}_{s}\left[D_{\mathsf{sw}}^{+}\Upsilon_{s}-P_{-}\Upsilon_{s+1}-P_{+}\Upsilon_{s-1}\right],
\end{equation}
where $\Upsilon$ refers to the staggered fermion field. Like in the
Wilson case we define
\begin{equation}
D_{\mathsf{sw}}^{\pm}=a_{d+1}D_{\mathsf{sw}}\left(-M_{0}\right)\pm\One.
\end{equation}
The chiral projectors are given by $P_{\pm}=\frac{1}{2}\left(\One\pm\epsilon\right)$,
where $\epsilon^{2}=\One$. Here we have $\epsilon\sim\gamma_{d+1}\otimes\xi_{d+1}$,
which reduces to $\epsilon\sim\gamma_{d+1}\otimes\One$ on the physical
species. One can easily verify the $R\epsilon$-Hermiticity of $D_{\mathsf{sdw}}$.
Note that we follow a sign convention in where our $D_{\mathsf{sdw}}$
is in full analogy to $D_{\mathsf{dw}}$, while in Ref.~\cite{Adams:2010gx}
a slightly different convention is used. The staggered domain wall
Dirac operator $D_{\mathsf{sdw}}$ can be constructed from $D_{\mathsf{dw}}$
by the replacement rule given in Ref.~\cite{Adams:2010gx}, which
we write down in a general $d$-dimensional form as
\begin{equation}
\gamma_{d+1}\to\epsilon,\quad D_{\mathsf{w}}\to D_{\mathsf{sw}}.\label{eq:ReplacementRule}
\end{equation}

Using the replacement rule in Eq.~\eqref{eq:ReplacementRule}, we
can also generalize Boriçi's and the optimal construction to the case
of a staggered Wilson kernel. This gives rise to previously not considered
truncated staggered domain wall fermions with the Dirac operator
\begin{multline}
\overline{\Upsilon}D_{\mathsf{sdw}}\Upsilon=\sum_{s=1}^{N_{s}}\overline{\Upsilon}_{s}D_{\mathsf{sw}}^{+}\Upsilon_{s}\\
+\sum_{s=1}^{N_{s}}\overline{\Upsilon}_{s}\left[D_{\mathsf{sw}}^{-}P_{-}\Upsilon_{s+1}+D_{\mathsf{sw}}^{-}P_{+}\Upsilon_{s-1}\right]
\end{multline}
as well as optimal staggered domain wall fermions
\begin{multline}
\overline{\Upsilon}D_{\mathsf{sdw}}\Upsilon=\sum_{s=1}^{N_{s}}\overline{\Upsilon}_{s}D_{\mathsf{sw}}^{+}\left(s\right)\Upsilon_{s}\\
+\sum_{s=1}^{N_{s}}\overline{\Upsilon}_{s}\left[D_{\mathsf{sw}}^{-}\left(s\right)P_{-}\Upsilon_{s+1}+D_{\mathsf{sw}}^{-}\left(s\right)P_{+}\Upsilon_{s-1}\right],
\end{multline}
where $D_{\mathsf{sw}}^{\pm}\left(s\right)=a_{d+1}\omega_{s}D_{\mathsf{sw}}\left(-M_{0}\right)\pm\One$
and the weight factors $\omega_{s}$ are given by Eq.~\eqref{eq:OptWeights}
for the kernel $H_{\mathsf{sw}}=\epsilon D_{\mathsf{sw}}\left(-M_{0}\right)$.

\section{Effective Dirac operator \label{sec:Effective-Dirac-operator}}

To understand the relation between the $\left(d+1\right)$-dimensional
fermions and the light $d$-dimensional fields $q$, $\overline{q}$
at the boundary, we introduce the effective $d$-dimensional Dirac
operator as derived in Refs.~\cite{Neuberger:1997bg,Kikukawa:1999sy,Kikukawa:1999dk,Borici:1999zw}
(see also Refs.~\cite{Shamir:1998ww,Edwards:2000qv}). In the following,
we give a short summary, following Refs.~\cite{Kikukawa:1999sy,Borici:1999zw,Edwards:2000qv}.

\subsection{Derivation}

The low energy effective $d$-dimensional action
\begin{equation}
S_{\mathsf{eff}}=\sum_{x}\overline{q}\left(x\right)D_{\mathsf{eff}}\,q\left(x\right)
\end{equation}
follows after integrating out the $N_{s}-1$ heavy modes. The effective
Dirac operator is defined via the propagator
\begin{equation}
D_{\mathsf{eff}}^{-1}\left(x,y\right)=\left\langle q\left(x\right)\overline{q}\left(y\right)\right\rangle .
\end{equation}
For a suitable choice of $M_{0}$, there is exactly one light and
$N_{s}-1$ heavy Dirac fermions.

In the chiral limit $N_{s}\to\infty$ (at fixed bare coupling $\beta$),
the contribution from the heavy fermions diverges. This bulk contribution
from the $\left(d+1\right)$-dimensional fermions can be canceled
by the introduction of suitable pseudofermion fields. One typically
chooses the fermion action with the replacement $m\to1$ as the action
for the pseudofermions.

Let us begin by defining the Hermitian operators
\begin{align}
H_{\mathsf{w}} & =\gamma_{d+1}D_{\mathsf{w}}\left(-M_{0}\right),\\
H_{\mathsf{m}} & =\gamma_{d+1}D_{\mathsf{m}}\left(-M_{0}\right),
\end{align}
where the kernel operator of standard domain wall fermions is given
by
\begin{equation}
D_{\mathsf{m}}\left(-M_{0}\right)=\frac{D_{\mathsf{w}}\left(-M_{0}\right)}{2\cdot\One+a_{d+1}D_{\mathsf{w}}\left(-M_{0}\right)}.
\end{equation}

The transfer matrix along the extra dimension is given by
\begin{equation}
T=\frac{T_{-}}{T_{+}},\quad T_{\pm}=\One\pm a_{d+1}H,
\end{equation}
where we use the notation
\begin{equation}
H=\begin{cases}
H_{\mathsf{m}} & \textrm{for standard constr.},\\
H_{\mathsf{w}} & \textrm{for Boriçi's constr.}
\end{cases}\label{eq:DefH}
\end{equation}
Then the effective operator can be written as
\begin{equation}
D_{\mathsf{eff}}=\frac{1+m}{2}\One+\frac{1-m}{2}\gamma_{d+1}\frac{T_{+}^{N_{s}}-T_{-}^{N_{s}}}{T_{+}^{N_{s}}+T_{-}^{N_{s}}}.\label{eq:DefDeff}
\end{equation}
Note that we can rewrite Eq.~\eqref{eq:DefDeff} as
\begin{equation}
D_{\mathsf{eff}}=\left(1-m\right)\left[D_{\mathsf{eff}}\left(0\right)+\frac{m}{1-m}\right],\label{eq:InducedMass}
\end{equation}
where $D_{\mathsf{eff}}\left(0\right)$ denotes the effective operator
$D_{\mathsf{eff}}$ at $m=0$. We can see that the parameter $m$
induces a bare fermion mass of $m/\left(1-m\right)$, see Ref.~\cite{Boyle:2016imm}.

Alternatively, one can also show \cite{Borici:1999zw,Edwards:2000qv}
the relation
\begin{equation}
D_{\mathsf{eff}}=\left(\mathcal{P}^{\intercal}D_{1}^{-1}D_{m}\mathcal{P}\right)_{1,1}\label{eq:ProjMethod}
\end{equation}
with the matrix $\mathcal{P}$ defined as
\begin{equation}
\mathcal{P}=\left(\begin{array}{ccccc}
P_{-} & P_{+}\\
 & P_{-} & P_{+}\\
 &  & \ddots & \ddots\\
 &  &  & P_{-} & P_{+}\\
P_{+} &  &  &  & P_{-}
\end{array}\right)
\end{equation}
and $\mathcal{P}^{-1}=\mathcal{P}^{\intercal}$. Here we used the
shorthand notation $D_{m}=D_{\mathsf{dw}}\left(m\right)$, while the
index stands for the $\left(1,1\right)$ $s$-block of the matrix.

The derivation of the effective operator for optimal domain wall fermions
follows Boriçi's construction after including the weight factors $\left\{ \omega_{s}\right\} $
appropriately \cite{Chiu:2002ir}. By construction the $\sign$-function
approximation equals the optimal rational approximation. It can be
either evaluated directly or via the projection method of Eq.~\eqref{eq:ProjMethod}.

\subsection{The $N_{s}\to\infty$ limit}

In the following, let us specialize to $m=0$. Note that we can rewrite
Eq.~\eqref{eq:DefDeff} using
\begin{equation}
\frac{T_{+}^{N_{s}}-T_{-}^{N_{s}}}{T_{+}^{N_{s}}+T_{-}^{N_{s}}}=\epsilon_{N_{s}/2}\left(a_{d+1}H\right),
\end{equation}
where $\epsilon_{N_{s}/2}$ is Neuberger’s polar decomposition approximation
\cite{Neuberger:1998my,higham1994matrix} of the $\sign$-function.
Therefore, we obtain an overlap operator in the $N_{s}\to\infty$
limit as follows,
\begin{align}
D_{\mathsf{ov}} & =\lim_{N_{s}\to\infty}D_{\mathsf{eff}}\nonumber \\
 & =\frac{1}{2}\One+\frac{1}{2}\gamma_{d+1}\sign H\nonumber \\
 & =\frac{1}{2}\left[\One+D_{-M_{0}}\left(D_{-M_{0}}^{\dagger}D_{-M_{0}}\right)^{-\frac{1}{2}}\right],\label{eq:DefDov}
\end{align}
with $H$ given in Eq.~\eqref{eq:DefH}, $D_{-M_{0}}=D\left(-M_{0}\right)$
and
\begin{equation}
D=\begin{cases}
D_{\mathsf{m}} & \textrm{for standard constr.},\\
D_{\mathsf{w}} & \textrm{for Boriçi's/Chiu's constr.}
\end{cases}\label{eq:DefD}
\end{equation}
The overlap operator satisfies the Ginsparg-Wilson equation
\begin{equation}
\left\{ \gamma_{d+1},D_{\mathsf{ov}}\right\} =2D_{\mathsf{ov}}\gamma_{d+1}D_{\mathsf{ov}}\label{eq:GWR}
\end{equation}
and allows for an exact chiral symmetry. Eq.~\eqref{eq:GWR} also
implies the normality of the overlap operator as can be easily verified.

Comparing Eq.~\eqref{eq:DefDov} to the standard definition of the
overlap operator
\begin{equation}
D_{\mathsf{ov}}=\rho\left[\One+D_{-\rho}\left(D_{-\rho}^{\dagger}D_{-\rho}\right)^{-\frac{1}{2}}\right]
\end{equation}
and using the relation for the effective negative mass parameter
\begin{equation}
\rho=\begin{cases}
M_{0}-\frac{a_{d+1}}{2}M_{0}^{2} & \textrm{for standard constr.},\\
M_{0} & \textrm{for Boriçi's/Chiu's constr.},
\end{cases}\label{eq:Defrho}
\end{equation}
we would obtain a restriction on the domain wall height $M_{0}$ from
$\rho=1/2$. This can be avoided by simply rescaling $D_{\mathsf{eff}}$
by a factor $\varrho=2\rho$, so that—up to discretization effects—the
low-lying eigenvalues of the kernel remain invariant under the effective
operator projection in the free-field case. This is also illustrated
in Fig.~\ref{fig:free-dov}, which we elaborate on in Sec.~\ref{subsec:Free-field-case}.
Consequently, we will employ this rescaling in all our numerical investigations.

\subsection{Approximate sign functions}

\begin{figure}[H]
\begin{centering}
\subfloat[$N_{s}=2$]{\includegraphics[width=1\columnwidth]{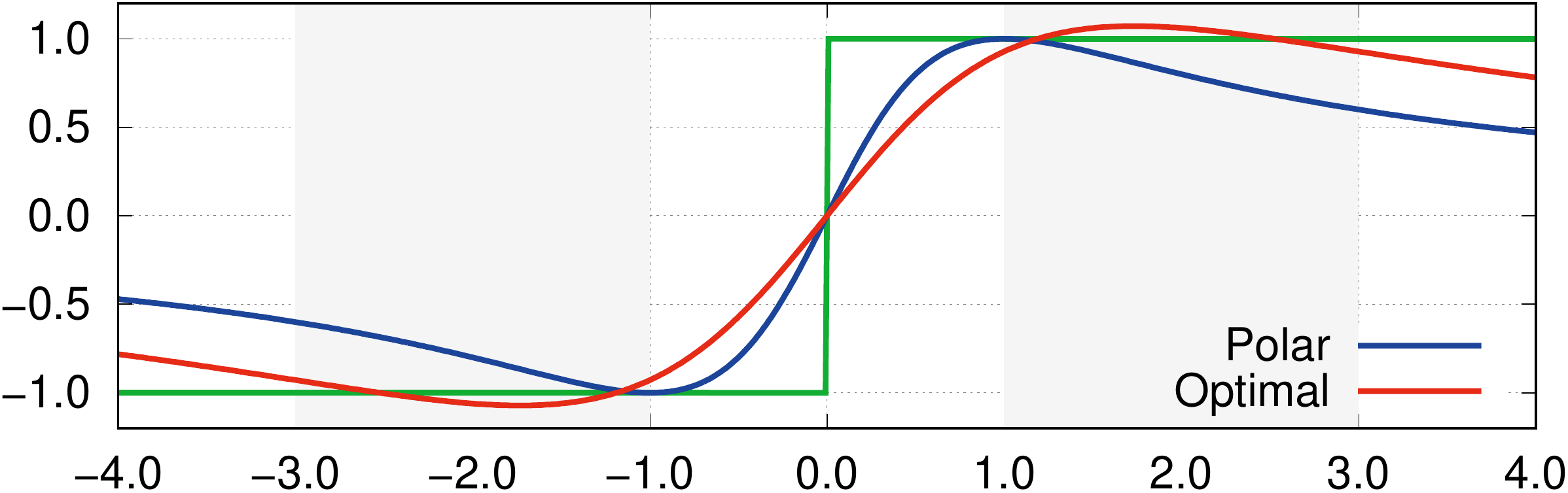}

}
\par\end{centering}
\begin{centering}
\subfloat[$N_{s}=6$]{\includegraphics[width=1\columnwidth]{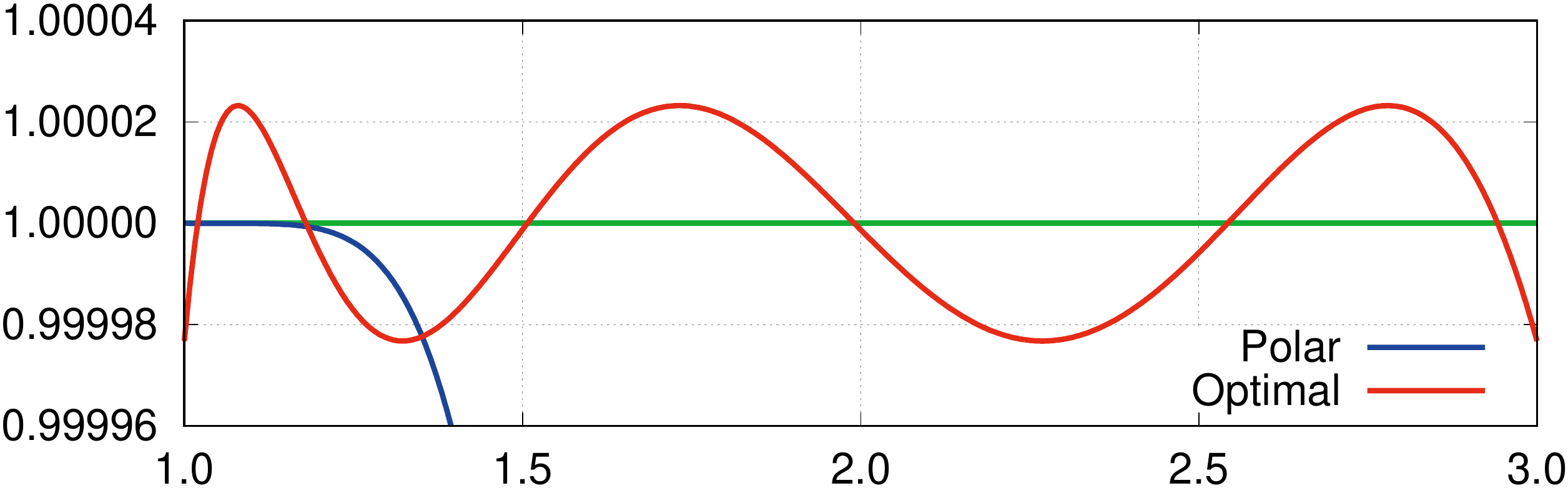}

}
\par\end{centering}
\caption{Approximation of $\protect\sign\left(z\right)$ by $r\left(z\right)$.
The optimal construction was done for $\lambda_{\mathsf{min}}=1$,
$\lambda_{\mathsf{max}}=3$. \label{fig:sign-approximations} }
\end{figure}

The effective Dirac operators in the various formulations are given
in terms of different $\sign$ function approximations. Explicitly,
these approximations of $\sign\left(z\right)$ read
\begin{equation}
r\left(z\right)=\frac{\Pi_{+}\left(z\right)-\Pi_{-}\left(z\right)}{\Pi_{+}\left(z\right)+\Pi_{-}\left(z\right)}\label{eq:SignApproximation}
\end{equation}
with
\begin{equation}
\Pi_{\pm}\left(z\right)=\begin{cases}
\left(1\pm z\right)^{N_{s}} & \textrm{for standard/Boriçi's constr.},\\
\prod_{s}\left(1\pm\omega_{s}z\right) & \textrm{for optimal constr.},
\end{cases}
\end{equation}
so that $r\left(z\right)\to\sign\left(z\right)$ for $N_{s}\to\infty$.
We illustrate these approximations in Fig.~\ref{fig:sign-approximations},
comparing the polar decomposition approximation in Boriçi's construction
with the optimal rational function approximation in Chiu's construction.
The coefficients $\left\{ \omega_{s}\right\} $ are directly linked
to Zolotarev's coefficients, cf.~Refs.~\cite{zolotarev1877application,vandenEshof:2002ms,Chiu:2002eh}.
Note that the $\sign$ function approximation for the standard construction
agrees with the one in Boriçi's construction, but is applied to $H_{\mathsf{m}}$
rather than $H_{\mathsf{w}}$.

\paragraph*{Staggered formulations.}

As previously suggested in Ref.~\cite{Adams:2010gx}, all central
equations in this section generalize to the case of staggered Wilson
fermions after the replacements given in Eq.~\eqref{eq:ReplacementRule}.
In particular, staggered domain wall fermions can be seen as a truncation
of Neuberger's overlap construction with staggered Wilson kernel \cite{Adams:2010gx}.

\section{Setting \label{sec:Setting}}

In the following, we elaborate on the setting of our numerical simulations.
In particular, we discuss our approach of comparing the chiral properties
of the different formulations.

\subsection{Choice of $M_{0}$}

In the free-field case, suitable choices for the domain wall height
$M_{0}$ are in the range $0<M_{0}<2$ for a one flavor theory. In
a gauge field background, in general this interval contracts. Intuitively,
the parameter $M_{0}$ has to be chosen in such a way, that the origin
is shifted sufficiently close to the center of the leftmost ``belly''
of the eigenvalue spectrum. While there is no unique optimal choice
even in the free-field case \cite{Durr:2005an}, one can use the canonical
choice $M_{0}=1$, which shifts the origin exactly to the center.
To be more precise, the mobility edge \cite{Golterman:2003qe,Golterman:2003ui,Golterman:2004cy,Golterman:2005fe}
is the decisive quantity determining valid choices of $M_{0}$.

In the Schwinger model, the interval of valid choices for $M_{0}$
remains close to the free-field case for reasonable values of the
inverse coupling $\beta$. In particular, $M_{0}=1$ remains a sensible
and simple choice, which we are consequently using for all numerical
work presented. This is in contrast to QCD in $3+1$ dimensions, where
one commonly sets $M_{0}=1.8$ (see e.g.~Ref.~\cite{Blum:2000kn}).

\subsection{Effective mass \label{subsec:Effective-mass}}

Two common approaches found in the literature to quantify the induced
effects of chiral symmetry breaking in domain wall fermions are the
determination of the residual mass $m_{\mathsf{res}}$ \cite{Furman:1994ky,Fleming:1999eq,Blum:2000kn,Blum:2001sr}
and the effective mass $m_{\mathsf{eff}}$ \cite{Gadiyak:2000kz,Jung:2000fh,Jung:2000ys}.
The former employs the explicit fermion mass dependence in the chiral
Ward-Takahashi identities, while the latter is given by the lowest
eigenvalue of the Hermitian operator in a topologically nontrivial
background field. Although the definitions are not equivalent, their
numerical values usually agree within a factor of $\mathcal{O}\left(1\right)$
and hence both are suitable to quantify the degree of chiral symmetry
breaking.

In this work, we use the effective mass $m_{\mathsf{eff}}$ due to
its conceptual simplicity. We hence define the effective mass for
a given Dirac operator $D$ with periodic boundary conditions in a
topologically nontrivial background field as the lowest eigenvalue
of the corresponding Hermitian version $H$ of the Dirac operator.
Noting that $H^{2}=D^{\dagger}D$, we define
\begin{equation}
m_{\mathsf{eff}}=\min_{\lambda\in\spec H}\left|\lambda\right|=\min_{\Lambda\in\spec D^{\dagger}D}\sqrt{\Lambda}.
\end{equation}
If $D$ is a normal operator, then $m_{\mathsf{eff}}=\min_{\lambda\in\spec D}\left|\lambda\right|$.
In the general case, however, there is no direct link between the
eigenvalues of $H$ and $D$.

On a topological nontrivial background configuration with topological
charge $Q\neq0$, the Atiyah-Singer index theorem \cite{AtiyahI,AtiyahIII,AtiyahIV,AtiyahV}
ensures the existence of zero modes of $H$ in the continuum. The
corresponding lattice version of this theorem \cite{Ginsparg:1981bj,Hasenfratz:1998ri,Luscher:1998pqa}
ensures that the overlap operator in Eq.~\eqref{eq:DefDov} has exact
zero modes as well. For domain wall fermions and their respective
effective operators these zero modes are recovered in the $N_{s}\to\infty$
limit. For finite $N_{s}$, however, these zero modes become approximate
and their deviation from zero can serve as a measure for the degree
of chiral symmetry breaking.

If $n_{\mp}$ refers to the number of left-handed and right-handed
zero modes, then the Atiyah-Singer index theorem states that
\begin{equation}
n_{-}-n_{+}=\left(-1\right)^{d/2}Q,
\end{equation}
see Ref.~\cite{Adams:2009eb}. A precise definition of $Q$ will
be given in Eq.~\eqref{eq:DefQ}. We note here that in $1+1$ dimensions
the Vanishing Theorem holds \cite{Kiskis:1977vh,Nielsen:1977aw,Ansourian:1977qe}.
That is, if $Q\neq0$, then either $n_{-}$ or $n_{+}$ vanishes.

\subsection{Normality and Ginsparg-Wilson relation}

In the continuum, the Dirac operator is normal. The same holds for
the naïve and staggered discretizations as well as for the overlap
operator. The (staggered) Wilson kernel and the (staggered) domain
wall fermion operators, on the other hand, are not normal.

As it has been shown that normality is necessary for chiral properties
\cite{Kerler:1999dk} (see also Refs.~\cite{Hip:2001hc,Hip:2001mh}),
the degree of violation of normality is an interesting quantity in
the context of chiral symmetry.

Let us recall that a normal operator $D$ satisfies $\left[D,D^{\dagger}\right]=0$
by definition. We then consider the quantity
\begin{equation}
\Delta_{\mathsf{N}}=\left\Vert \left[D,D^{\dagger}\right]\right\Vert _{\infty},
\end{equation}
where $\left\Vert \cdot\right\Vert _{\infty}$ is the by the $L_{\infty}$-norm
induced matrix norm. We know that $\Delta_{\mathsf{N}}$ has to vanish
for the effective operators introduced in Sec.~\ref{sec:Effective-Dirac-operator}
in the limit $N_{s}\to\infty$.

Similarly, we consider violations of the Ginsparg-Wilson relation
given in Eq.~\eqref{eq:GWR}. The quantity
\begin{equation}
\Delta_{\mathsf{GW}}=\left\Vert \left\{ \gamma_{3},D\right\} -\rho^{-1}D\gamma_{3}D\right\Vert _{\infty},
\end{equation}
has to vanish in the limit $N_{s}\to\infty$ as well. As before, we
replace $\gamma_{3}$ by $\epsilon$ in the case of a staggered Wilson
kernel. As previously already considered in Refs.~\cite{Durr:2005an,Durr:2010ch},
$\Delta_{\mathsf{N}}$ and $\Delta_{\mathsf{GW}}$ will give us a
measure for the degree of chiral symmetry violation of the Dirac operators
under consideration.

\subsection{Topological charge}

We determine the topological charge of the gauge configurations via
both the standard overlap definition,
\begin{equation}
Q=\frac{1}{2}\Tr\left(H_{\mathsf{w}}/\sqrt{H_{\mathsf{w}}^{2}}\right),
\end{equation}
and its staggered counterpart,
\begin{equation}
Q=\frac{1}{2}\Tr\left(H_{\mathsf{sw}}/\sqrt{H_{\mathsf{sw}}^{2}}\right),\label{eq:DefQ}
\end{equation}
with $H_{\mathsf{sw}}=\epsilon D_{\mathsf{sw}}\left(-M_{0}\right)$
as derived in Ref.~\cite{Adams:2009eb}. On the small sample of gauge
configurations considered in this paper, they were found to be in
exact agreement. Although a more careful investigation of the continuum
limit would be needed, this observation is consistent with analytical
results \cite{Adams:2013lpa} and other numerical studies \cite{deForcrand:2012bm,Azcoiti:2014pfa}.

\section{Numerical results \label{sec:Numerical-results}}

We are now moving to the numerical part of this work. We calculate
the complete eigenvalue spectra of all Dirac operators introduced
in the previous sections, both with a Wilson and staggered Wilson
kernel, for the $\left(1+1\right)$-dimensional Schwinger model. We
consider the free-field case, thermalized gauge configurations and
the smooth topological configurations constructed in Ref.~\cite{Smit:1986fn}.

In the following, we set the lattice spacing to $a=a_{d+1}=1$ and
Wilson parameter to $r=1$. The extent in the extra dimension will
be varied in the range $2\leq N_{s}\leq8$. We use periodic boundary
conditions in both space and time direction, so that the determination
of the effective mass $m_{\mathsf{eff}}$ as defined in Sec.~\ref{subsec:Effective-mass}
applies.

Extremal eigenvalues are determined with \textsc{arpack} \cite{Lehoucq97arpackusers},
while complete spectra are computed with \textsc{lapack} \cite{LAPACK}.
Calculations are carried out in double precision. In all figures,
the abbreviation ``std'' refers to the standard construction, ``Bor''
to Boriçi's construction and ``opt'' to Chiu's optimal construction.
With respect to the overlap constructions, ``DW'' refers to the
overlap operator with kernel $H_{\mathsf{m}}$, ``Neub'' to Neuberger's
overlap with kernel $H_{\mathsf{w}}$ and ``Adams'' to Adams' staggered
overlap with kernel $H_{\mathsf{sw}}$.

\subsection{Free-field case \label{subsec:Free-field-case}}

We begin with the free-field case. Here we can employ a momentum space
representation of the kernel. In particular, the Wilson kernel can
be represented as a $2\times2$ linear map
\begin{equation}
D_{\mathsf{w}}=\left(m_{\mathsf{f}}+2r\right)\One+\ii\sum_{\mu}\gamma_{\mu}\sin p_{\mu}-r\sum_{\nu}\cos p_{\nu}\One,
\end{equation}
where $p_{\mu}=2\pi n_{\mu}/N_{\mu}$ with $n_{\mu}=0,1,\ldots,N_{\mu}-1$
and $N_{\mu}$ is the number of slices in $\mu$-direction. The staggered
Wilson kernel takes the form of the $4\times4$ linear map
\begin{multline}
D_{\mathsf{sw}}=m_{\mathsf{f}}\left(\One\otimes\One\right)+\ii\sum_{\mu}\sin p_{\mu}\left(\gamma_{\mu}\otimes\One\right)\\
+r\One\otimes\left(\One+\xi_{3}\prod_{\nu}\cos p_{\nu}\right)
\end{multline}
with $n_{\mu}=0,1,\ldots,N_{\mu}/2-1$ (see also Refs.~\cite{Adams:2013lpa,deForcrand:2012bm}).

In the three-dimensional operators, we keep the extra dimension in
the position space formulation, as in general it is lacking periodicity.
Besides reducing the dimensionality of the eigenvalue problem by choosing
a momentum space representation for the kernel, this also avoids some
numerical instabilities of the free-field case encountered in a purely
position space based formulation.

In the following, all numerical results are for the case of a $N_{s}\times N_{t}=20\times20$
lattice.

\paragraph{Kernel operators.}

\begin{figure}[!]
\begin{centering}
\includegraphics[width=0.9\columnwidth]{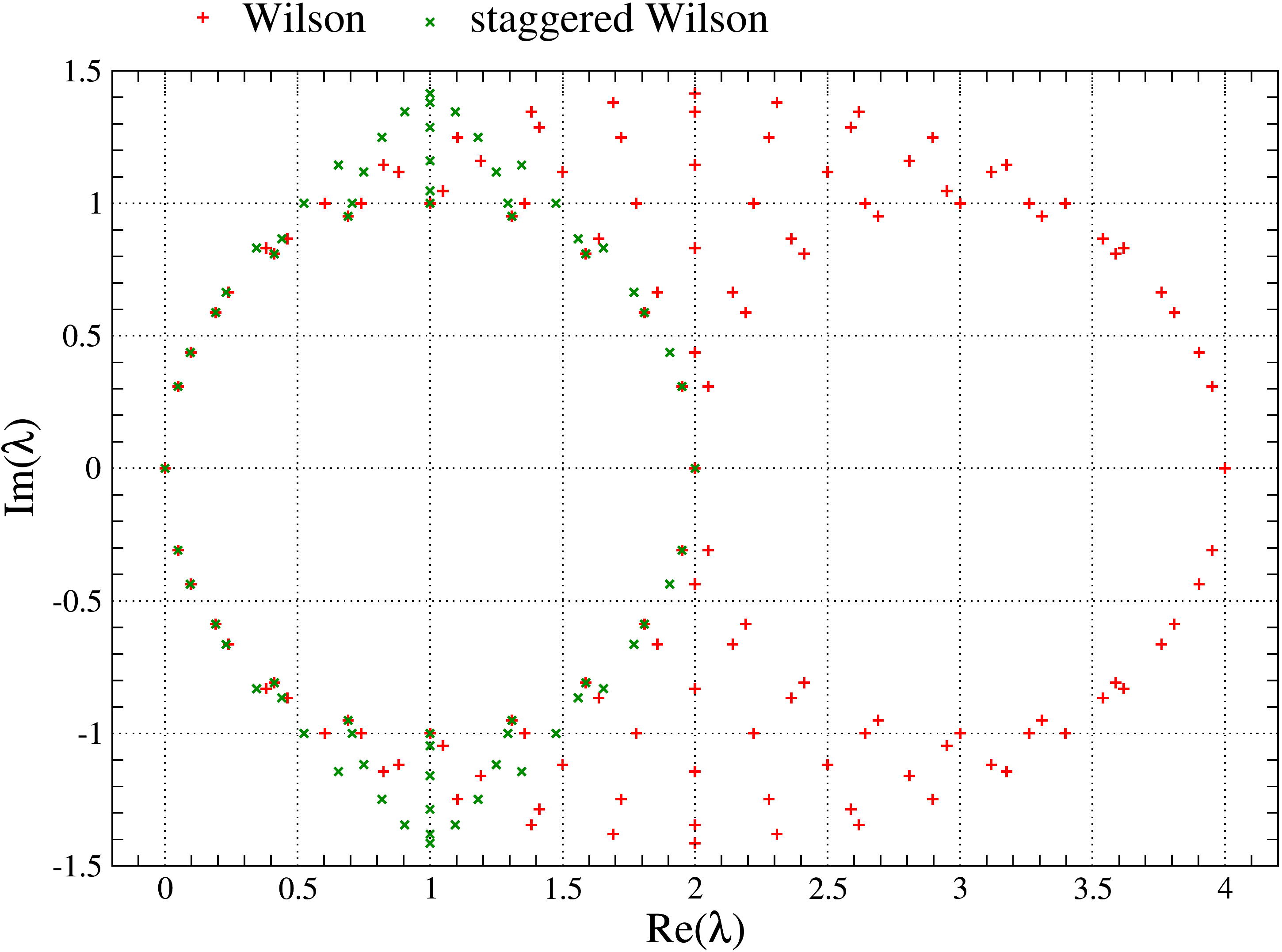} 
\par\end{centering}
\caption{Free-field spectrum of kernel operators. \label{fig:free-kernel} }
\end{figure}

In Fig.~\ref{fig:free-kernel}, we can find the well known free spectra
of the kernel operators. In $1+1$ dimensions the Wilson Dirac operator
has only two doubler branches in the eigenvalue spectrum due to the
reduced number of fermion species in this low-dimensional setting.
Like in $3+1$ dimensions, the staggered Wilson Dirac operator has
a single doubler branch due to the splitting of positive and negative
flavor-chirality species.

One can see that the free-field spectrum of the staggered Wilson kernel
is closer to the Ginsparg-Wilson circle compared to the Wilson kernel.
One can then hope for a better performance of chiral formulations
with this kernel, at least on sufficiently smooth configurations.

The bulk and effective operators, which we are going to discuss in
the following, use either a Wilson or a staggered Wilson kernel. Comparing
both cases, we note that the spectra of these operators differ mostly
due to the different ultraviolet parts of the respective kernel spectra.
Although the low-lying parts of the kernel spectrum in the physical
branch are alike, the ultraviolet modes will alter the resulting spectrum
of the bulk and effective operators differently and have an impact
on the efficiency and chiral properties.

\paragraph{Bulk operators.}

\begin{figure*}[!]
\begin{centering}
\subfloat[Standard construction]{\includegraphics[width=0.32\textwidth]{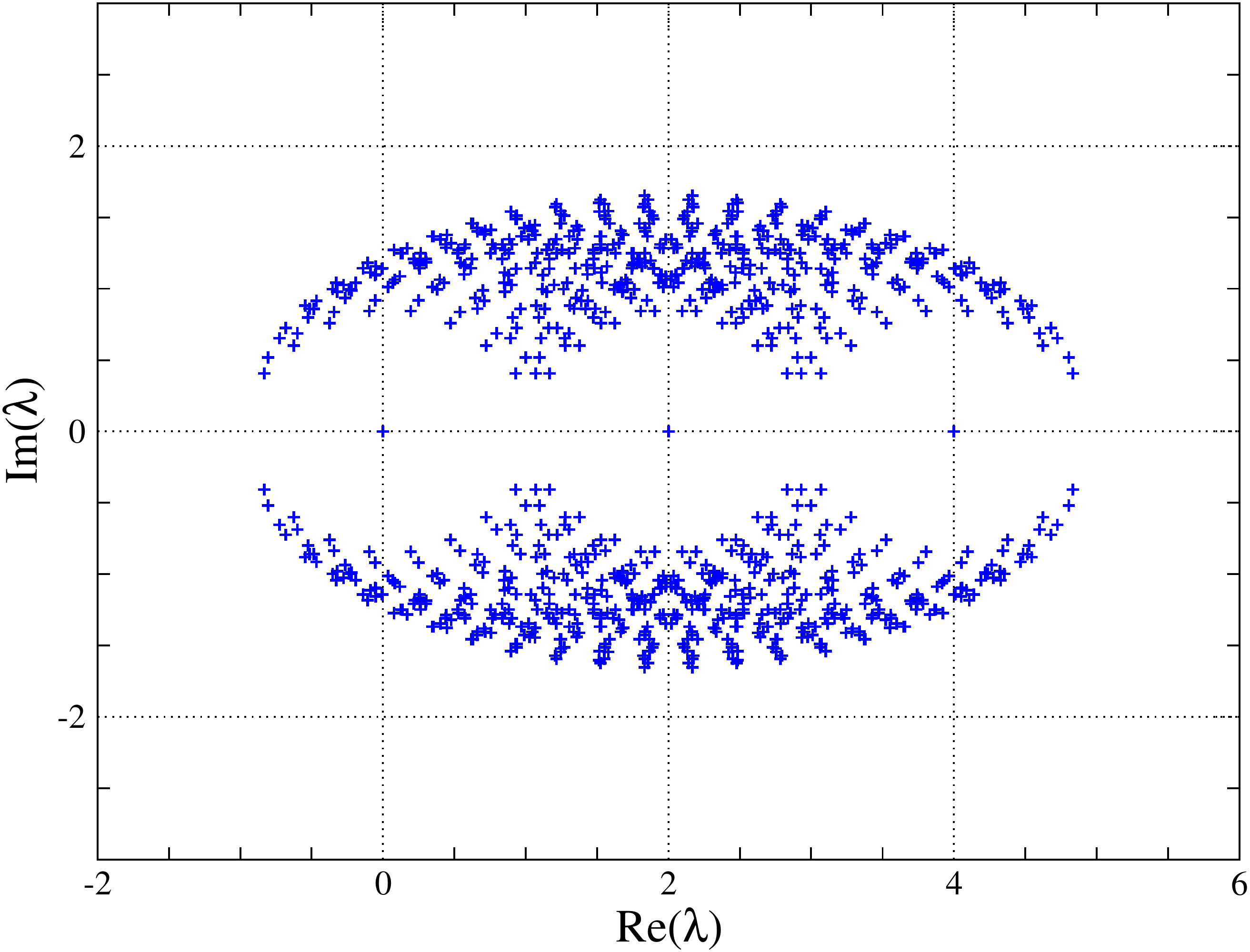}

}\hfill{}\subfloat[Boriçi's construction]{\includegraphics[width=0.32\textwidth]{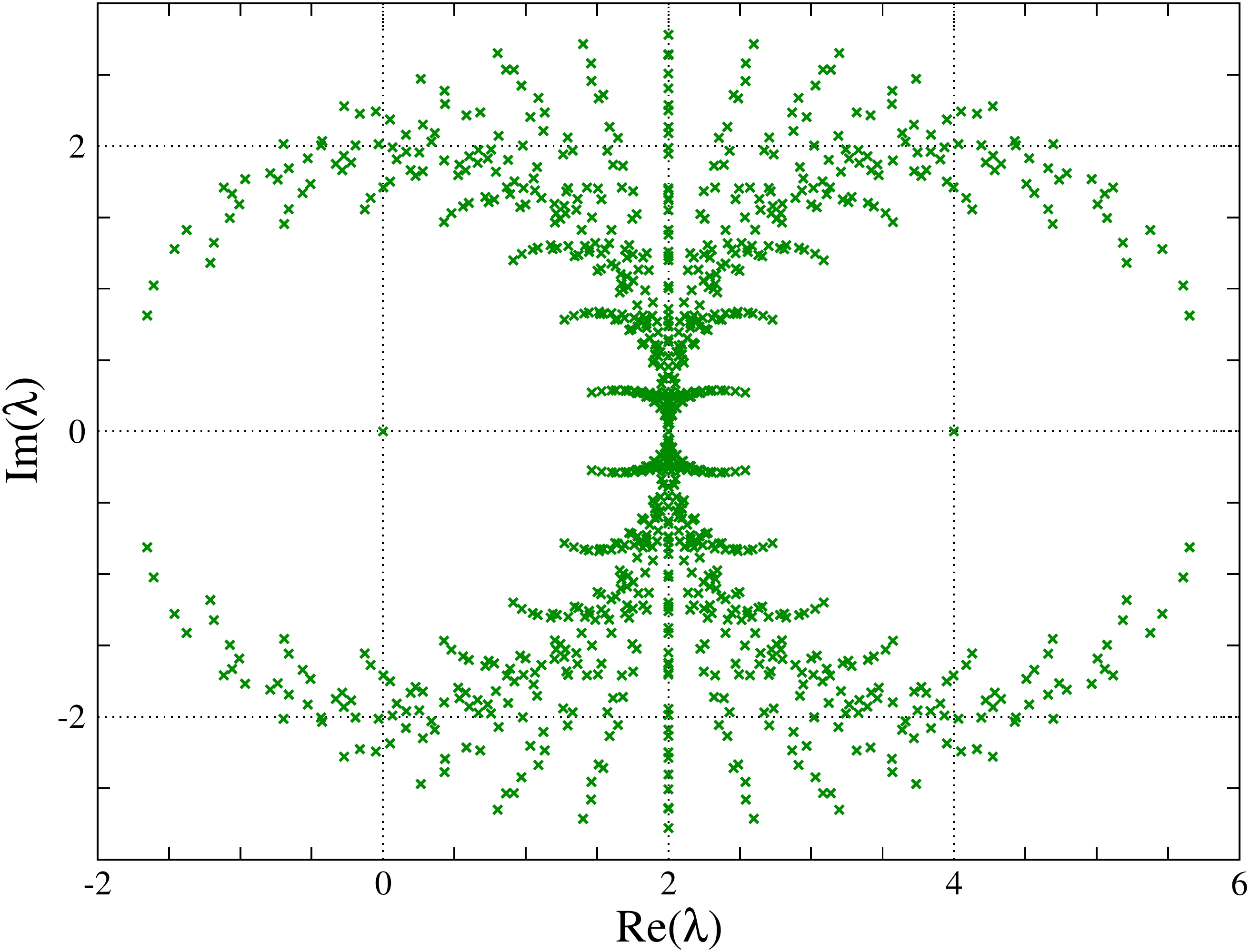}

}\hfill{}\subfloat[Optimal construction]{\includegraphics[width=0.32\textwidth]{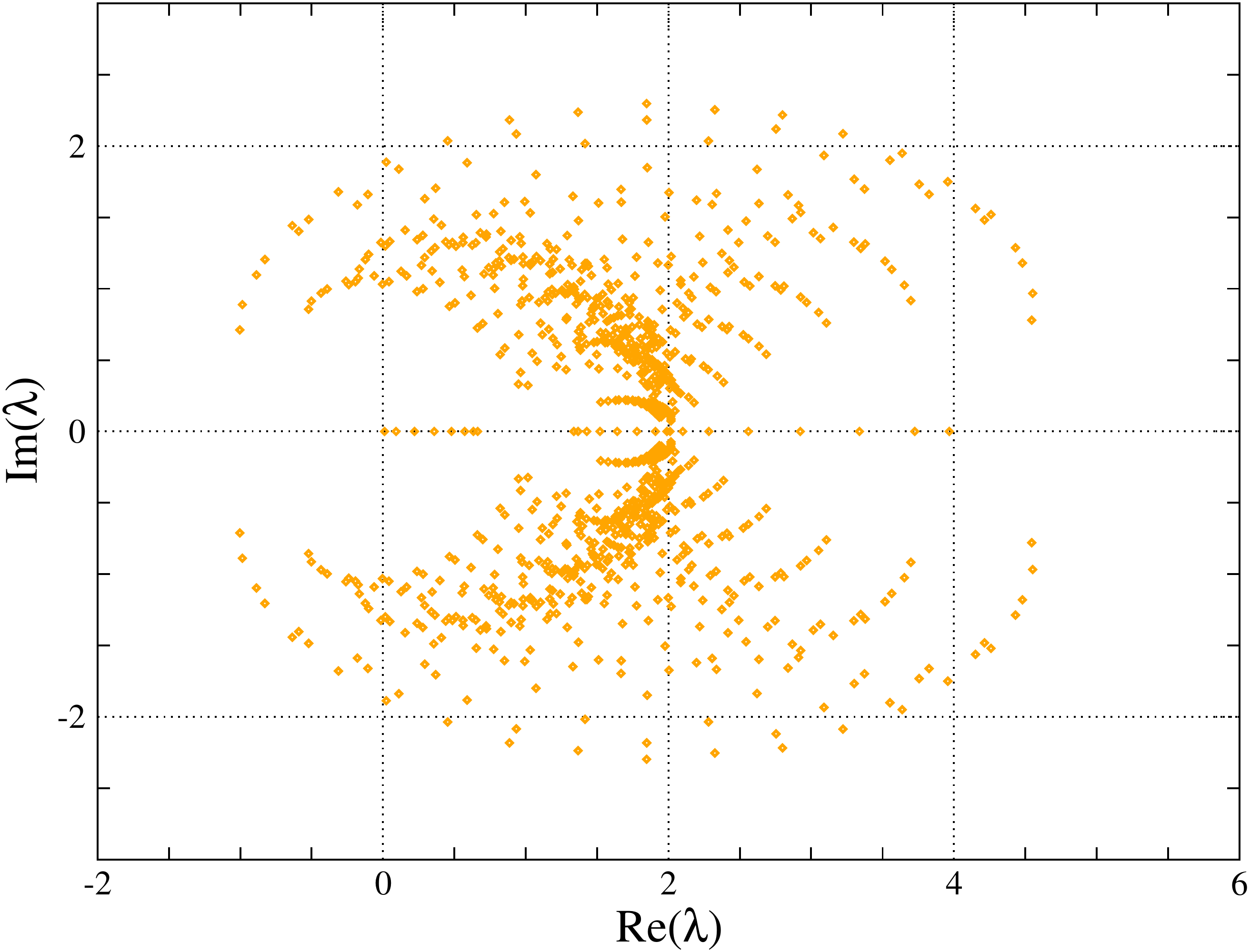}

}
\par\end{centering}
\caption{Free-field spectrum of $D_{\mathsf{dw}}$ with Wilson kernel for $m=0$
at $N_{s}=8$. \label{fig:free-wilson-bulk}}
\end{figure*}
\begin{figure*}[!]
\begin{centering}
\subfloat[Standard construction]{\includegraphics[width=0.32\textwidth]{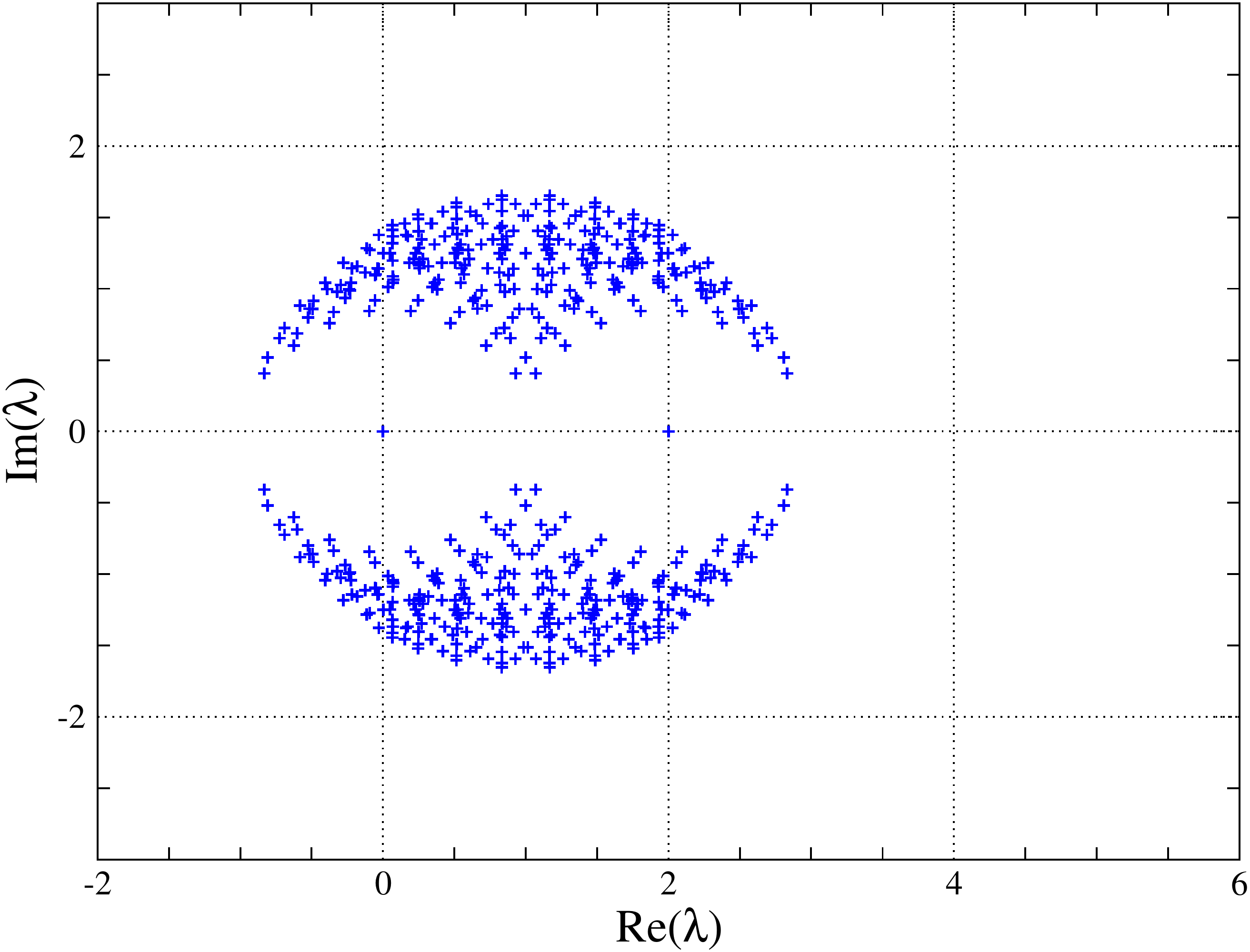}

}\hfill{}\subfloat[Boriçi's construction]{\includegraphics[width=0.32\textwidth]{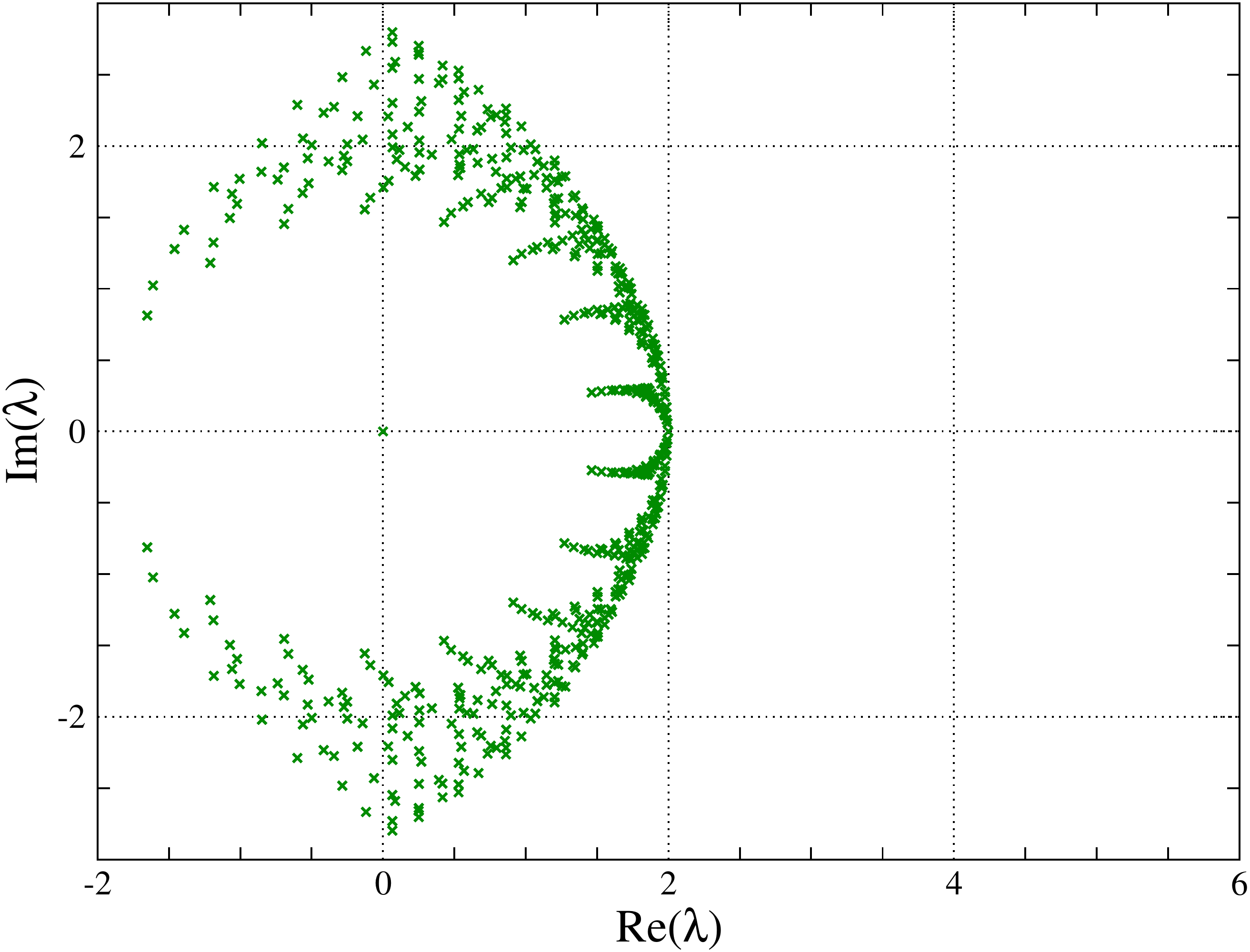}

}\hfill{}\subfloat[Optimal construction]{\includegraphics[width=0.32\textwidth]{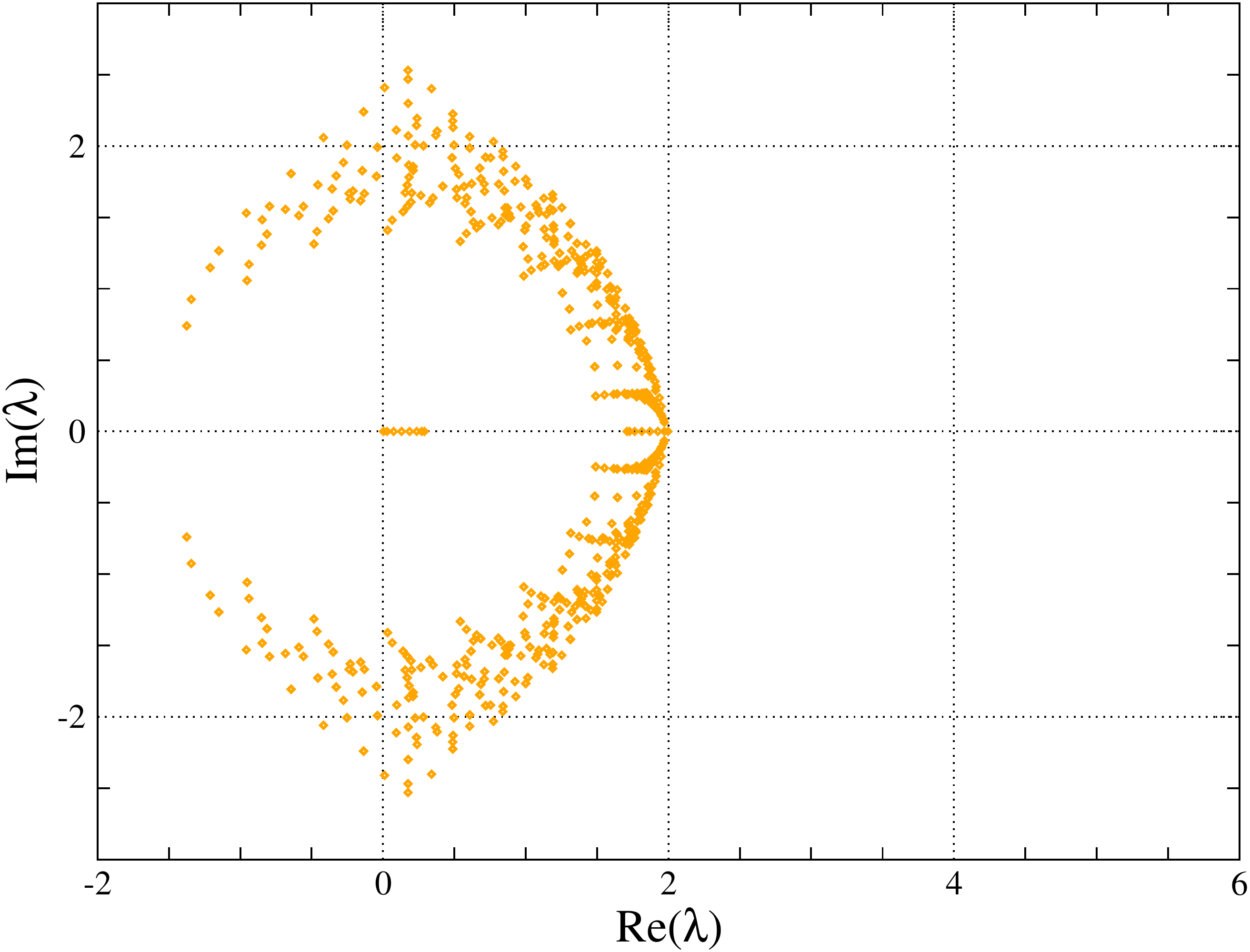}

}
\par\end{centering}
\caption{Free-field spectrum of $D_{\mathsf{sdw}}$ with staggered Wilson kernel
for $m=0$ at $N_{s}=8$. \label{fig:free-stw-bulk}}
\end{figure*}
\begin{figure*}[!]
\begin{centering}
\hfill{}\subfloat[Periodic case ($m=-1$)]{\includegraphics[width=0.32\textwidth]{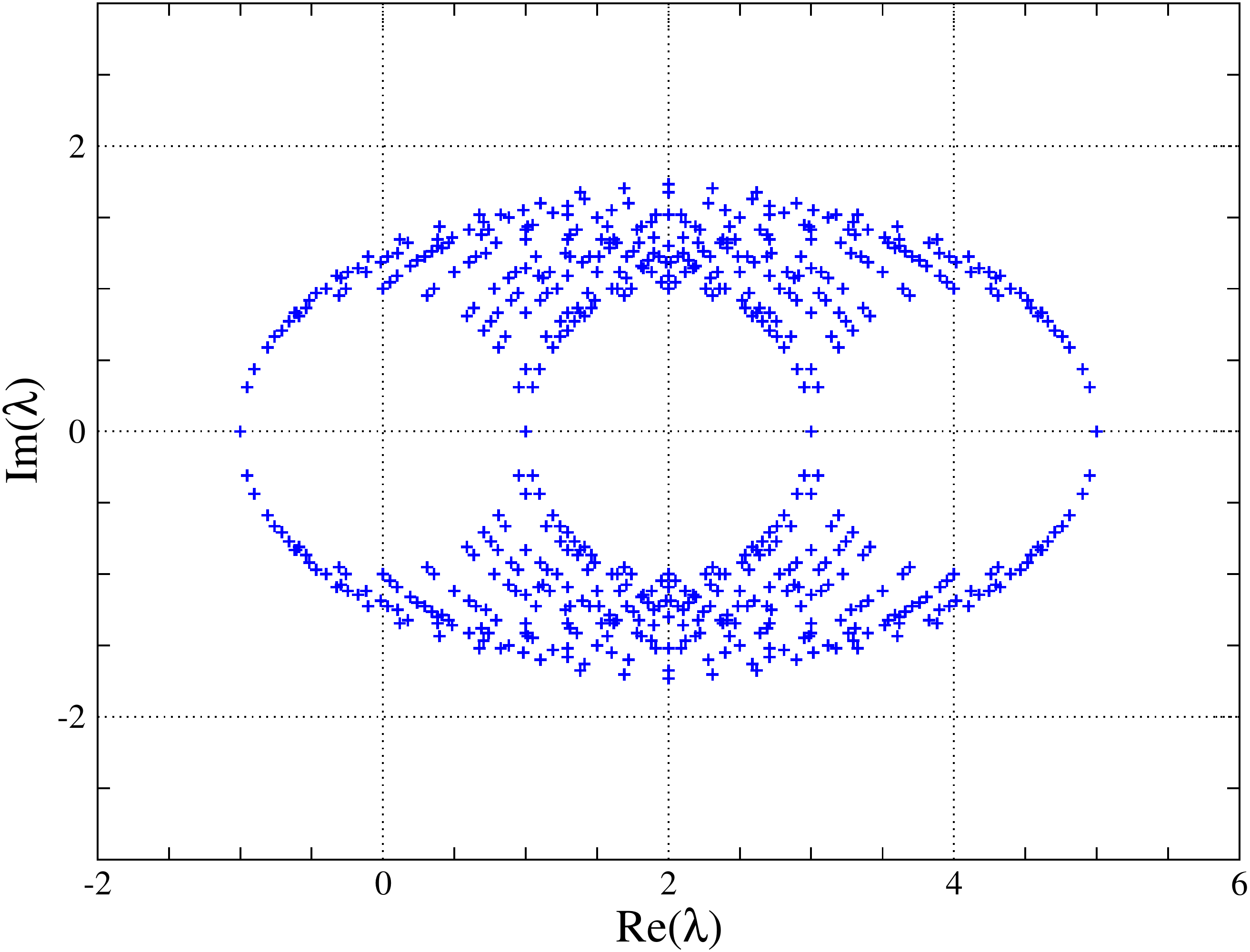}

}\hfill{}\subfloat[Antiperiodic case ($m=1$)]{\includegraphics[width=0.32\textwidth]{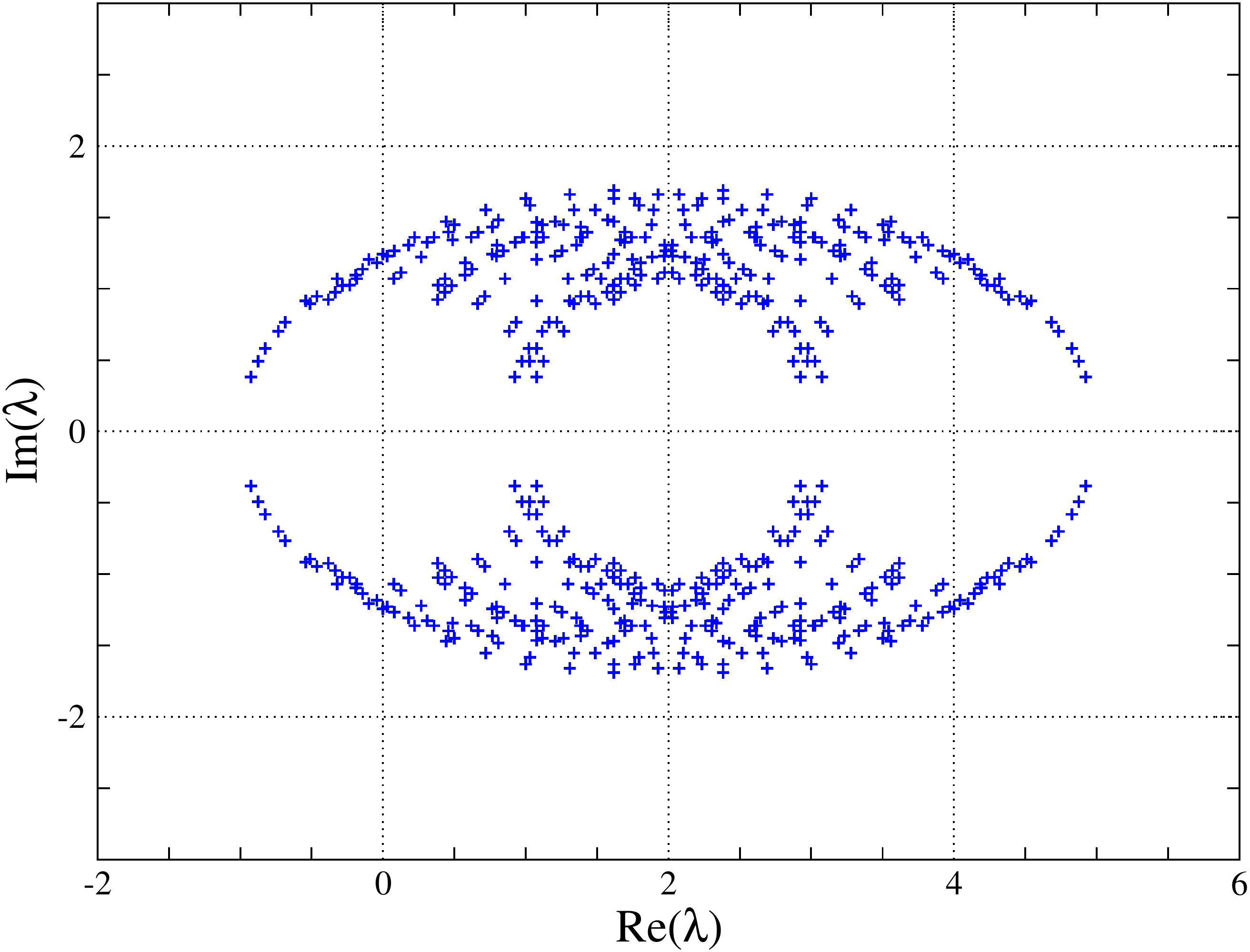}

}\hfill{} 
\par\end{centering}
\caption{Free-field spectrum of $D_{\mathsf{dw}}$ with Wilson kernel in the
standard construction\protect \protect \\
 with different boundary conditions at $N_{s}=8$ (cf.~Fig.~\ref{fig:free-wilson-bulk})
\label{fig:free-wilson-bulk-bcs}}
\end{figure*}

In Figs.~\ref{fig:free-wilson-bulk} and \ref{fig:free-stw-bulk},
we show the spectrum of the $\left(2+1\right)$-dimensional bulk operator
in the standard, Boriçi's and the optimal construction. In Fig.~\ref{fig:free-wilson-bulk-bcs},
we show periodic ($m=-1$) and antiperiodic ($m=1$) boundary conditions
in the extra dimension to compare with the Dirichlet ($m=0)$ case.

We can observe that the bulk spectra for Boriçi's and the optimal
construction have lost their resemblance to a Wilson operator in three
dimensions. Specifically, while for the standard construction we find
(two) three doubler branches in the spectrum with the (staggered)
Wilson kernel, in Boriçi's construction one less branch is visible.
In the standard and Boriçi's construction, we find $2N_{s}$ exact
zero modes in the Dirichlet case \cite{Shamir:1993zy,Gadiyak:2000kz},
which disappear for $m\neq0$. In the optimal construction, we notice
how the corresponding eigenvalues get spread out along the real axis
and we are left with only two approximate zero modes.

\paragraph{Effective operators.}

\begin{figure*}[!]
\begin{centering}
\hfill{}\subfloat[Wilson kernel]{\includegraphics[width=0.4\textwidth]{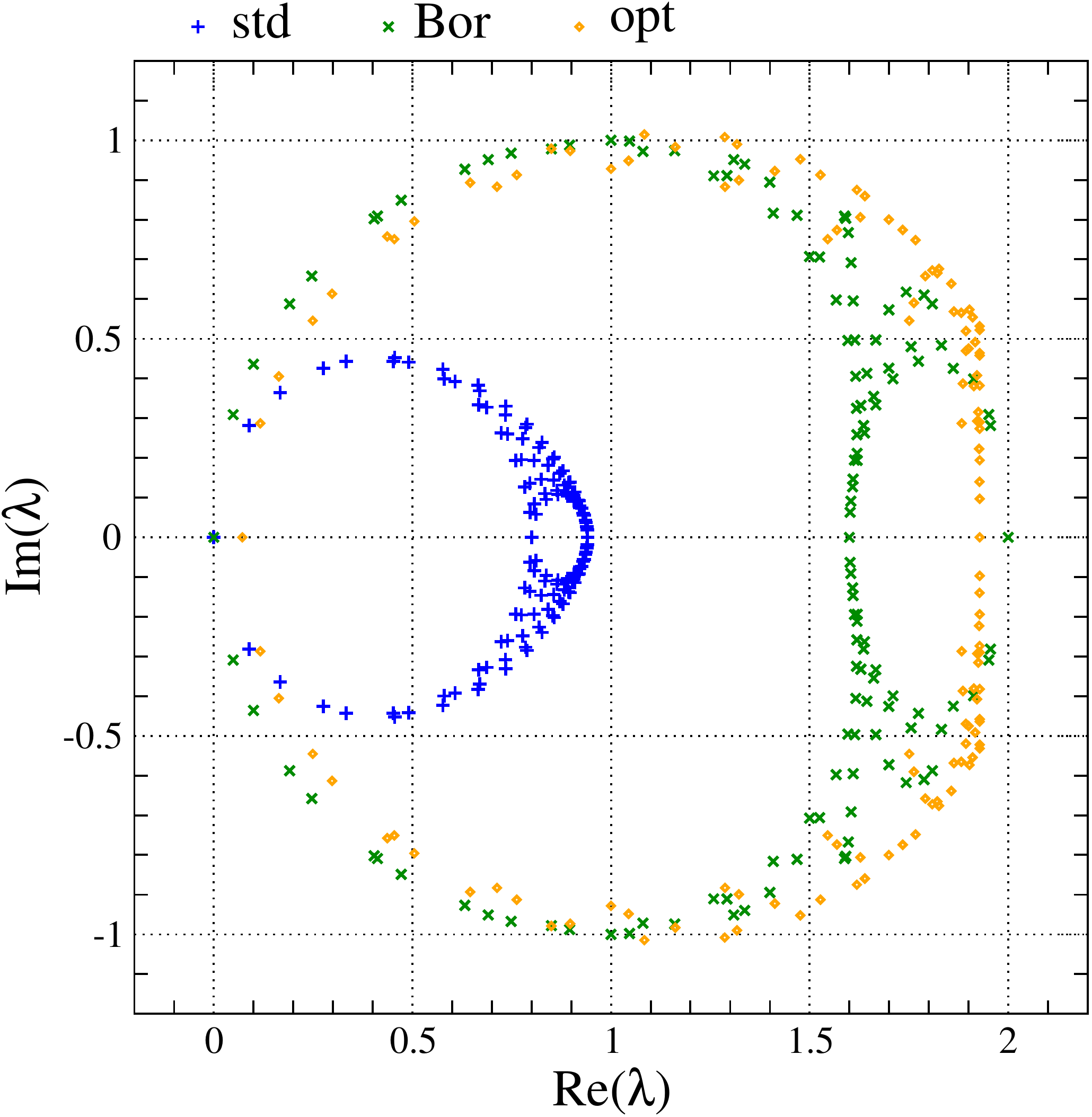}

}\hfill{}\subfloat[Staggered Wilson kernel]{\includegraphics[width=0.4\textwidth]{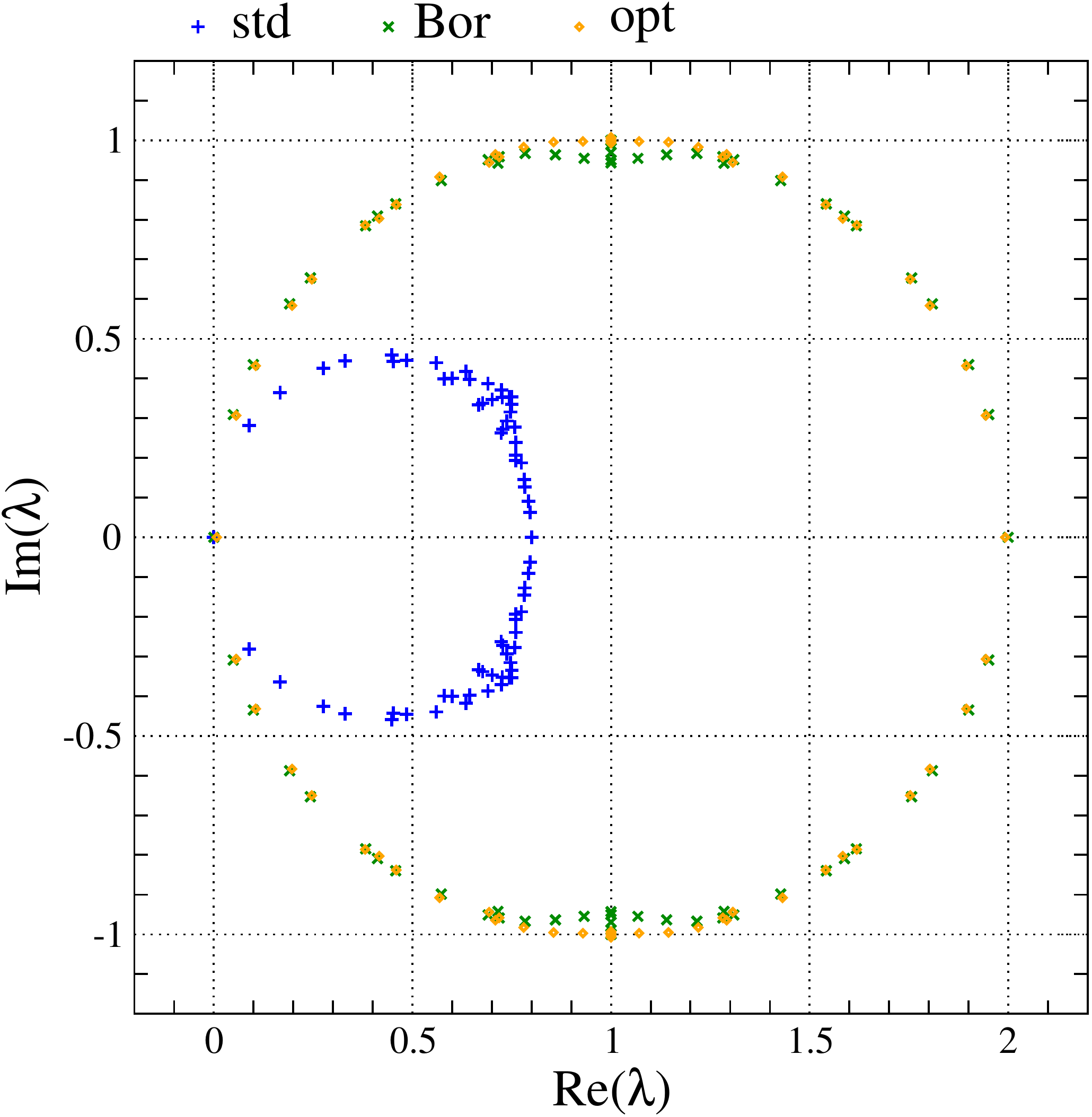}

}\hfill{} 
\par\end{centering}
\caption{Spectrum of $\varrho D_{\mathsf{eff}}$ at $N_{s}=2$ for the standard
(std), Boriçi (Bor) and optimal (opt) construction. \label{fig:free-deff-Ns2}}
\end{figure*}
\begin{figure*}[!]
\begin{centering}
\hfill{}\subfloat[Wilson kernel]{\includegraphics[width=0.4\textwidth]{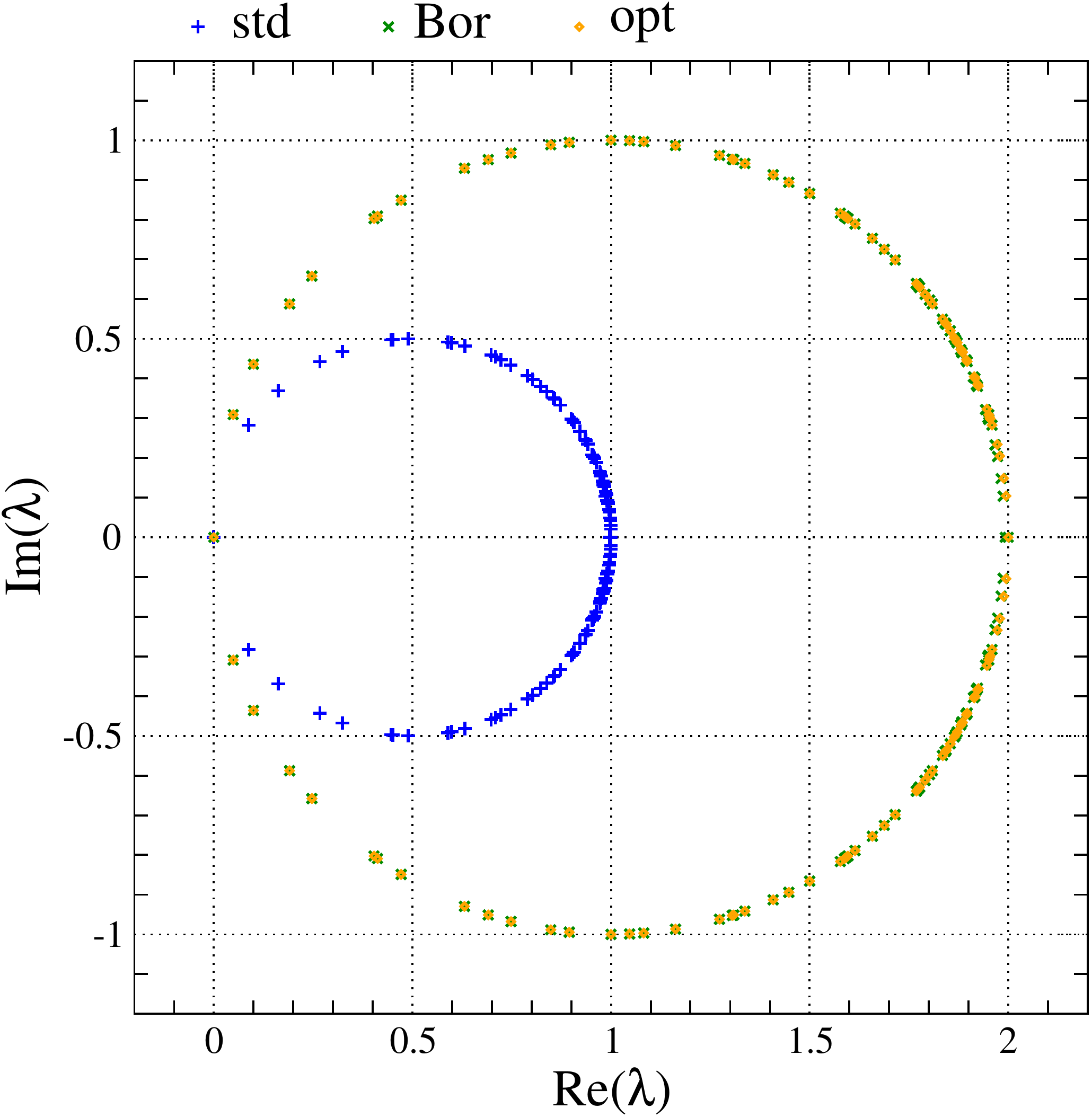}

}\hfill{}\subfloat[Staggered Wilson kernel]{\includegraphics[width=0.4\textwidth]{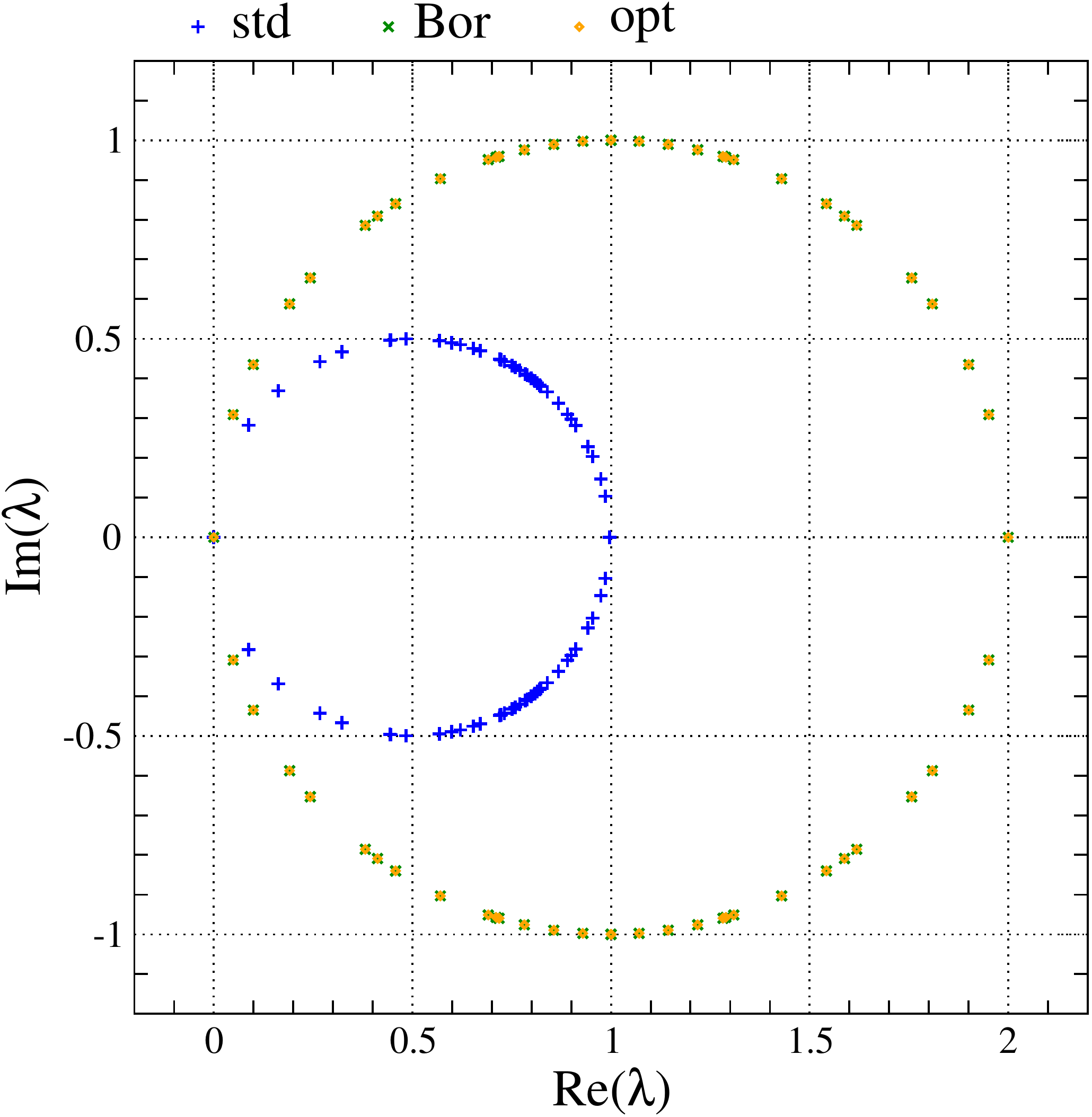}

}\hfill{} 
\par\end{centering}
\caption{Spectrum of $\varrho D_{\mathsf{eff}}$ at $N_{s}=8$ for the standard
(std), Boriçi (Bor) and optimal (opt) construction. \label{fig:free-deff-Ns8}}
\end{figure*}
\begin{figure*}[!]
\begin{centering}
\hfill{}\subfloat[Wilson kernel]{\includegraphics[width=0.4\textwidth]{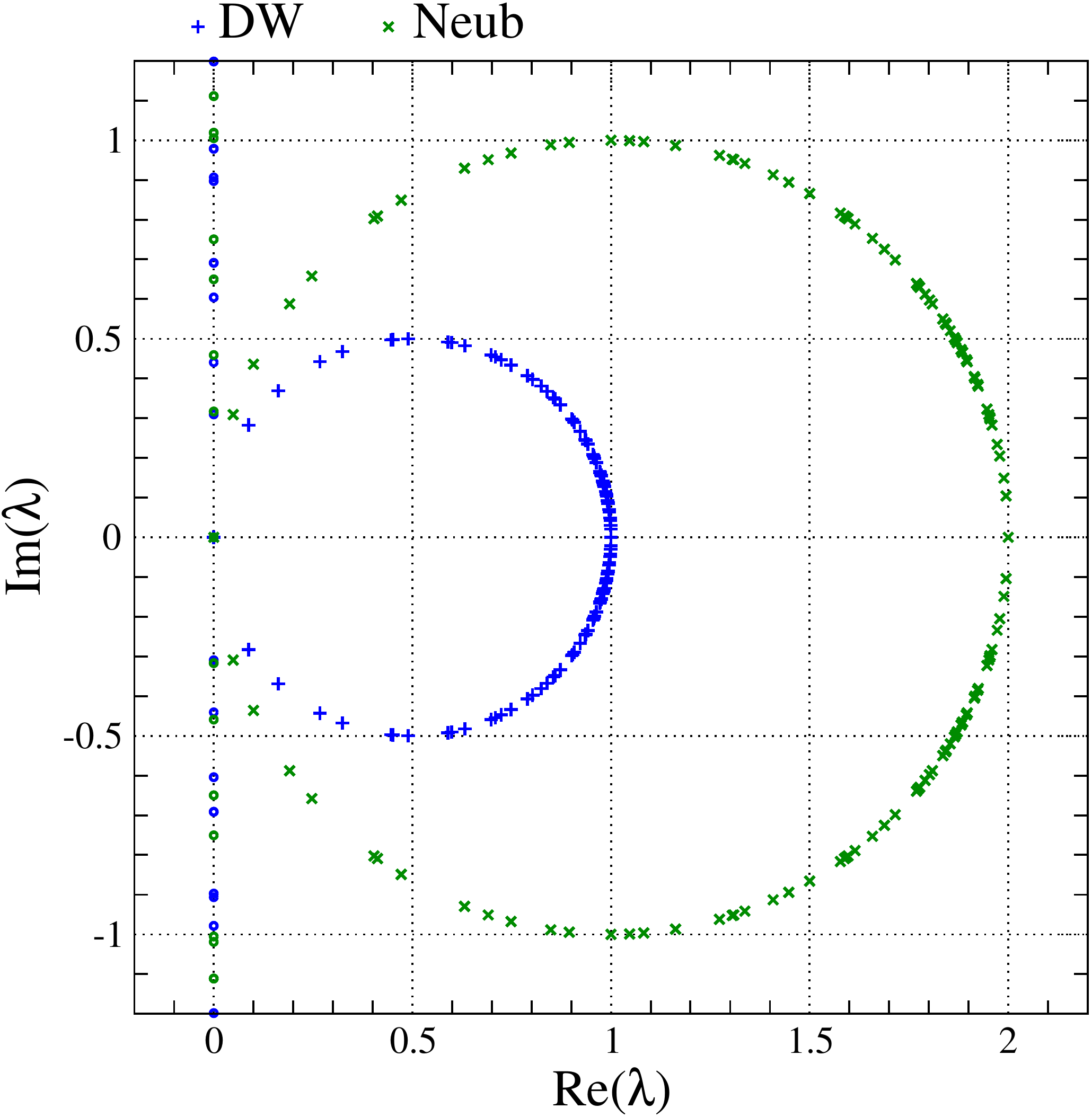}

}\hfill{}\subfloat[Staggered Wilson kernel]{\includegraphics[width=0.4\textwidth]{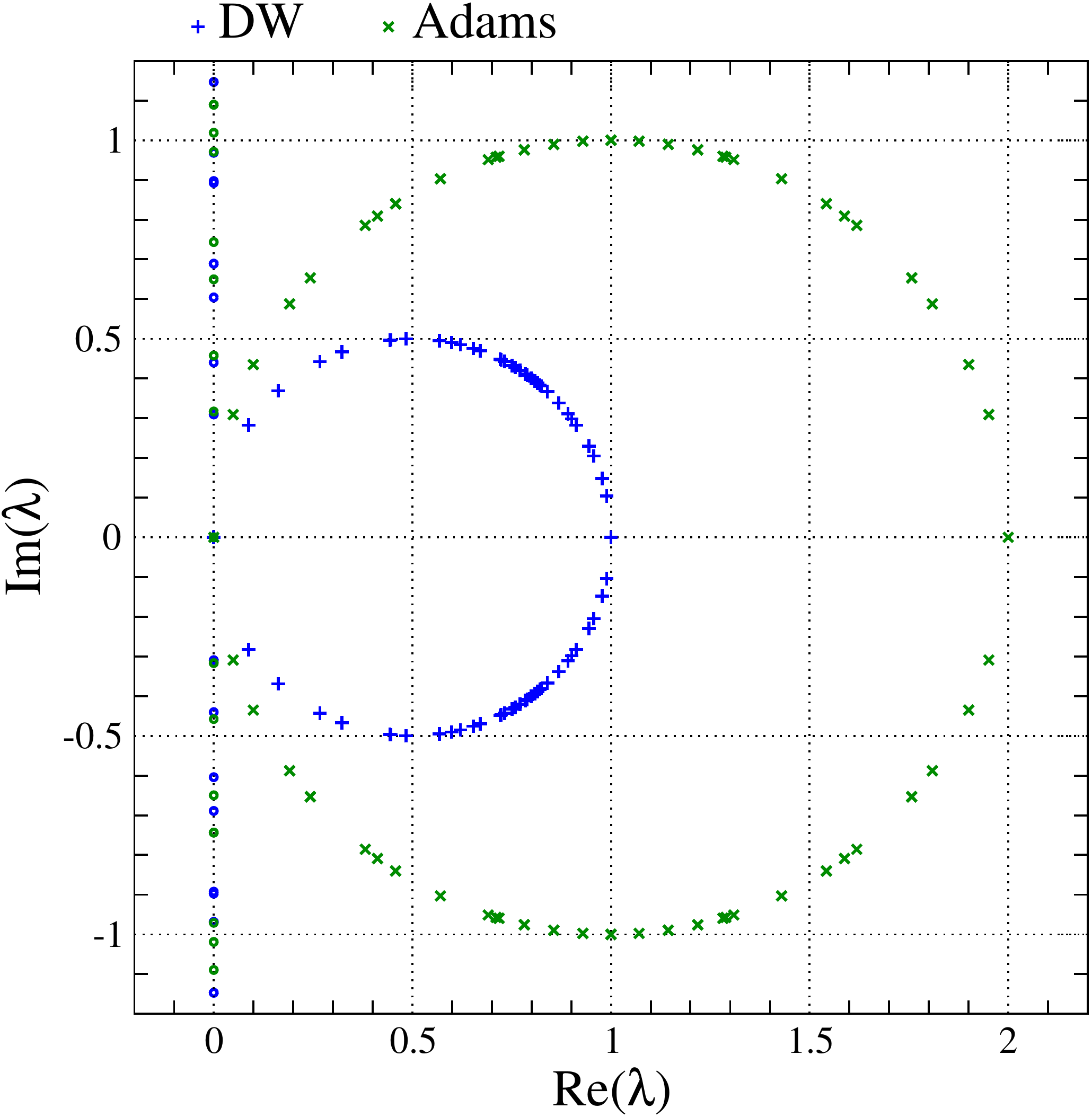}

}\hfill{} 
\par\end{centering}
\caption{Spectrum of $\varrho D_{\mathsf{ov}}$ with stereographic projection
for domain wall (DW)\protect \\
and standard (Neub/Adams) kernels. \label{fig:free-dov} }
\end{figure*}

We now move on to the effective operators $\varrho D_{\mathsf{eff}}$
as defined in Eq.~\eqref{eq:DefDeff}. In Figs.~\ref{fig:free-deff-Ns2}
and  \ref{fig:free-deff-Ns8}, we show the respective eigenvalue spectra
with a Wilson and staggered Wilson kernel.

As we can see, the spectra approach the Ginsparg-Wilson circle rapidly
for increasing values of $N_{s}$. This fast convergence is of course
expected on smooth configurations like the free field. Already for
$N_{s}=8$ the spectrum is close to the spectrum of the corresponding
overlap operator in the $N_{s}\to\infty$ limit. In particular, we
note the rapid convergence of Boriçi's and Chiu's construction with
a staggered Wilson kernel.

The effective Dirac operator in the optimal construction shows a significantly
improved convergence. Let us recall that for a given interval $I=\left[\lambda_{\mathsf{min}},\lambda_{\mathsf{max}}\right]$
the optimal rational function approximation $r_{\mathsf{opt}}\left(z\right)$
of the $\sign$ function minimizes the maximal deviation
\begin{align}
\delta_{\mathsf{max}} & =\max_{z\in-I\cup I}\left|\sign\left(z\right)-r_{\mathsf{opt}}\left(z\right)\right|\nonumber \\
 & =1\mp r_{\mathsf{opt}}\left(\pm\lambda_{\mathsf{min}}\right)\label{eq:DeltaMax}
\end{align}
on the domain $-I\cup I$. As expected, we observe that all eigenvalues
lie within a tube with diameter $2\delta_{\mathsf{max}}$ around the
Ginsparg-Wilson circle. That is, for all eigenvalues $\lambda$ we
find $\left|\,\left|\lambda\right|-1\,\right|\leq\delta_{\mathsf{max}}$.
Noting that the $\sign$ function has a point of maximal deviation
at both $\lambda_{\mathsf{min}}$, we observe the absence of an exact
zero mode in contrast to the standard and Boriçi's construction. However,
due to the rapid convergence of the rational function approximation,
the approximate zero mode is of small magnitude for already moderate
values of $N_{s}$.

\paragraph{Overlap operators.}

In the $N_{s}\to\infty$ limit, the effective operators can be formulated
as overlap operators defined in Eq.~\eqref{eq:DefDov} with the kernel
$H$ given in Eq.~\eqref{eq:DefH}. In Fig.~\ref{fig:free-dov},
we can find the spectra of $\varrho D_{\mathsf{ov}}$ together with
the stereographic projection $\pi$ of the eigenvalues onto the imaginary
axis via
\begin{equation}
\pi\left(\lambda\right)=\frac{\lambda}{1-\frac{1}{\varrho}\lambda}.
\end{equation}

We also point out the high degree of symmetry of the spectrum in the
case of Adams' overlap. As noted before, the effective Dirac operators
in Boriçi's and the optimal construction converge towards Neuberger's
and Adams' overlap operator for $N_{s}\to\infty$, while in the standard
construction we find a modified overlap kernel.

\subsection{$\protect\Uone$ gauge field case}

While the free field is an interesting case, our main interest is
the performance of the Dirac operators in non-trivial background fields.
Dealing with the Schwinger model, for the rest of the work we consider
$\Uone$ gauge fields. We use quenched thermalized gauge fields following
the setup of Refs.~\cite{Durr:2003xs,Durr:2004ta}. Note that while
these gauge fields originate from a quenched ensemble, there is no
problem in the Schwinger model to reweight them to an arbitrary mass
unquenched ensemble~\cite{Giusti:2001cn,Durr:2003xs,Durr:2004ta,Durr:2006ze}.

In this section, we start our investigation by focusing on a few individual
$20^{2}$ configurations at an inverse coupling of $\beta=5$ to illustrate
the qualitative features of the spectra.

\paragraph{Kernel operators.}

\begin{figure}[!]
\begin{centering}
\includegraphics[width=0.9\columnwidth]{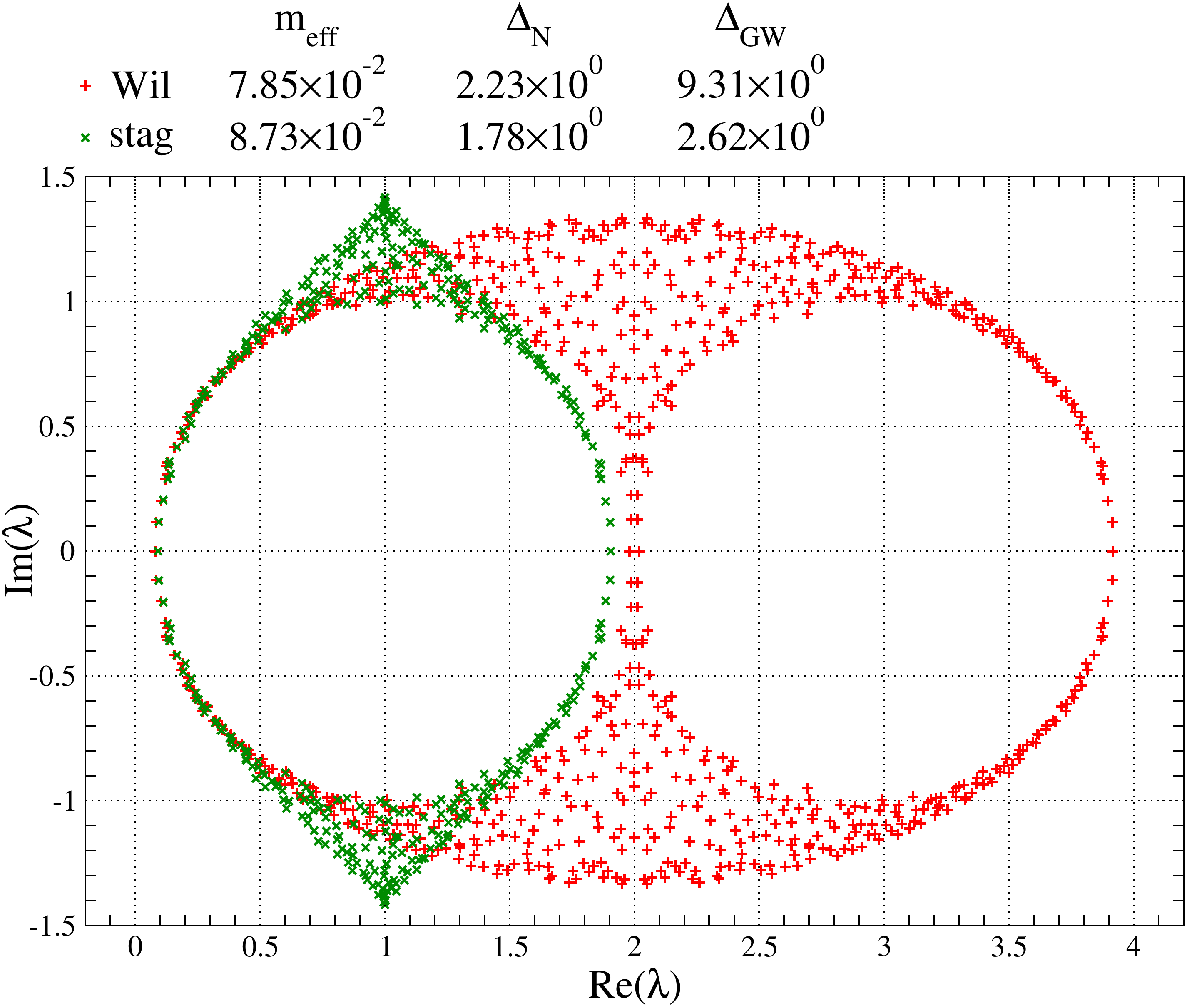} 
\par\end{centering}
\caption{Spectrum of kernel operators. \label{fig:gauge-kernel}}
\end{figure}

In Fig.~\ref{fig:gauge-kernel}, we can see the kernel spectra in
a gauge background with $Q=1$. As expected in the Schwinger model,
the branches stay much sharper and well separated compared to the
$\left(3+1\right)$-dimensional QCD case \cite{deForcrand:2011ak,deForcrand:2012bm,Durr:2013gp,Adams:2013tya}.

\paragraph{Bulk operators.}

\begin{figure*}[!]
\begin{centering}
\subfloat[Standard construction]{\includegraphics[width=0.32\textwidth]{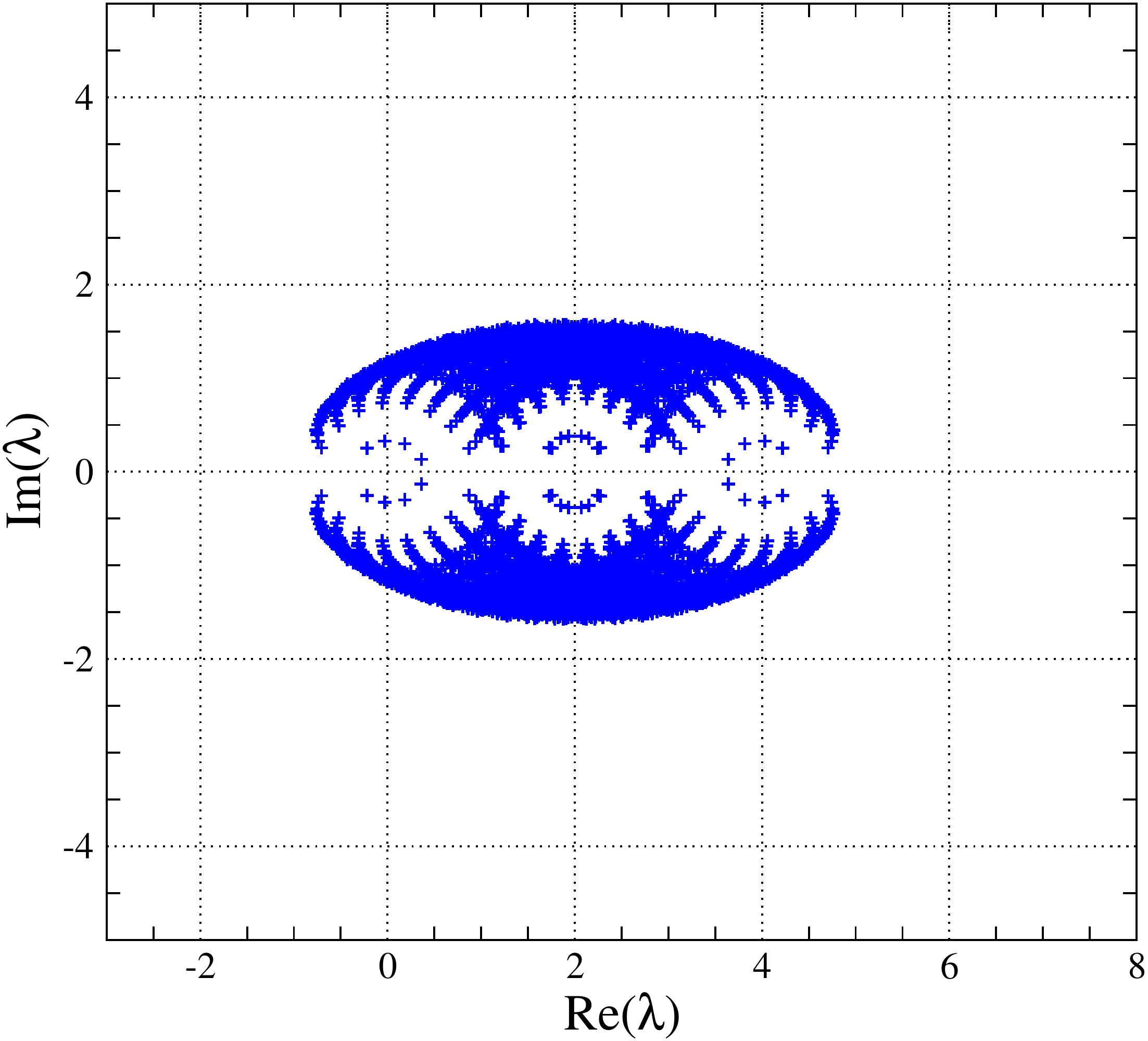}

}\hfill{}\subfloat[Boriçi's construction]{\includegraphics[width=0.32\textwidth]{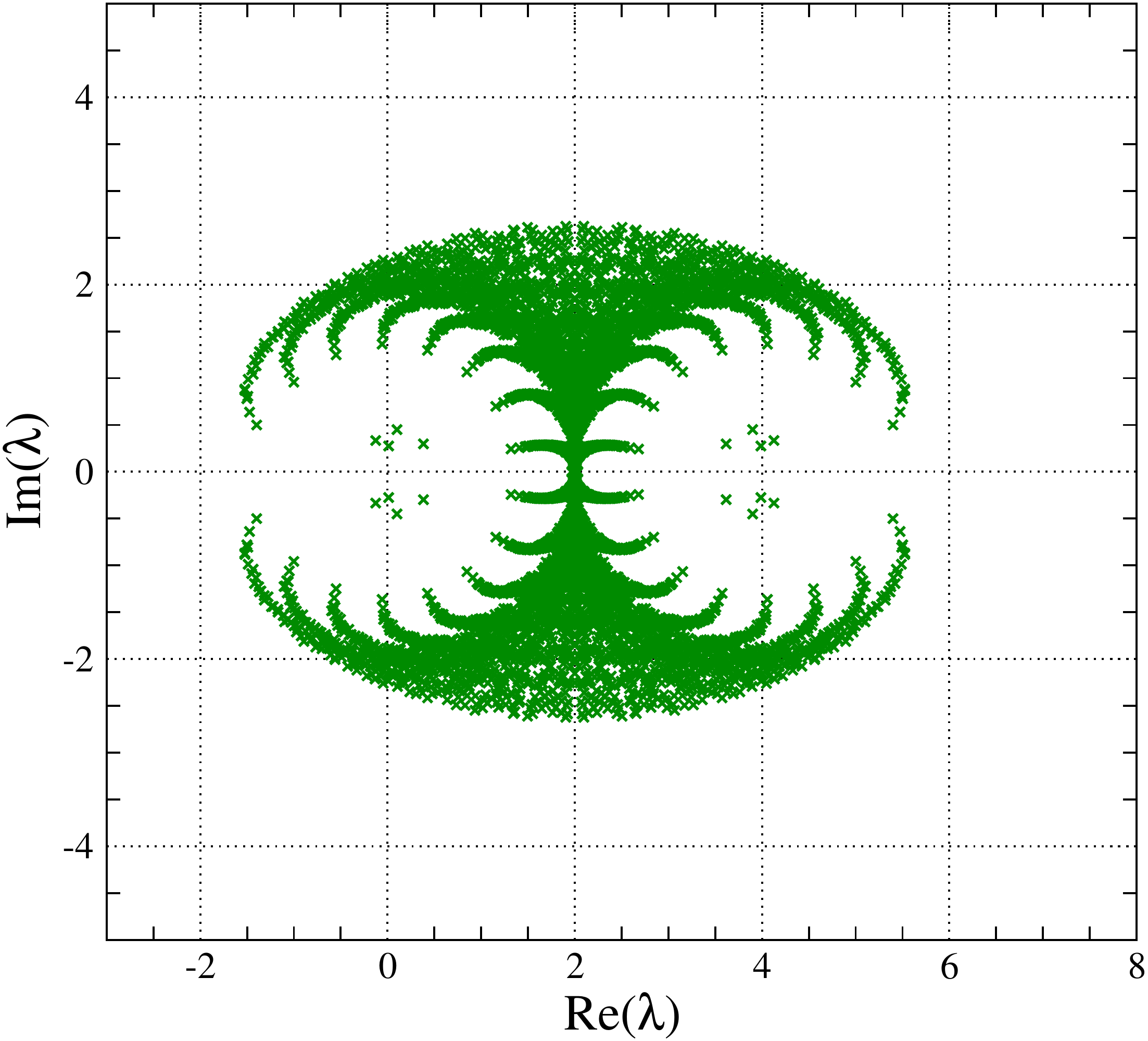}

}\hfill{}\subfloat[Optimal construction]{\includegraphics[width=0.32\textwidth]{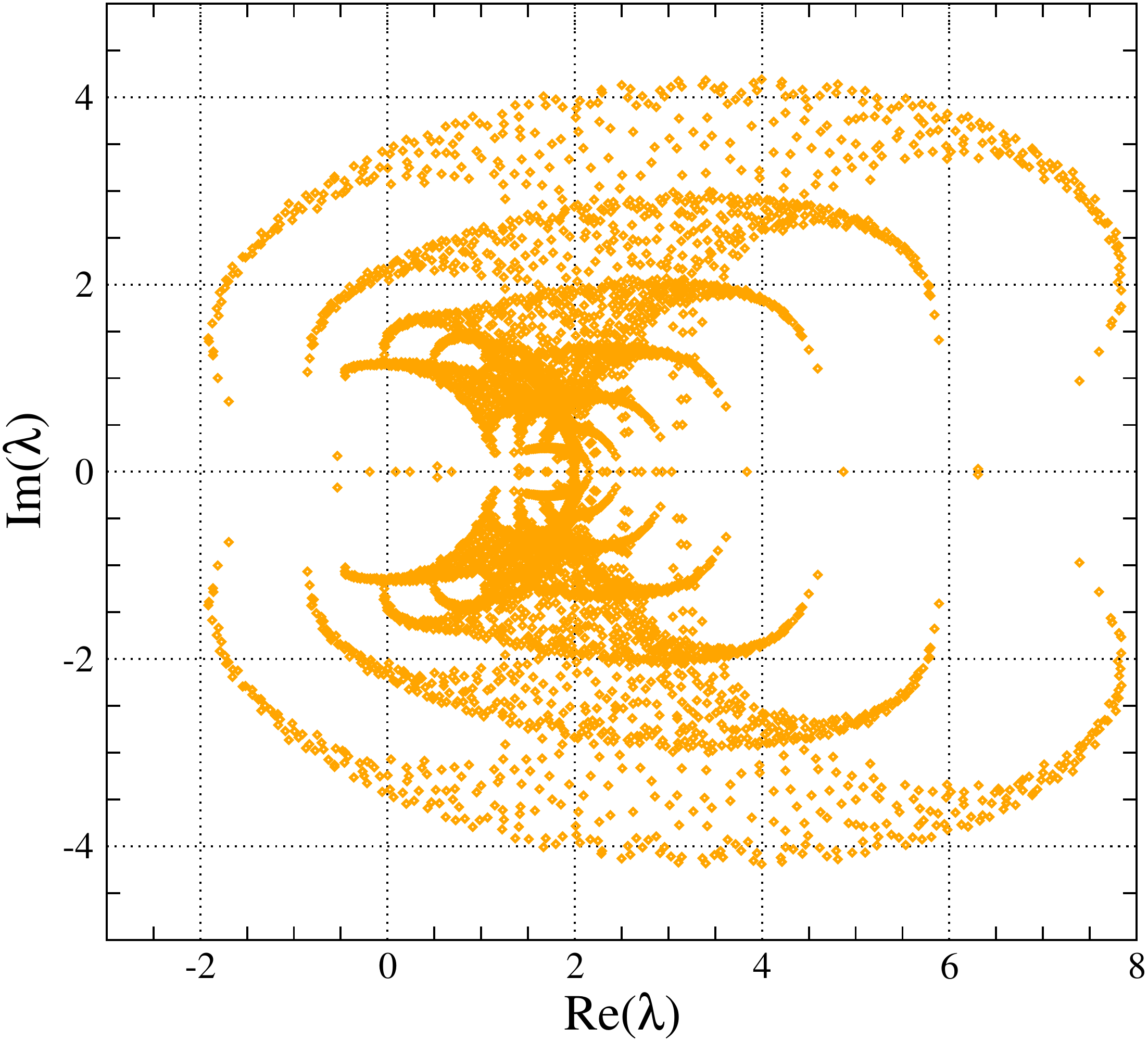}

}
\par\end{centering}
\caption{Spectrum of $D_{\mathsf{dw}}$ with Wilson kernel for $m=0$ at $N_{s}=8$.
\label{fig:gauge-wilson-dbulk}}
\end{figure*}
\begin{figure*}[!]
\begin{centering}
\subfloat[Standard construction]{\includegraphics[width=0.32\textwidth]{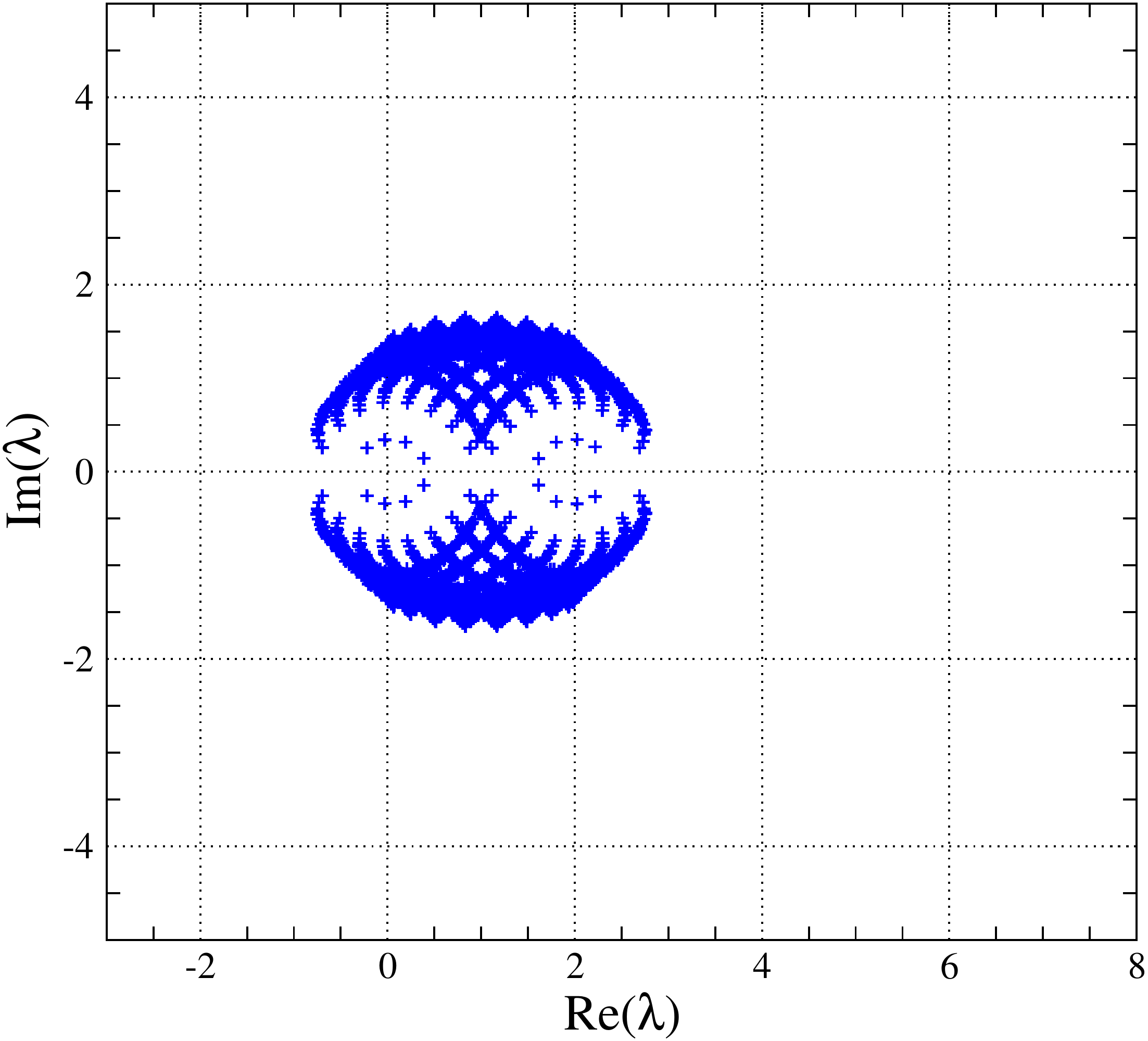}

}\hfill{}\subfloat[Boriçi's construction]{\includegraphics[width=0.32\textwidth]{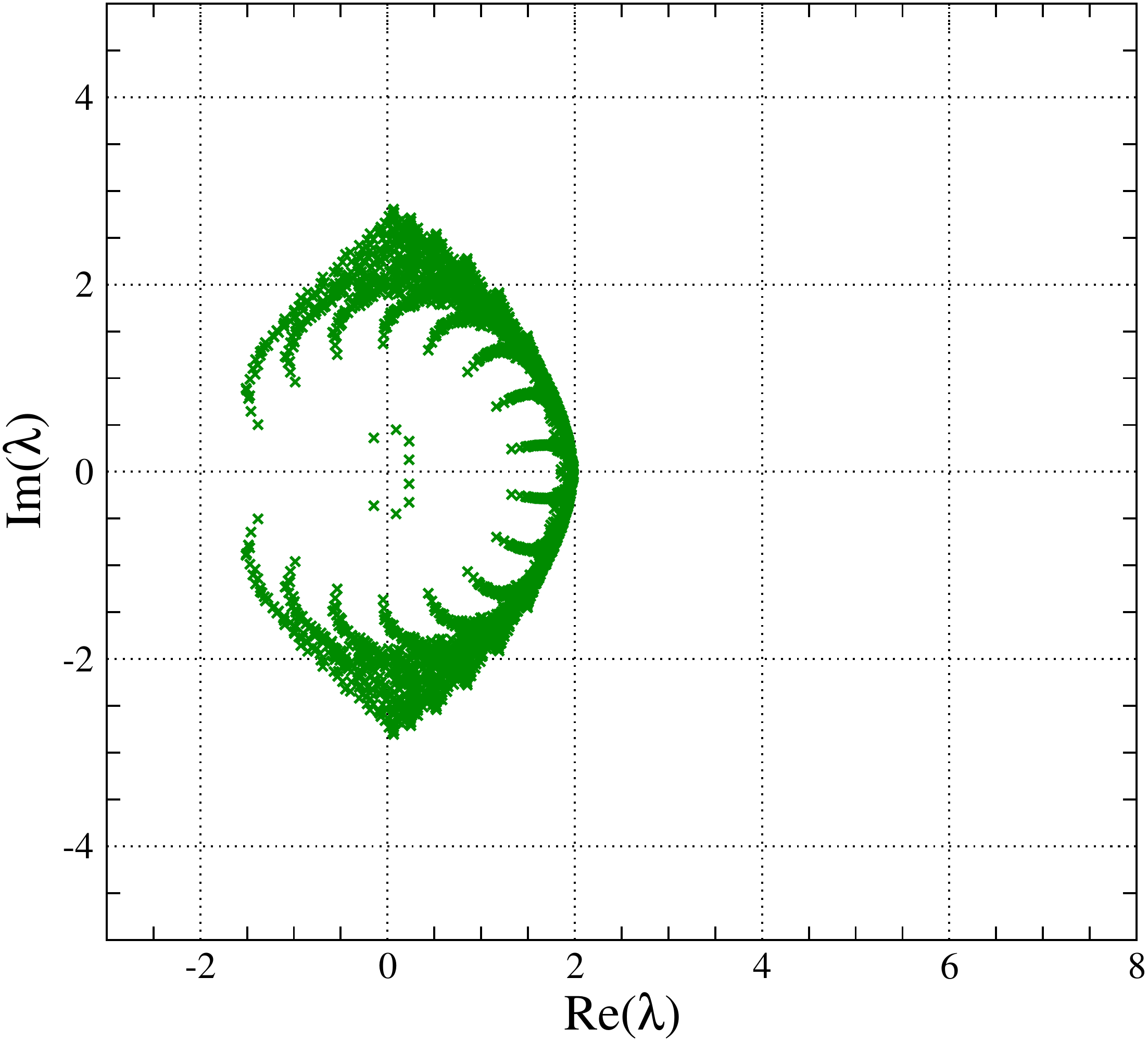}

}\hfill{}\subfloat[Optimal construction]{\includegraphics[width=0.32\textwidth]{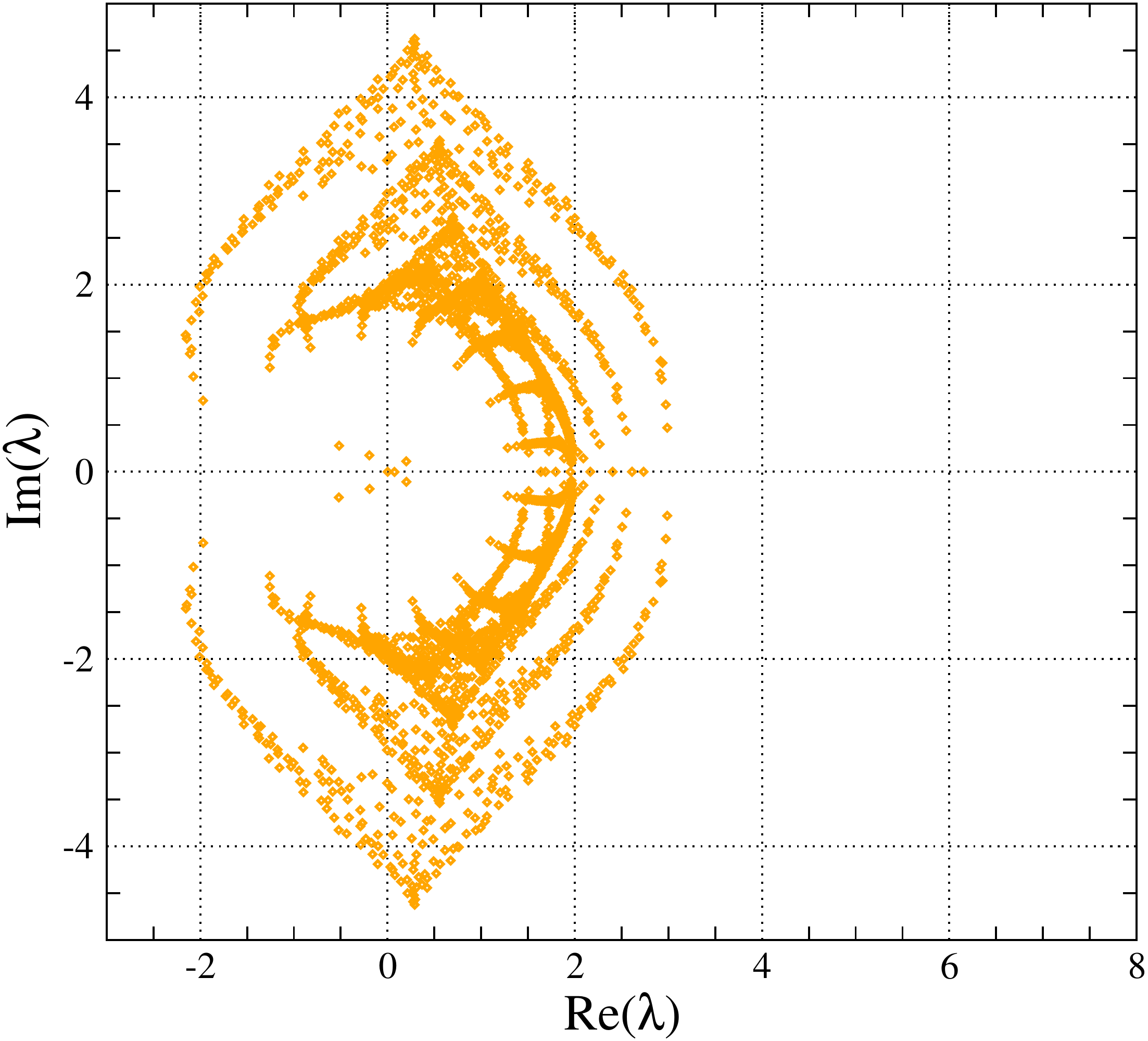}

}
\par\end{centering}
\caption{Spectrum of $D_{\mathsf{sdw}}$ with staggered Wilson kernel for $m=0$
at $N_{s}=8$. \label{fig:gauge-stw-dbulk}}
\end{figure*}

In Figs.~\ref{fig:gauge-wilson-dbulk} and \ref{fig:gauge-stw-dbulk},
we show the spectra of the bulk operators on the same gauge configuration
as used in Fig.~\ref{fig:gauge-kernel}. Due to the use of a gauge
configuration with $Q\neq0$, the effective operator is guaranteed
to have $\left|Q\right|$ exact zero modes in the limit $N_{s}\to\infty$.
In this setting, we find $N_{s}\cdot\left|Q\right|$ eigenvalues in
the vicinity of the origin in the bulk spectrum, which are linked
to these zero modes. We can also see how the optimal construction
distorts the spectrum, effectively improving chiral properties and
resulting in a reduced $m_{\mathsf{eff}}$.

\paragraph{Effective operators.}

\begin{figure*}[t]
\begin{centering}
\hfill{}\subfloat[Wilson kernel]{\includegraphics[width=0.4\textwidth]{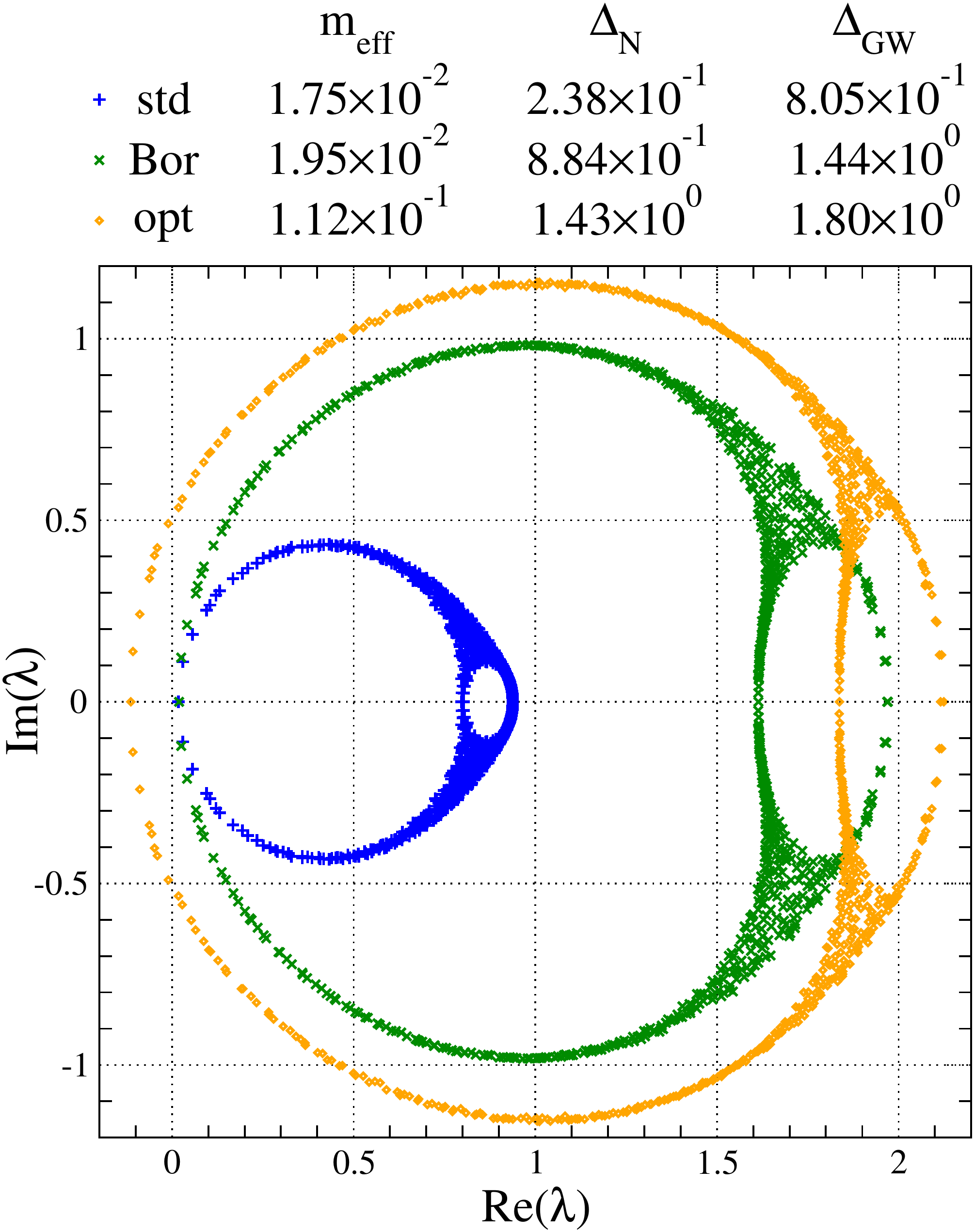}

}\hfill{}\subfloat[Staggered Wilson kernel]{\includegraphics[width=0.4\textwidth]{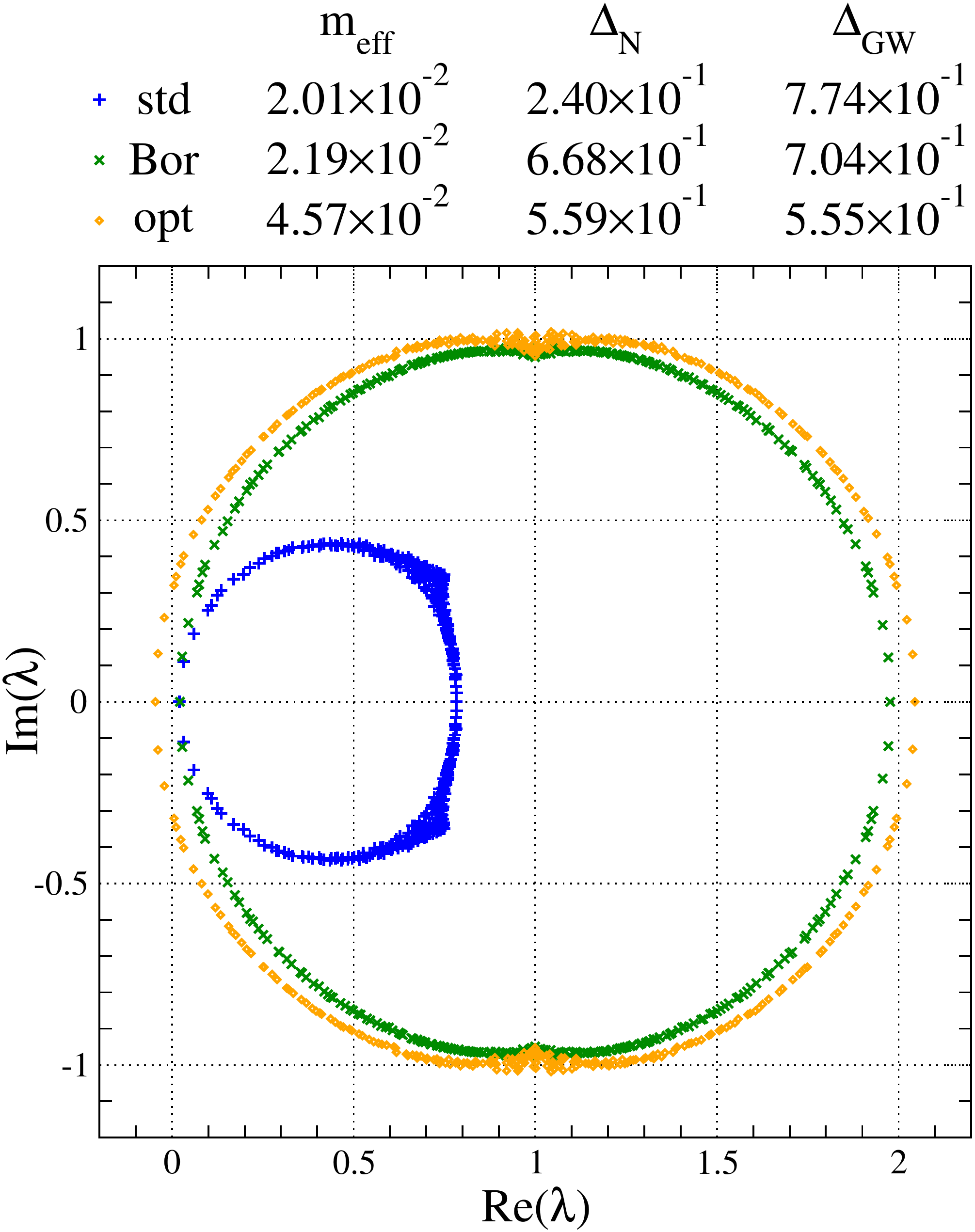}

}\hfill{} 
\par\end{centering}
\caption{Spectrum of $\varrho D_{\mathsf{eff}}$ at $N_{s}=2$ for the standard
(std), Boriçi (Bor) and optimal (opt) construction. \label{fig:gauge-deff-Ns2}}
\end{figure*}
\begin{figure*}[!]
\begin{centering}
\hfill{}\subfloat[Wilson kernel]{\includegraphics[width=0.4\textwidth]{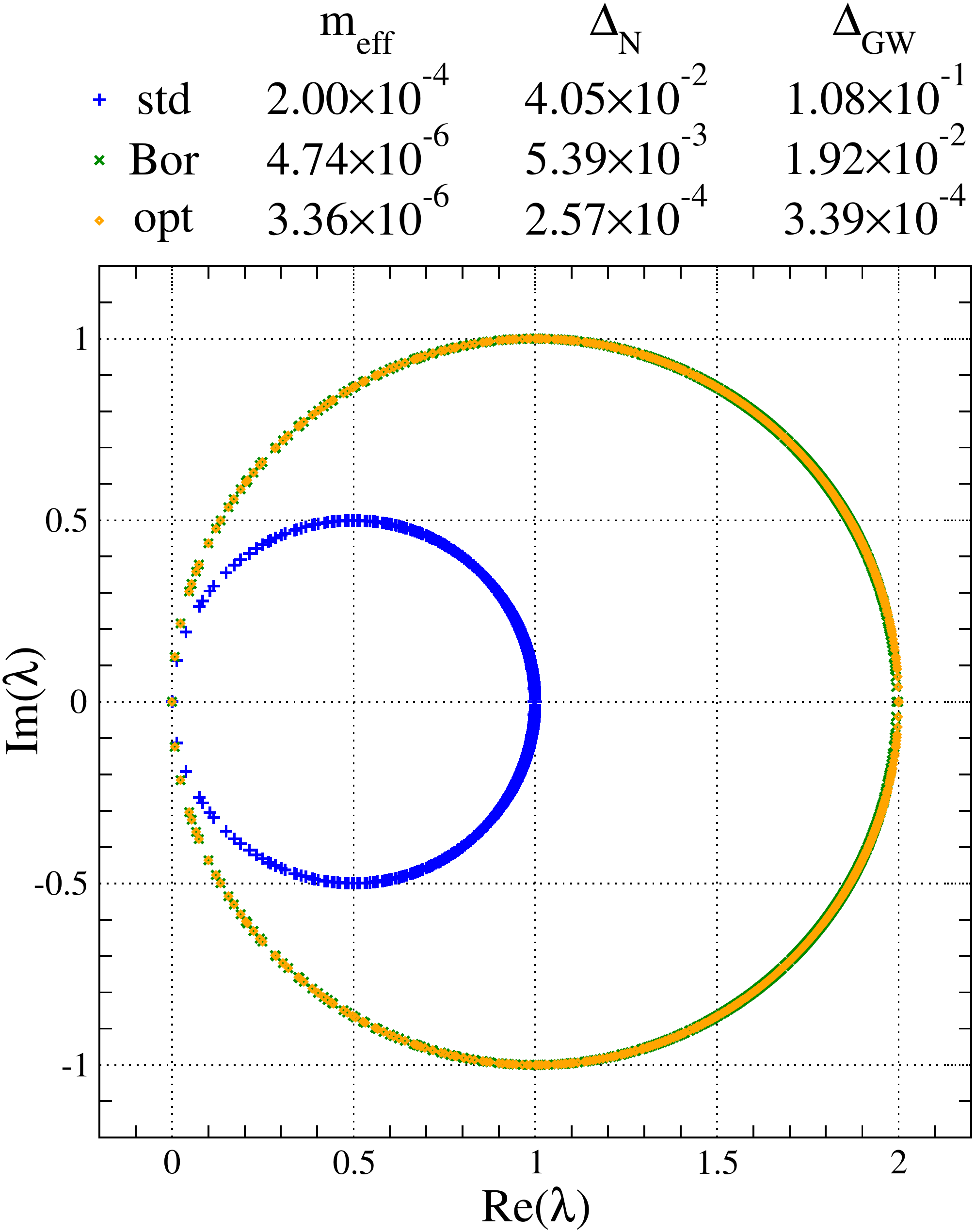}

}\hfill{}\subfloat[Staggered Wilson kernel \label{fig:gauge-deff-Ns8-b}]{\includegraphics[width=0.4\textwidth]{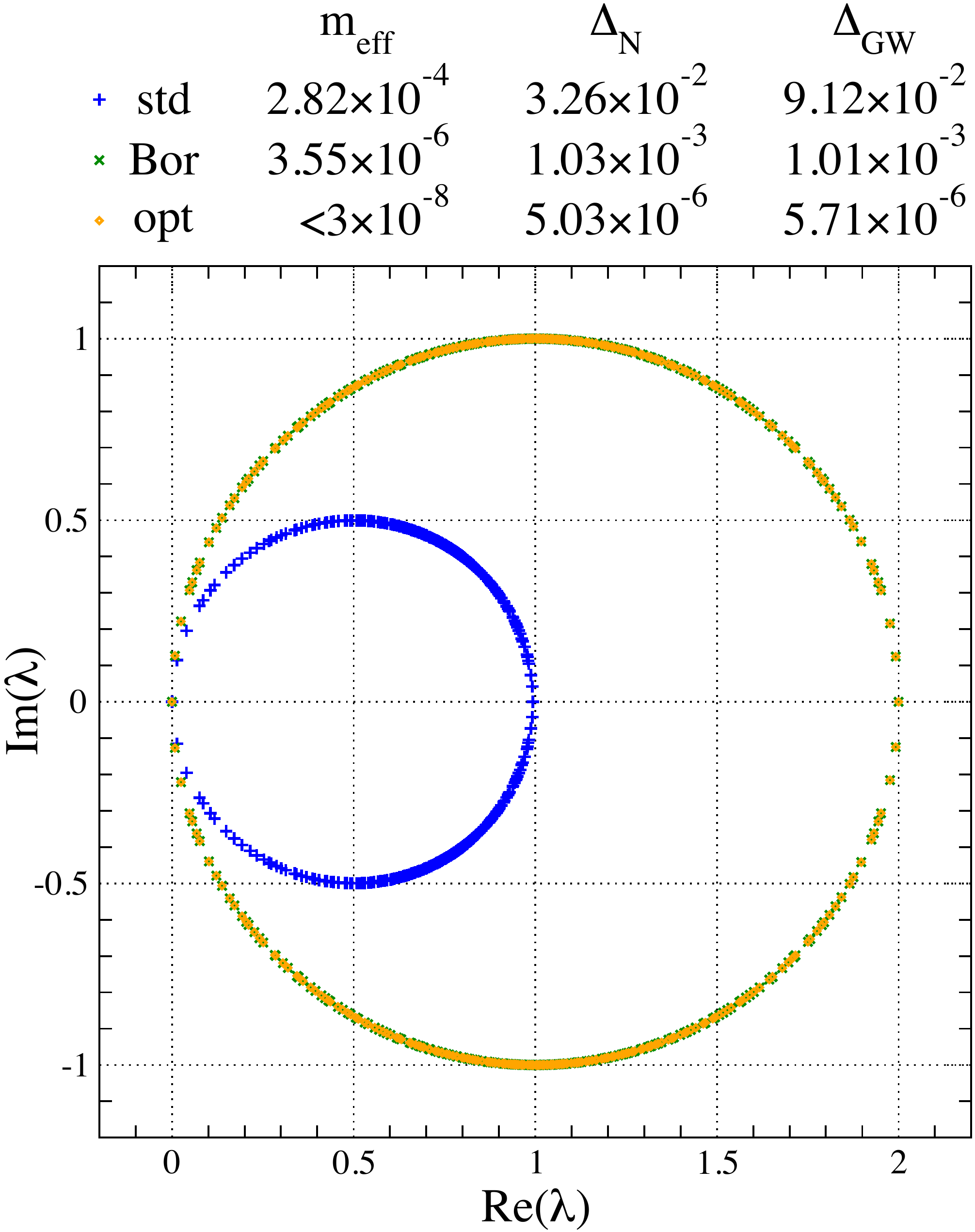}

}\hfill{} 
\par\end{centering}
\caption{Spectrum of $\varrho D_{\mathsf{eff}}$ at $N_{s}=8$ for the standard
(std), Boriçi (Bor) and optimal (opt) construction. \label{fig:gauge-deff-Ns8}}
\end{figure*}
\begin{figure*}[!]
\begin{centering}
\hfill{}\subfloat[Wilson kernel]{\includegraphics[width=0.4\textwidth]{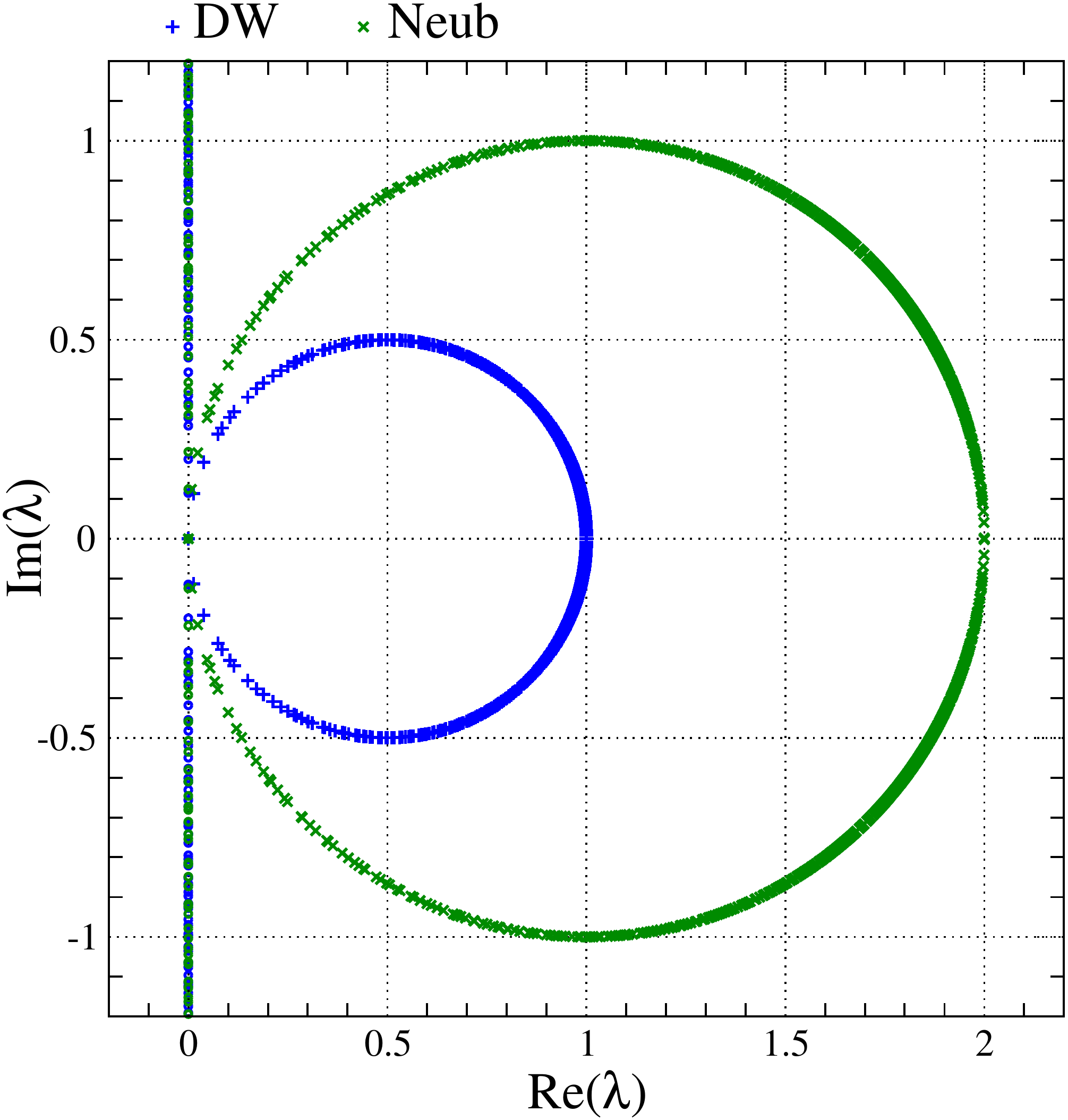}

}\hfill{}\subfloat[Staggered Wilson kernel]{\includegraphics[width=0.4\textwidth]{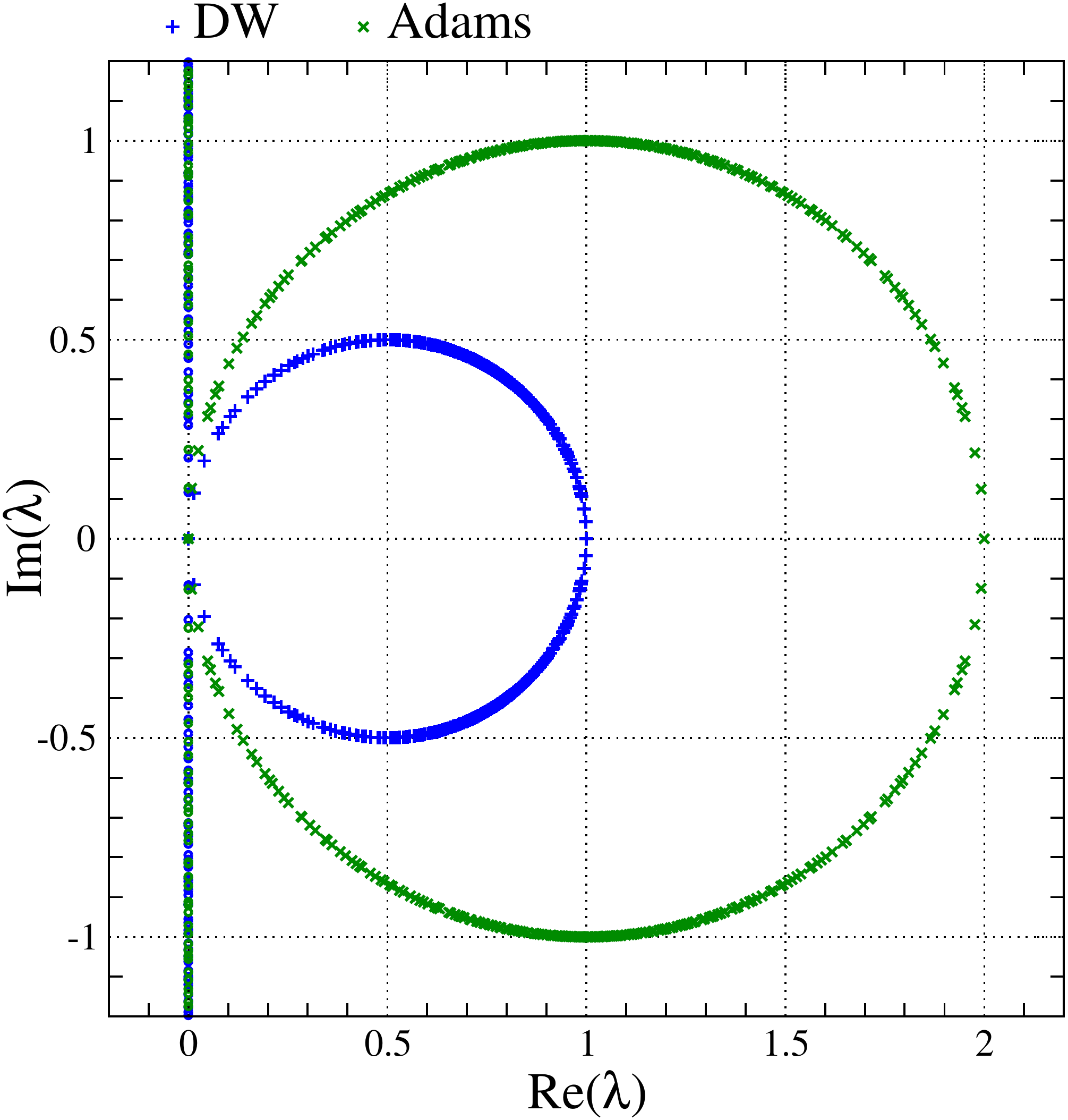}

}\hfill{} 
\par\end{centering}
\caption{Spectrum of $\varrho D_{\mathsf{ov}}$ with stereographic projection
for domain wall (DW)\protect \\
and standard (Neub/Adams) kernels. \label{fig:gauge-dov} }
\end{figure*}
\begin{figure*}[!]
\begin{centering}
\hfill{}\subfloat[No smearing]{\includegraphics[width=0.4\textwidth]{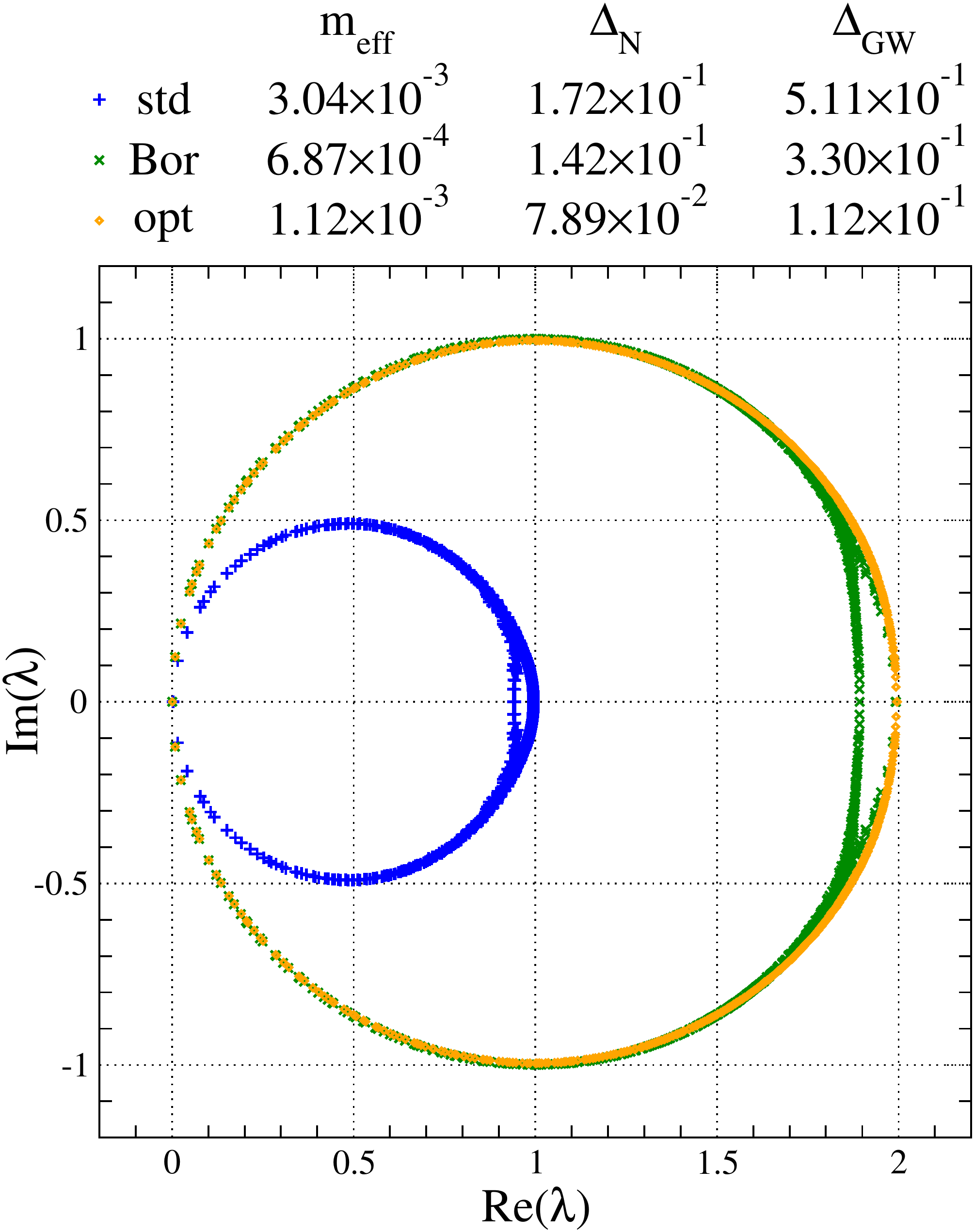}

}\hfill{}\subfloat[Three smearing iterations]{\includegraphics[width=0.4\textwidth]{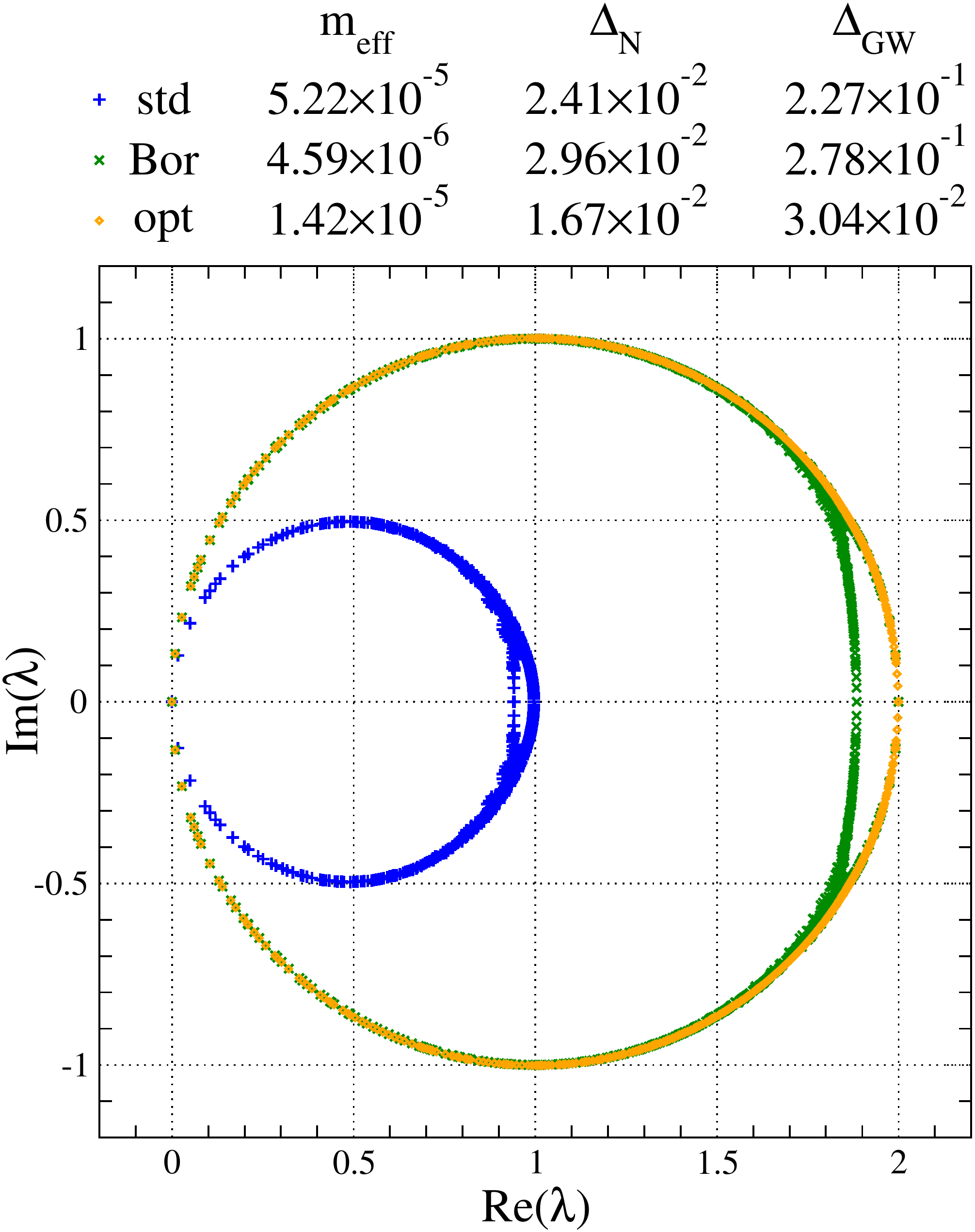}

}\hfill{} 
\par\end{centering}
\caption{Spectrum of $\varrho D_{\mathsf{eff}}$ with Wilson kernel and smearing
parameter $\alpha=0.5$ at $N_{s}=4$\protect \\
for the standard (std), Boriçi (Bor) and optimal (opt) construction.\label{fig:gauge-wilson-smearing}}
\end{figure*}
\begin{figure*}[t]
\begin{centering}
\hfill{}\subfloat[No smearing]{\includegraphics[width=0.4\textwidth]{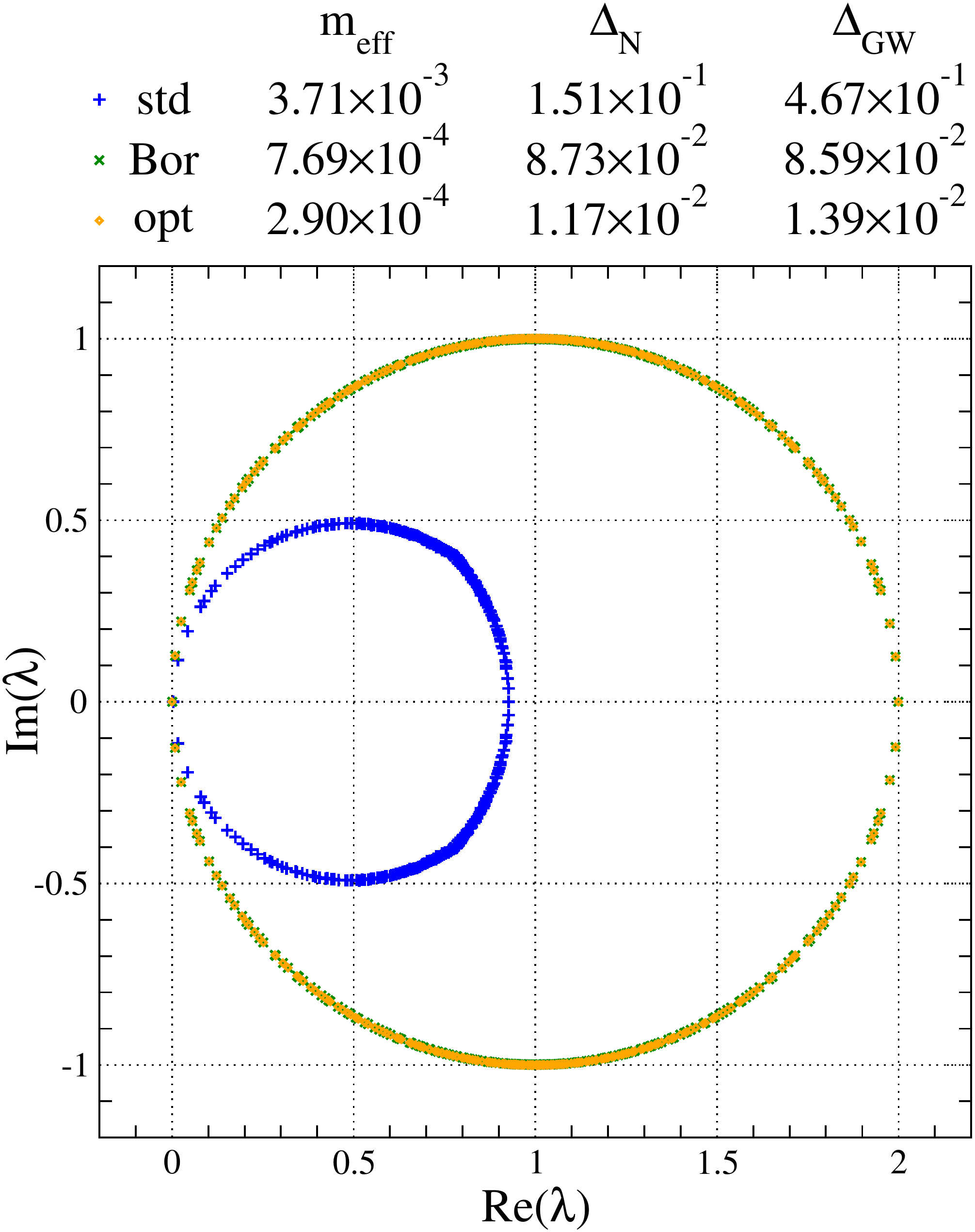}

}\hfill{}\subfloat[Three smearing iterations]{\includegraphics[width=0.4\textwidth]{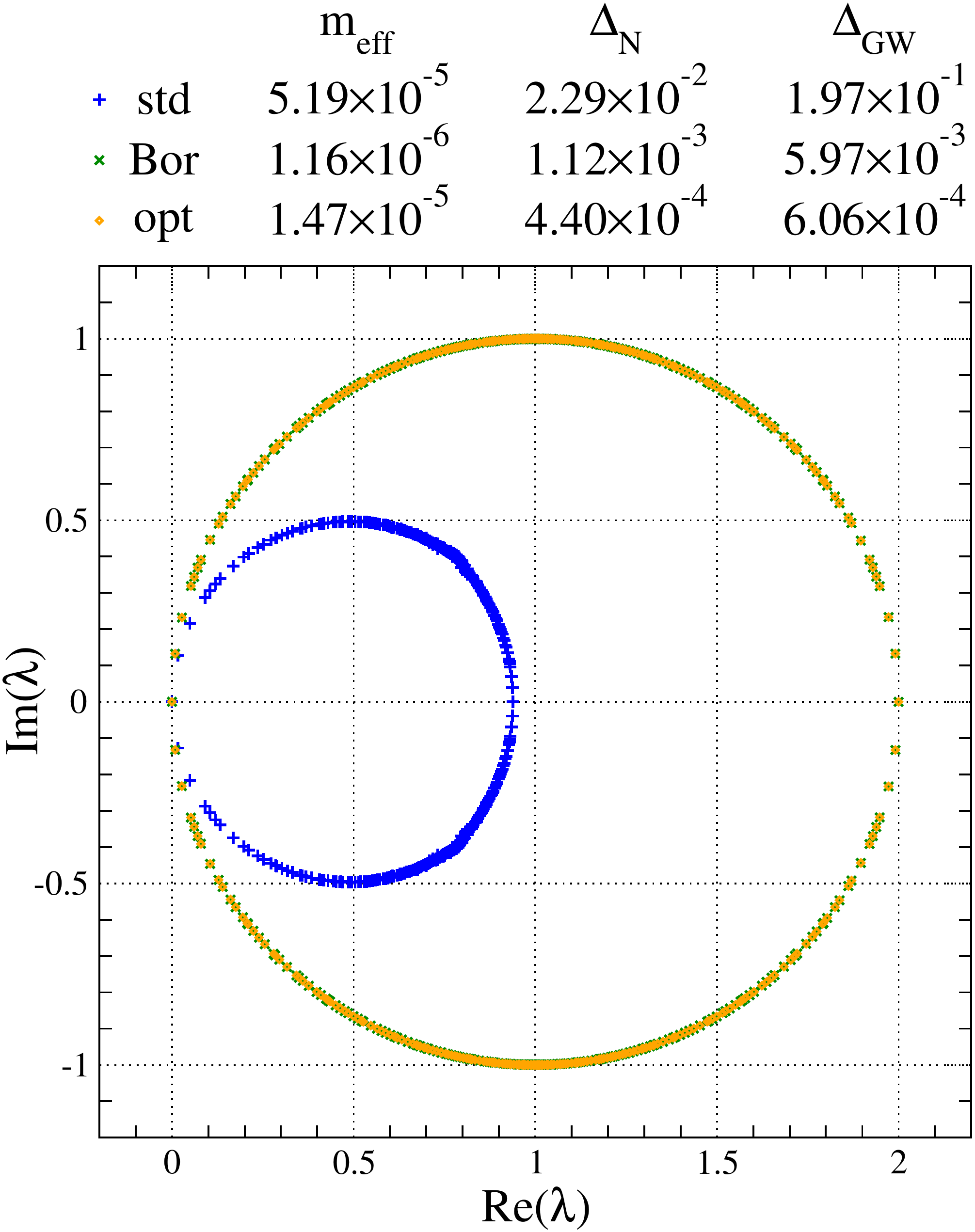}

}\hfill{} 
\par\end{centering}
\caption{Spectrum of $\varrho D_{\mathsf{eff}}$ with staggered Wilson kernel
and smearing parameter $\alpha=0.5$ at $N_{s}=4$\protect \\
for the standard (std), Boriçi (Bor) and optimal (opt) construction.
\label{fig:gauge-stw-smearing}}
\end{figure*}
\begin{figure*}[t]
\subfloat[Configuration with $Q=0$]{\includegraphics[width=0.32\textwidth]{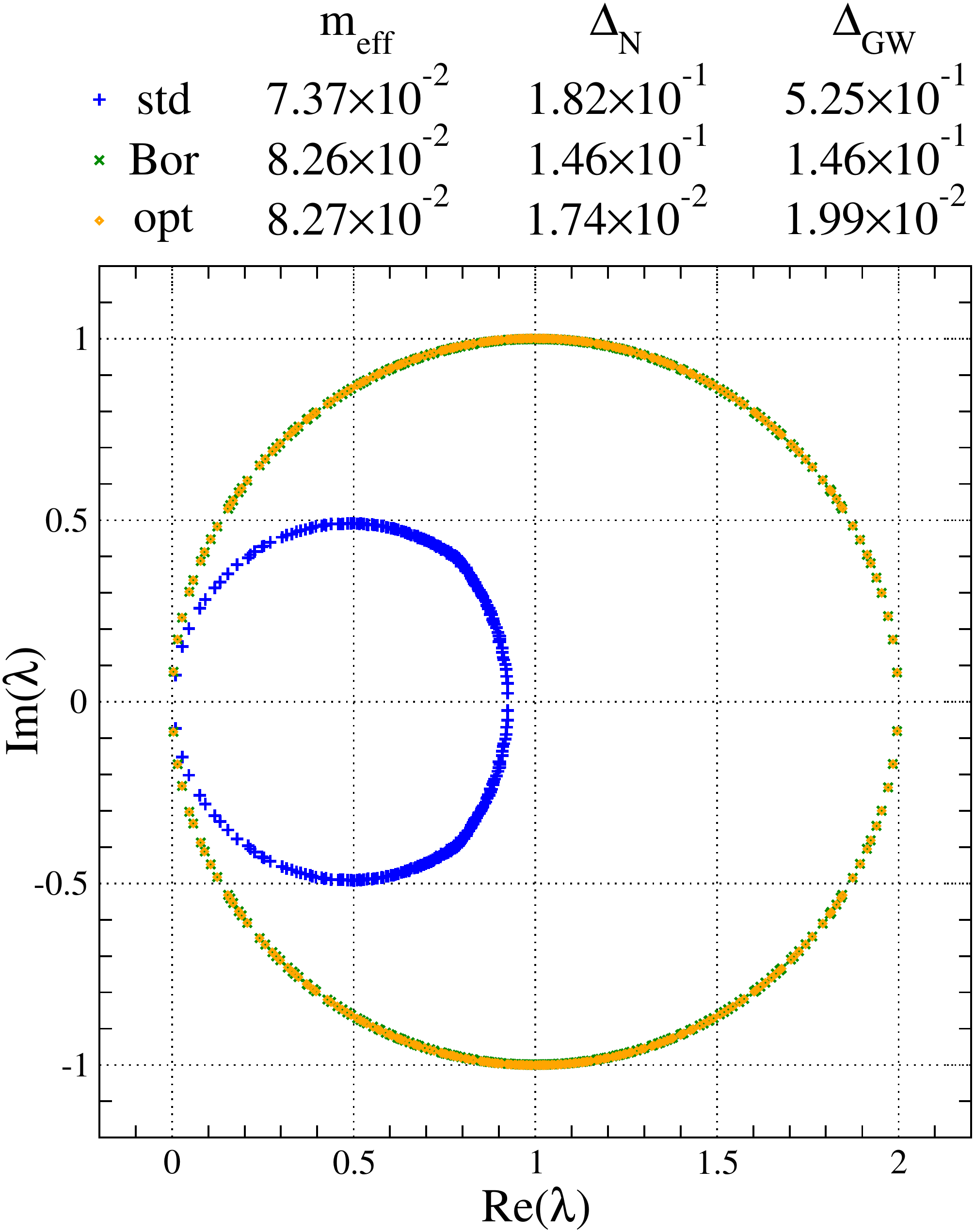}

}\hfill{}\subfloat[Configuration with $Q=1$]{\includegraphics[width=0.32\textwidth]{gauge_field/schw_20_20_5_1009_0.65_0/stw_deff_4}

}\hfill{}\subfloat[Configuration with $Q=3$]{\includegraphics[width=0.32\textwidth]{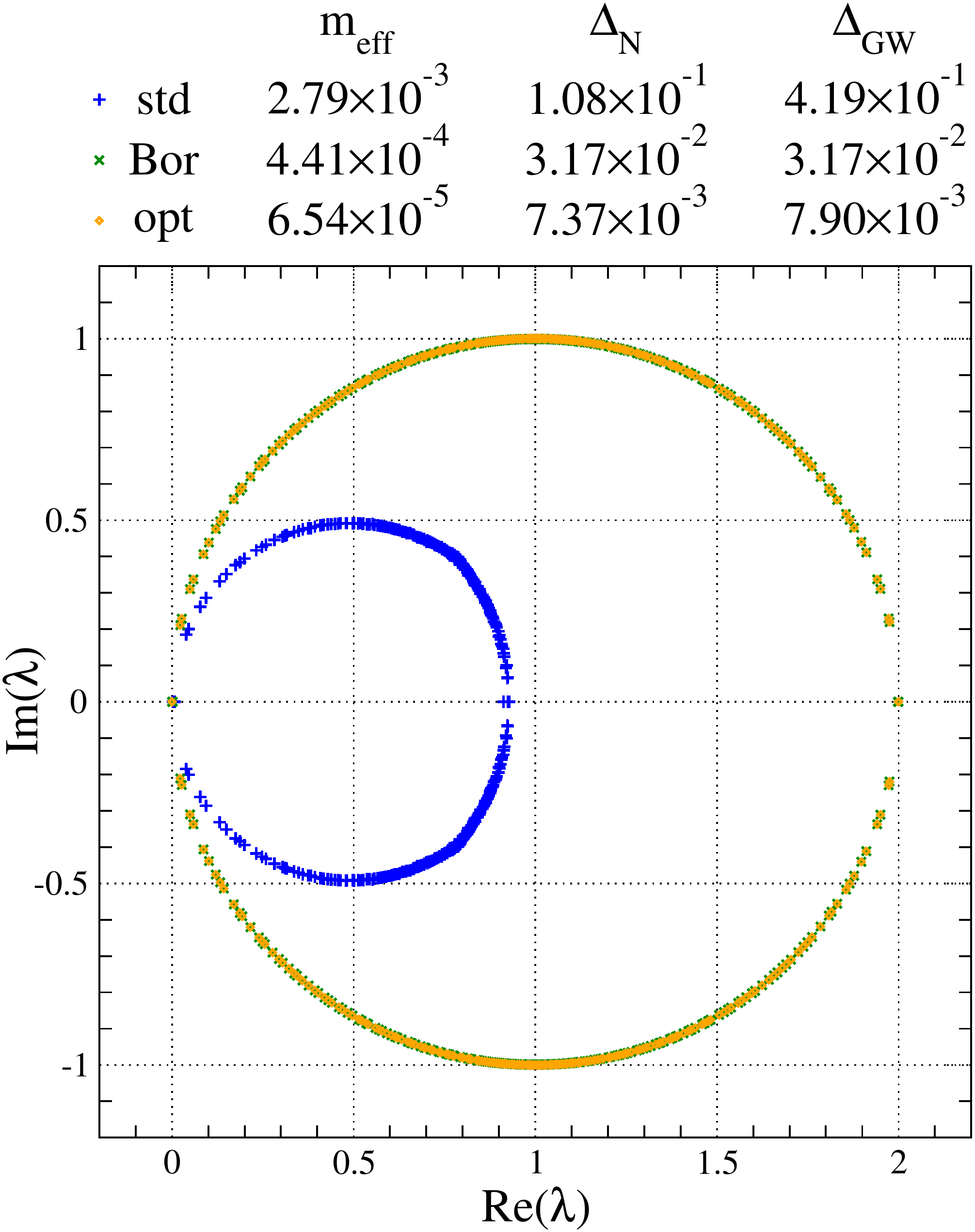}

}

\caption{Spectrum of $\varrho D_{\mathsf{eff}}$ with staggered Wilson kernel
for various topological charges $Q$ at $N_{s}=4$\protect \\
for the standard (std), Boriçi (Bor) and optimal (opt) construction.
\label{fig:gauge-stw-topology}}
\end{figure*}

In Figs.~\ref{fig:gauge-deff-Ns2} and \ref{fig:gauge-deff-Ns8},
we plot the spectra of the effective operators on the same gauge field
background as used in Fig.~\ref{fig:gauge-kernel}. We note that
for $N_{s}\geq4$ Boriçi's construction outperforms the standard construction
with respect to all measures $m_{\mathsf{eff}}$, $\Delta_{\mathsf{N}}$
and $\Delta_{\mathsf{GW}}$.

The optimal construction decreases most of these numbers even further.
In Fig.~\ref{fig:gauge-deff-Ns8-b} we see that $m_{\mathsf{eff}}$
is already comparable to the round-off error and, hence, we only quote
an upper bound. Note that as in a $\Uone$ background field the kernel
operator is in general not normal, the inequality $\left|\,\left|\lambda\right|-1\,\right|\leq\delta_{\mathsf{max}}$
does not have to be saturated. Moreover it is evident that a smaller
maximum deviation $\delta_{\mathsf{max}}$ from the $\sign$ function
on a given interval does not necessarily translate to a smaller $\Delta_{\mathsf{GW}}$,
although there is a strong correlation. For larger $N_{s}$ this problem
is cured by the fast convergence of optimal domain wall fermions.

Let us remark that optimal domain wall fermions are optimal in a very
particular sense, namely the minimization of $\delta_{\mathsf{max}}$
as defined in Eq.~\eqref{eq:DeltaMax}. As Ref.~\cite{Brower:2005qw}
suggests, they are not optimal with respect to e.g.~the number of
iterations needed for solving a linear system. In principle, one could
also formulate domain wall fermions optimized with respect to other
measures, such as the minimization of $\Delta_{\mathsf{GW}}$ (which,
however, might require more knowledge about the spectrum).

Comparing domain wall fermions with a Wilson and staggered Wilson
kernel, we can see that in the case of the standard construction $m_{\mathsf{eff}}$,
$\Delta_{\mathsf{N}}$ and $\Delta_{\mathsf{GW}}$ are usually of
the same magnitude. However, for Boriçi's and the optimal construction
a staggered Wilson kernel seems to outperform the usual Wilson kernel
in terms of chiral symmetry violations in the $\Uone$ background
fields under consideration.

For the rather artificial case $N_{s}=2$ we can make some interesting
observations. While for a staggered Wilson kernel the relative performances
of all formulations under consideration is comparable, for a Wilson
kernel the standard formulation performs better than Boriçi's and
markedly better than optimal.

\paragraph{Overlap operators.}

In Fig.~\ref{fig:gauge-dov}, we show the corresponding overlap operators
together with the stereographic projection of the spectrum. All of
the quantities $m_{\mathsf{eff}}$, $\Delta_{\mathsf{N}}$ and $\Delta_{\mathsf{GW}}$
vanish in exact arithmetic and are, therefore, omitted in the figure
labels.

\subsection{Smearing}

In realistic simulations, smearing is a commonly employed technique.
As suggested in Ref.~\cite{Durr:2013gp}, it is expected to be especially
beneficial when employing a staggered Wilson kernel (in $3+1$ dimensions,
this is more pronounced). Hence, we consider three-step \textsc{ape}
smeared \cite{Falcioni:1984ei,Albanese:1987ds} gauge field backgrounds
with a smearing parameter $\alpha=0.5$, which is the maximal value
within the perturbatively reasonable range in two dimensions \cite{Capitani:2006ni}.

As a first test on the impact on the performance of staggered domain
wall fermions, we can find a direct comparison at $N_{s}=4$ in Figs.~\ref{fig:gauge-wilson-smearing}
and \ref{fig:gauge-stw-smearing} between an unsmeared and a smeared
version of a configuration both for a Wilson and a staggered Wilson
kernel. As we can see, with three smearing steps, $m_{\mathsf{eff}}$,
$\Delta_{\mathsf{N}}$ and $\Delta_{\mathsf{GW}}$ get reduced significantly—in
some cases of up to $2$ orders of magnitude—with the exception of
$m_{\mathsf{eff}}$ for standard domain wall fermions with Wilson
kernel.

\subsection{Topology \label{subsec:Topology}}

\begin{figure*}[!t]
\begin{centering}
\hfill{}\subfloat[Wilson kernel]{\includegraphics[height=0.35\textwidth]{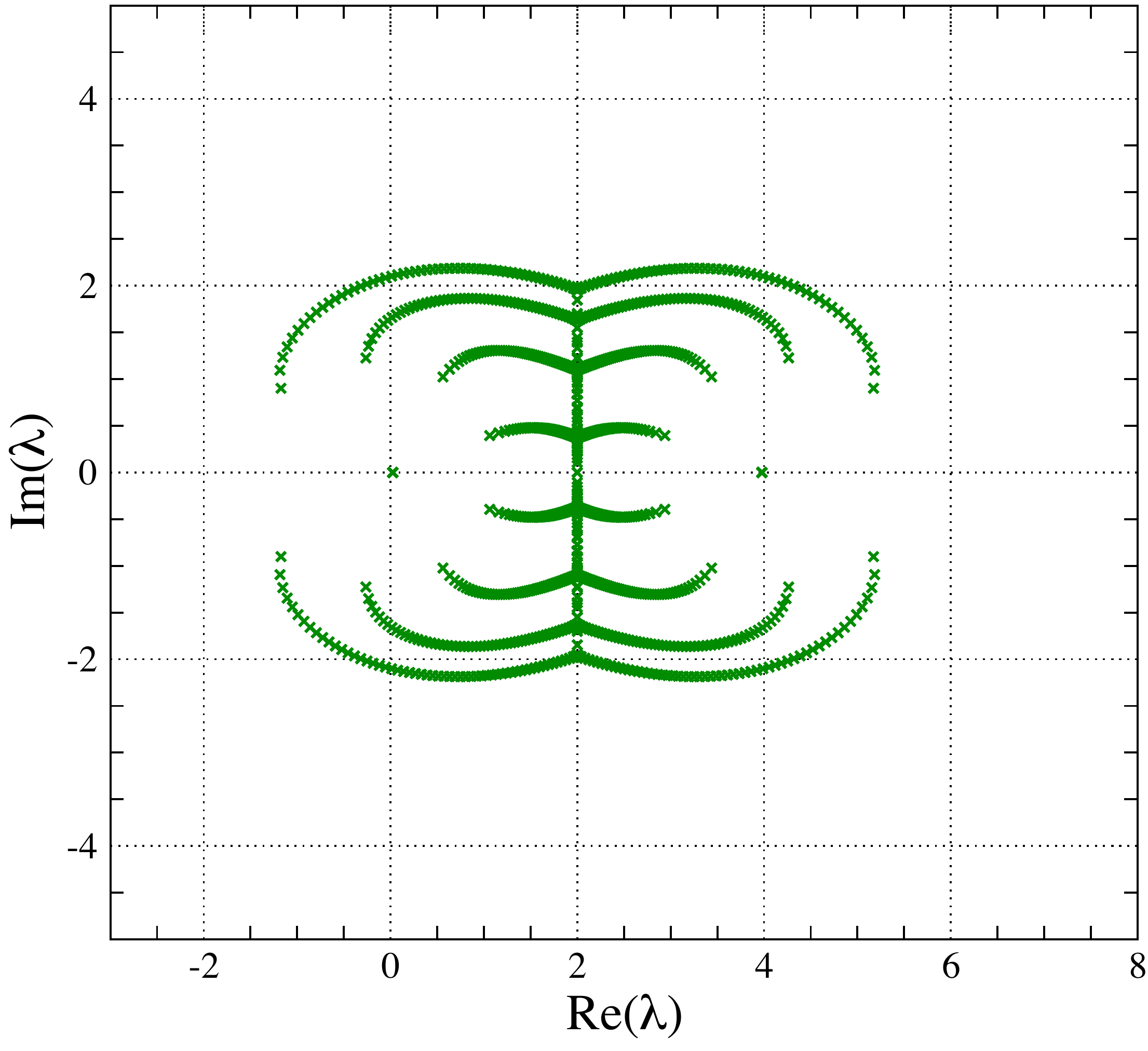}

}\hfill{}\subfloat[Staggered Wilson kernel]{\includegraphics[height=0.35\textwidth]{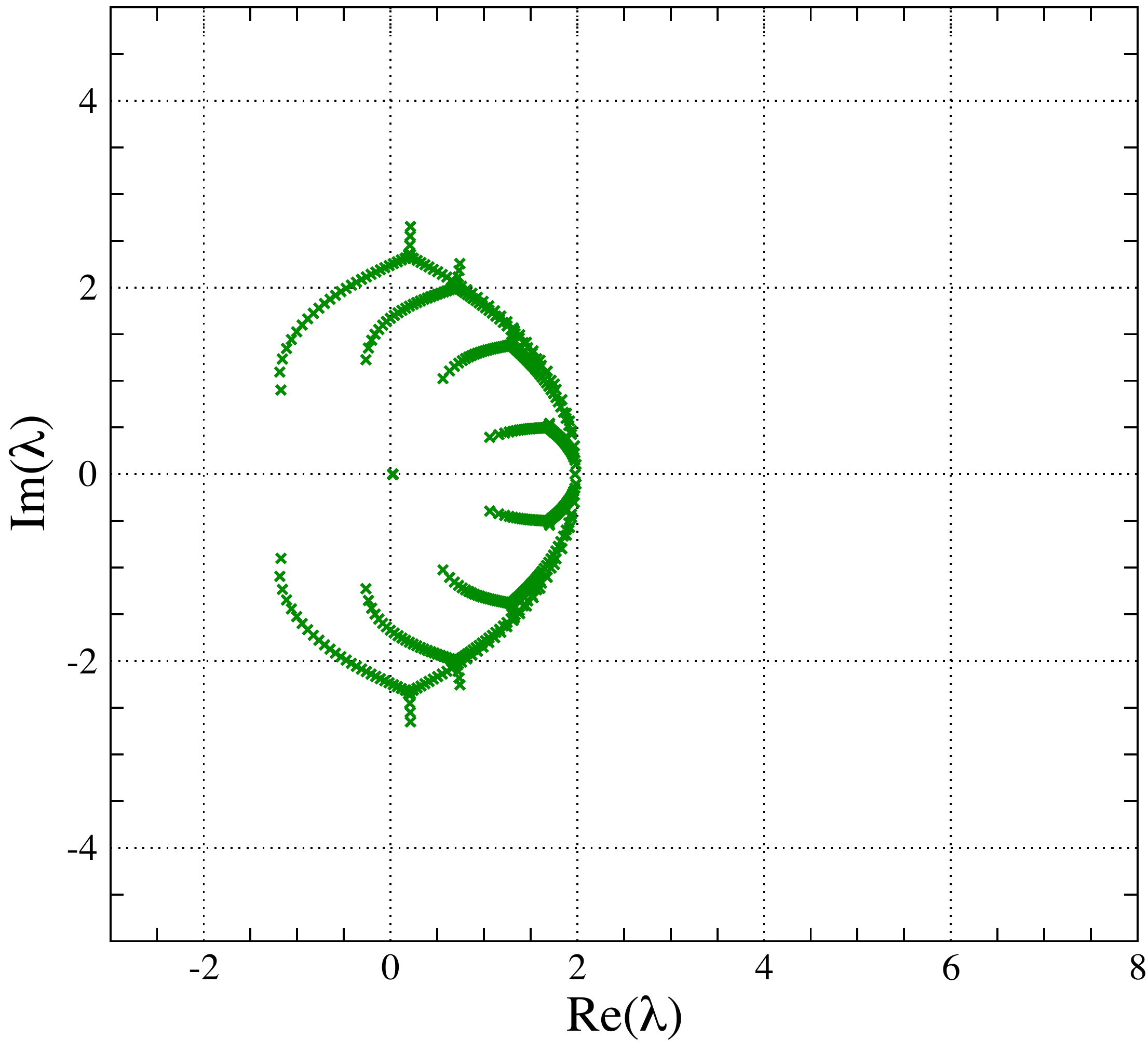}

}\hfill{} 
\par\end{centering}
\caption{Spectrum of $D_{\mathsf{dw}}$ in Boriçi's construction at $N_{s}=4$
for a topological configuration with $Q=3$. \label{fig:topological-configs}}
\end{figure*}

As discussed earlier, for topological charges $Q\neq0$, the Atiyah-Singer
index theorem guarantees the existence of zero modes of the continuum
Dirac operator. On the lattice one can show the same for the overlap
operators defined in Eq.~\eqref{eq:DefDov}. As a consequence we
observe approximate zero modes in the eigenvalue spectra of the effective
operators $\varrho D_{\mathsf{eff}}$ as illustrated in Fig.~\ref{fig:gauge-stw-topology}.
As the Vanishing Theorem holds in $1+1$ dimensions, we find these
modes with a multiplicity of $\left|Q\right|$.

In addition, we can study topological aspects by employing the method
in Ref.~\cite{Smit:1986fn} to construct gauge configurations with
given topological charge $Q$. These smooth configurations are fixed
points with respect to the \textsc{ape} smearing prescription. We
construct these configurations for a wide range of topological charges
$Q$ and evaluate the measures $m_{\mathsf{eff}}$, $\Delta_{\mathsf{N}}$
and $\Delta_{\mathsf{GW}}$ for the effective operators. In this setting,
these measures are equal within numerical rounding errors for topological
charge $\pm Q$ and, thus, only depend on $\left|Q\right|$.

We find that $m_{\mathsf{eff}}$, $\Delta_{\mathsf{N}}$ and $\Delta_{\mathsf{GW}}$
are very small in magnitude on these specific configurations compared
to thermalized configurations. Moreover, they increase with larger
values of $\left|Q\right|$. In Fig.~\ref{fig:topological-configs},
we can see two examples of bulk spectra on a $20^{2}$ lattice, which
reveal a very clear structure on these smooth background fields.

\subsection{Spectral flow}

Another tool to investigate topological aspects on the lattice, such
as the index theorem, is the spectral flow \cite{Itoh:1987iy,Narayanan:1994gw}.

\begin{figure*}[!t]
\subfloat[Unsmeared configuration]{\includegraphics[width=0.32\textwidth]{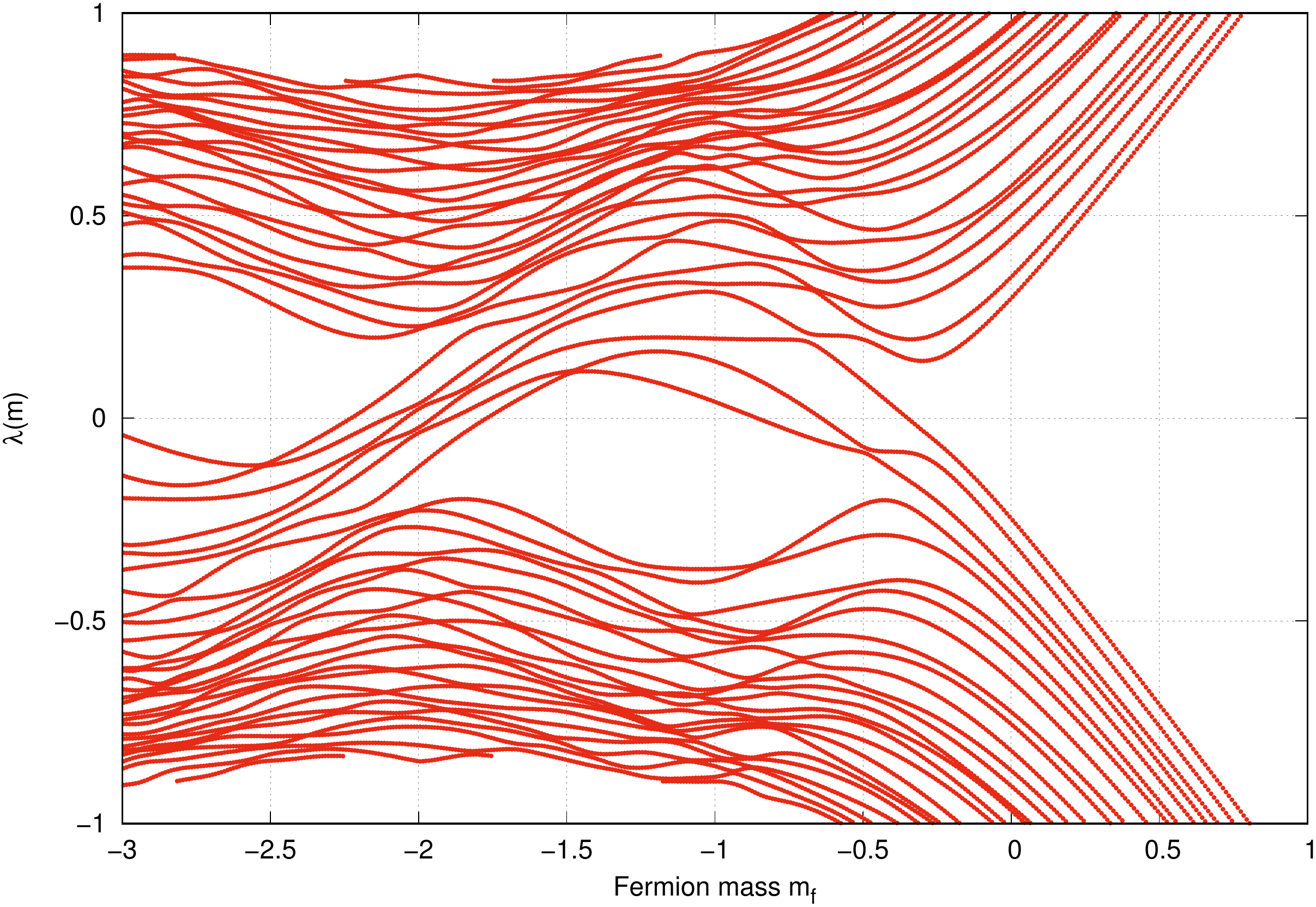}

}\hfill{}\subfloat[Smeared configuration]{\includegraphics[width=0.32\textwidth]{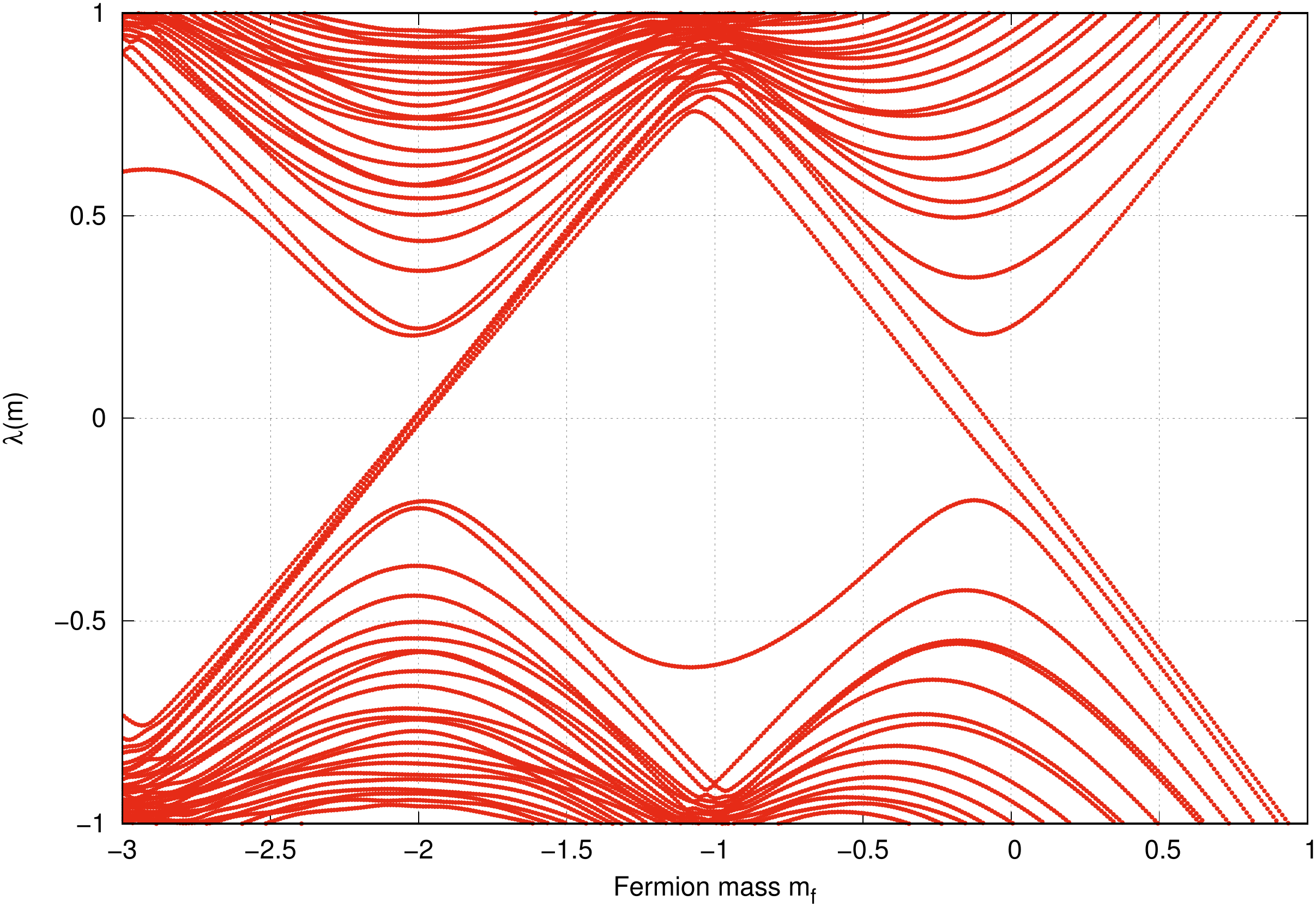}

}\hfill{}\subfloat[Topological configuration]{\includegraphics[width=0.32\textwidth]{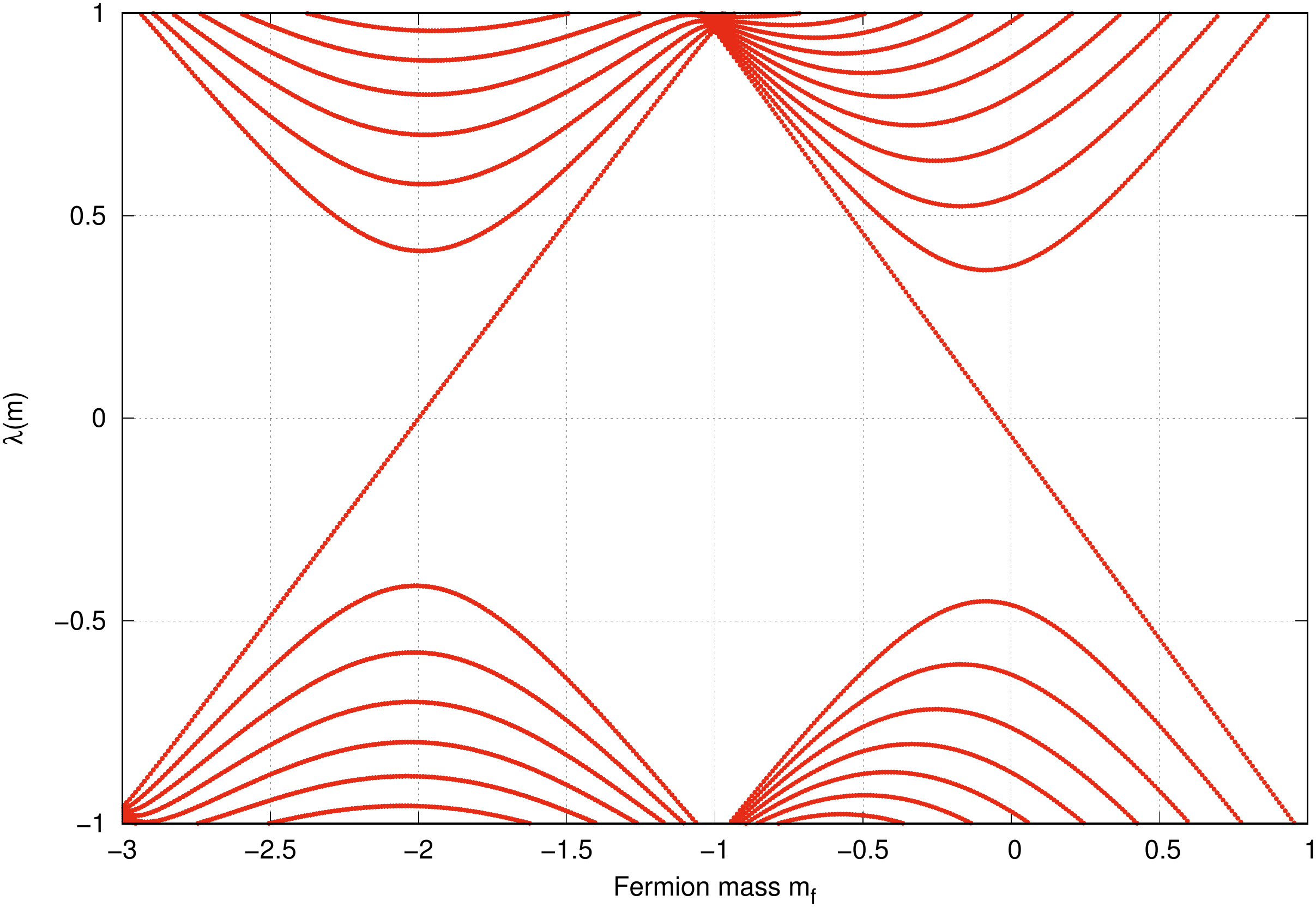}

}

\caption{Spectrum of the Wilson kernel for gauge fields with $Q=2$. \label{fig:spectral-flow-wilson}}
\end{figure*}
\begin{figure*}[!t]
\subfloat[Unsmeared configuration]{\includegraphics[width=0.32\textwidth]{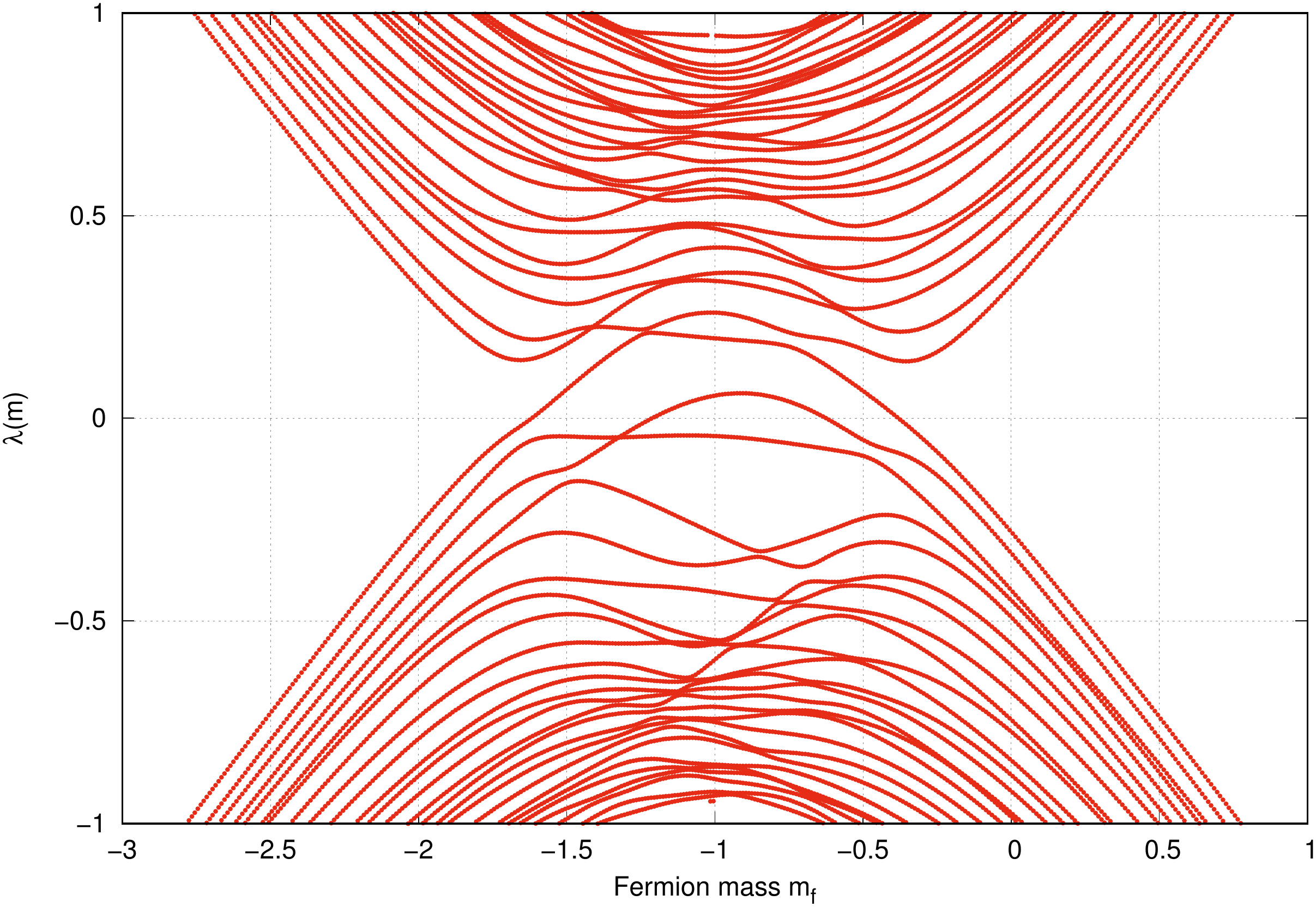}

}\hfill{}\subfloat[Smeared configuration]{\includegraphics[width=0.32\textwidth]{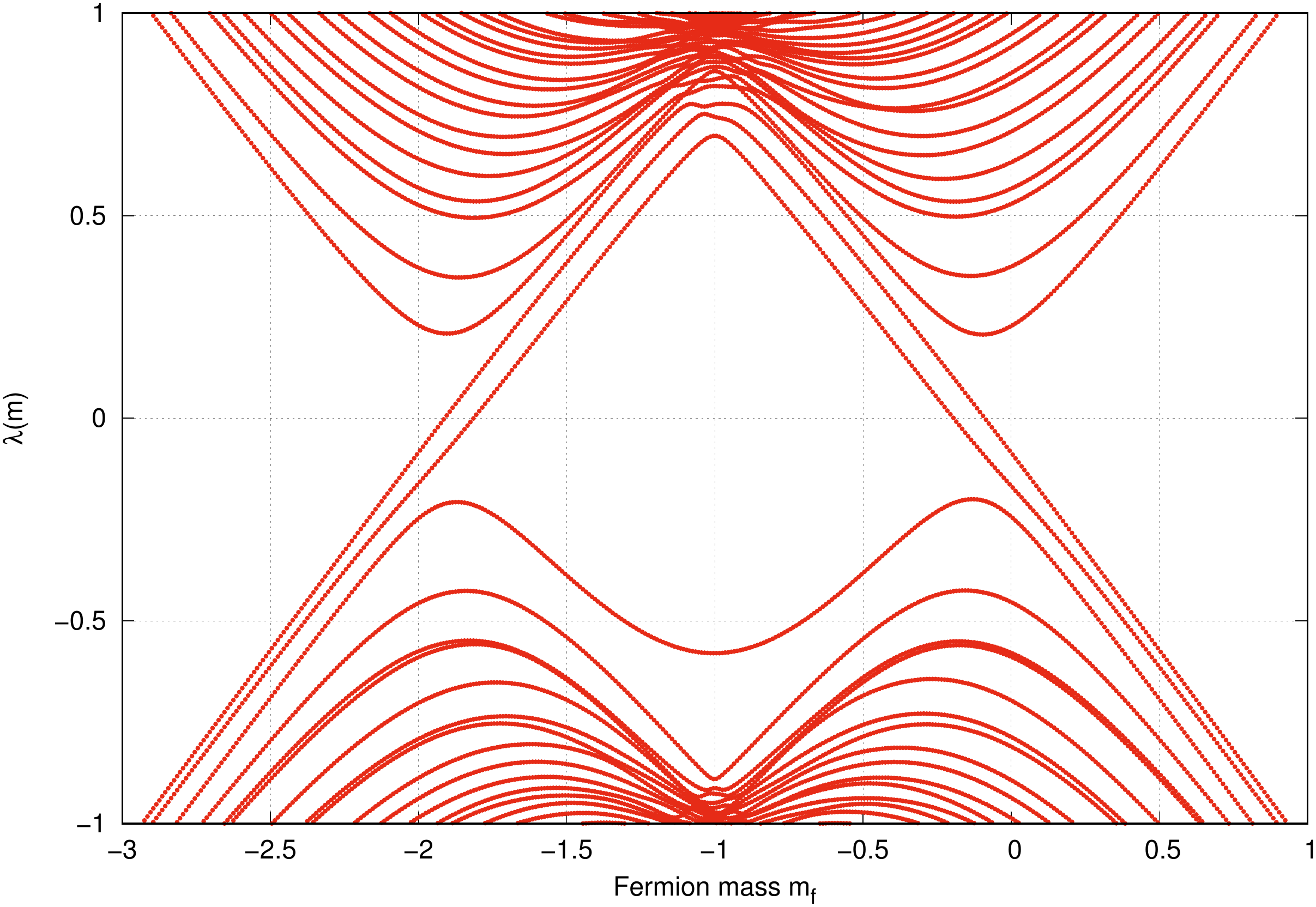}

}\hfill{}\subfloat[Topological configuration]{\includegraphics[width=0.32\textwidth]{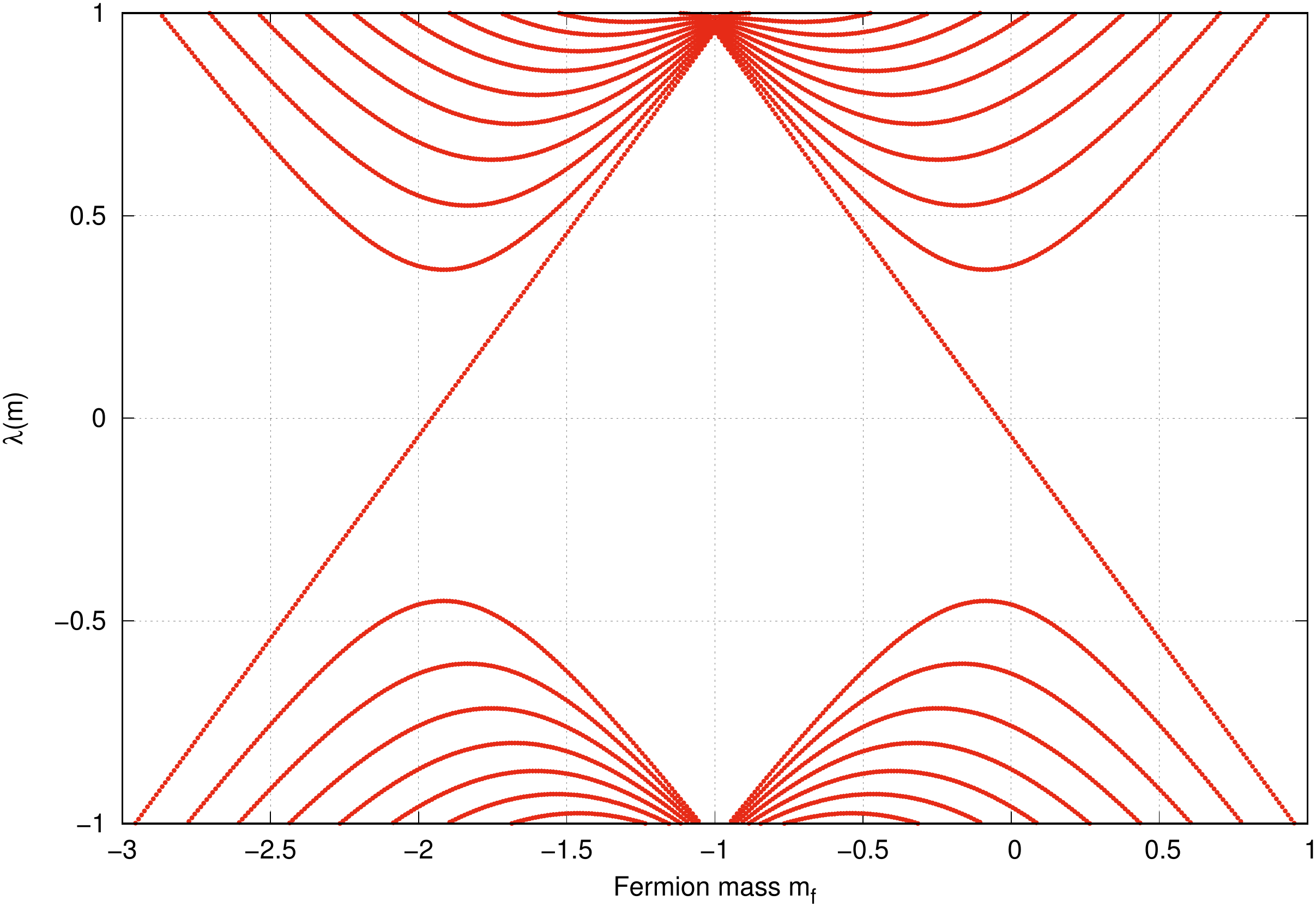}

}

\caption{Spectrum of the staggered Wilson kernel for gauge fields with $Q=2$.
\label{fig:spectral-flow-stw}}
\end{figure*}

In the case of a Wilson kernel, one considers the eigenvalues $\left\{ \lambda\left(m_{\mathsf{f}}\right)\right\} $
of the Hermitian operator $H_{\mathsf{w}}\left(m_{\mathsf{f}}\right)$
as a function of $m_{\mathsf{f}}$. One can show that there is a one-to-one
correspondence between the eigenvalue crossings of $H_{\mathsf{w}}\left(m_{\mathsf{f}}\right)$
and the real eigenvalues of $D_{\mathsf{w}}\left(m_{\mathsf{f}}\right)$.
Furthermore the low-lying real eigenvalues of $D_{\mathsf{w}}\left(m_{\mathsf{f}}\right)$
correspond to the would-be zero modes \cite{Smit:1986fn} and one
can show that for these modes the slope of the eigenvalue crossings
in the vicinity of $m_{\mathsf{f}}=0$ equals minus the chirality.
By identifying the eigenvalues crossings occurring at small values
of $m_{\mathsf{f}}$ as well as their slopes, one can then infer the
topological charge $Q$ of the gauge field.

While originally the spectral flow was used for a Wilson kernel, eventually
the applicability to the staggered case could be shown as well \cite{Adams:2009eb,deForcrand:2011ak,Follana:2011kh,deForcrand:2012bm,Azcoiti:2014pfa}.
Spectral flows of both the Wilson and staggered Wilson kernel have
been investigated in the literature before. Here we want to illustrate
the effectiveness of smearing.

In Figs.~\ref{fig:spectral-flow-wilson} and \ref{fig:spectral-flow-stw},
we show the eigenvalue flow $\lambda\left(m_{\mathsf{f}}\right)$
of $H_{\mathsf{w}}\left(m_{\mathsf{f}}\right)$ and $H_{\mathsf{sw}}\left(m_{\mathsf{f}}\right)$
for the lowest $50$ eigenvalues. We consider a $12^{2}$ gauge configuration
at $\beta=1.8$ with topological charge $Q=2$. We calculate the eigenvalue
flow on the unsmeared and three-step \textsc{ape} smeared configuration
with smearing parameter $\alpha=0.5$. For comparison, we also show
the corresponding topological gauge configuration described in Sec.~\ref{subsec:Topology}.
As one can see, the use of smearing allows the unambiguous determination
of the topological charge $Q$, which on the rough configuration without
smearing is otherwise difficult. Finally, the corresponding topological
charge for $Q=2$ is so smooth that the two eigenvalue crossings close
to the origin lie on top of each other. This shows how beneficial
the use of smearing is when studying topological aspect using spectral
flows.

\subsection{Approaching the continuum}

\begin{table*}[t]
\begin{centering}
\begin{tabular*}{0.9\textwidth}{@{\extracolsep{\fill}}cccccc}
\toprule 
Kernel & Construction & $N_{s}$ & $m_{\mathsf{eff}}$ & $\Delta_{\mathsf{N}}$ & $\Delta_{\mathsf{GW}}$\tabularnewline
\midrule
 &  & 2 & $5.47\times10^{-3}$ & $1.53\times10^{-1}$ & $6.72\times10^{-1}$\tabularnewline
\cmidrule{3-6} 
 & Standard & 4 & $5.90\times10^{-4}$ & $6.87\times10^{-2}$ & $3.10\times10^{-1}$\tabularnewline
\cmidrule{3-6} 
 &  & 6 & $1.00\times10^{-4}$ & $2.30\times10^{-2}$ & $1.02\times10^{-1}$\tabularnewline
\cmidrule{2-6} 
 &  & 2 & $5.73\times10^{-3}$ & $3.71\times10^{-1}$ & $1.16\times10^{0\phantom{-}}$\tabularnewline
\cmidrule{3-6} 
Wilson & Boriçi & 4 & $8.56\times10^{-5}$ & $6.64\times10^{-2}$ & $3.00\times10^{-1}$\tabularnewline
\cmidrule{3-6} 
 &  & 6 & $5.42\times10^{-6}$ & $1.57\times10^{-2}$ & $7.60\times10^{-2}$\tabularnewline
\cmidrule{2-6} 
 &  & 2 & $8.30\times10^{-3}$ & $7.51\times10^{-1}$ & $1.16\times10^{0\phantom{-}}$\tabularnewline
\cmidrule{3-6} 
 & Optimal & 4 & $2.11\times10^{-3}$ & $3.59\times10^{-2}$ & $4.78\times10^{-2}$\tabularnewline
\cmidrule{3-6} 
 &  & 6 & $8.71\times10^{-6}$ & $1.21\times10^{-3}$ & $1.74\times10^{-3}$\tabularnewline
\midrule
 &  & 2 & $6.22\times10^{-3}$ & $1.48\times10^{-1}$ & $6.49\times10^{-1}$\tabularnewline
\cmidrule{3-6} 
 & Standard & 4 & $7.18\times10^{-4}$ & $5.88\times10^{-2}$ & $2.88\times10^{-1}$\tabularnewline
\cmidrule{3-6} 
 &  & 6 & $1.34\times10^{-4}$ & $2.01\times10^{-2}$ & $9.72\times10^{-2}$\tabularnewline
\cmidrule{2-6} 
 &  & 2 & $6.38\times10^{-3}$ & $2.02\times10^{-1}$ & $2.92\times10^{-1}$\tabularnewline
\cmidrule{3-6} 
Staggered Wilson & Boriçi & 4 & $5.13\times10^{-5}$ & $5.45\times10^{-3}$ & $8.68\times10^{-3}$\tabularnewline
\cmidrule{3-6} 
 &  & 6 & $5.91\times10^{-7}$ & $1.53\times10^{-4}$ & $2.69\times10^{-4}$\tabularnewline
\cmidrule{2-6} 
 &  & 2 & $1.72\times10^{-2}$ & $2.32\times10^{-1}$ & $2.66\times10^{-1}$\tabularnewline
\cmidrule{3-6} 
 & Optimal & 4 & $3.35\times10^{-5}$ & $2.23\times10^{-3}$ & $2.63\times10^{-3}$\tabularnewline
\cmidrule{3-6} 
 &  & 6 & $5.02\times10^{-8}$ & $2.01\times10^{-5}$ & $2.36\times10^{-5}$\tabularnewline
\bottomrule
\end{tabular*}
\par\end{centering}
\caption{Median values for the chiral symmetry violations on unsmeared $32^{2}$
configurations at $\beta=12.8$. \label{tab:cont-limit-values}}
\end{table*}
\begin{figure*}[!]
\begin{centering}
\hfill{}\subfloat[Wilson kernel]{\includegraphics[width=0.4\textwidth]{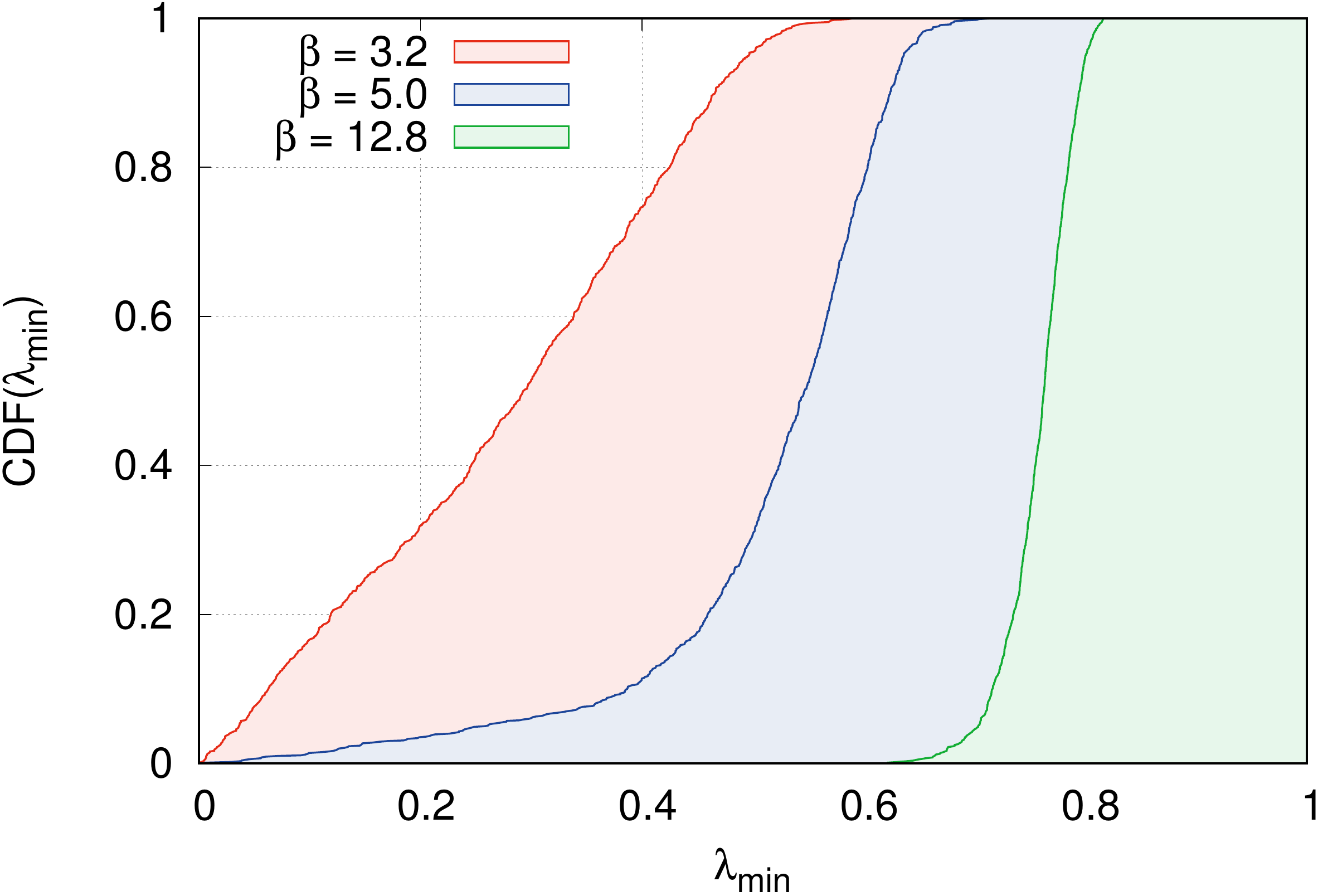}

}\hfill{}\subfloat[Staggered Wilson kernel]{\includegraphics[width=0.4\textwidth]{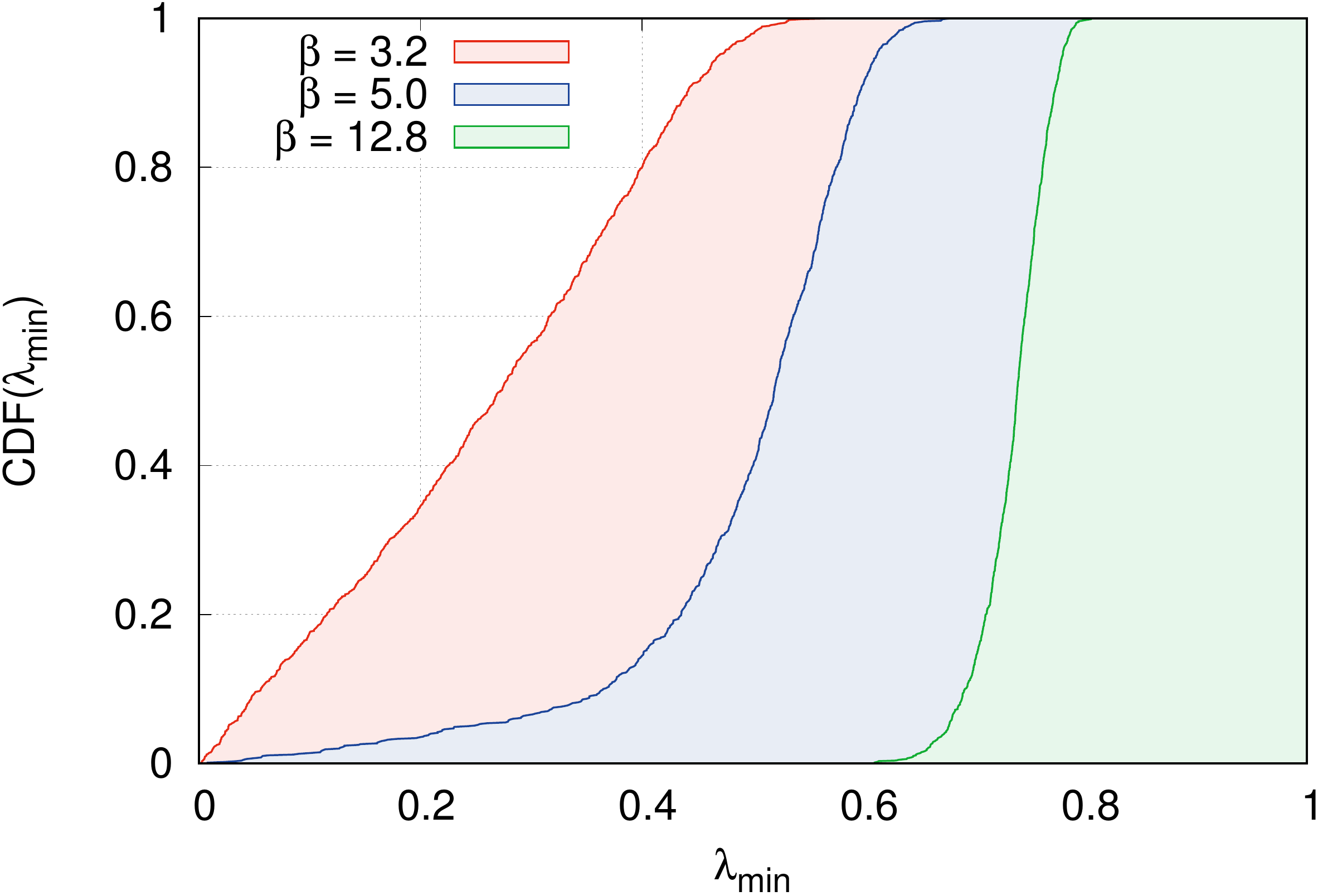}

}\hfill{} 
\par\end{centering}
\caption{Cumulative distribution function $\mathsf{CDF}$ for $\lambda_{\mathsf{min}}$
of $H_{\mathsf{w}}$ and $H_{\mathsf{sw}}$. \label{fig:cdf-lmin}}
\end{figure*}
\begin{figure*}[!]
\begin{centering}
\hfill{}\subfloat[Without smearing]{\includegraphics[width=0.45\textwidth]{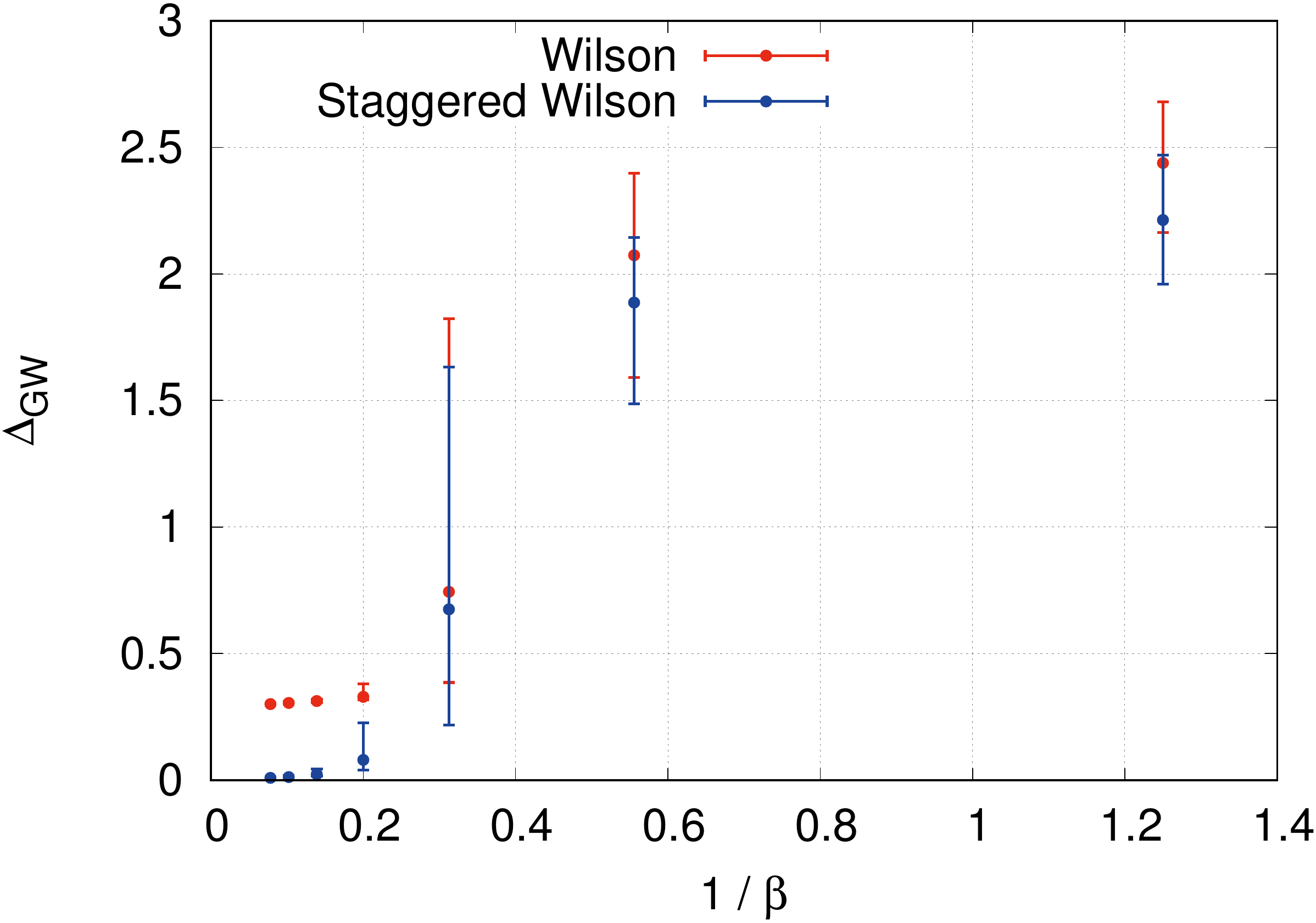}

}\hfill{}\subfloat[With smearing]{\includegraphics[width=0.45\textwidth]{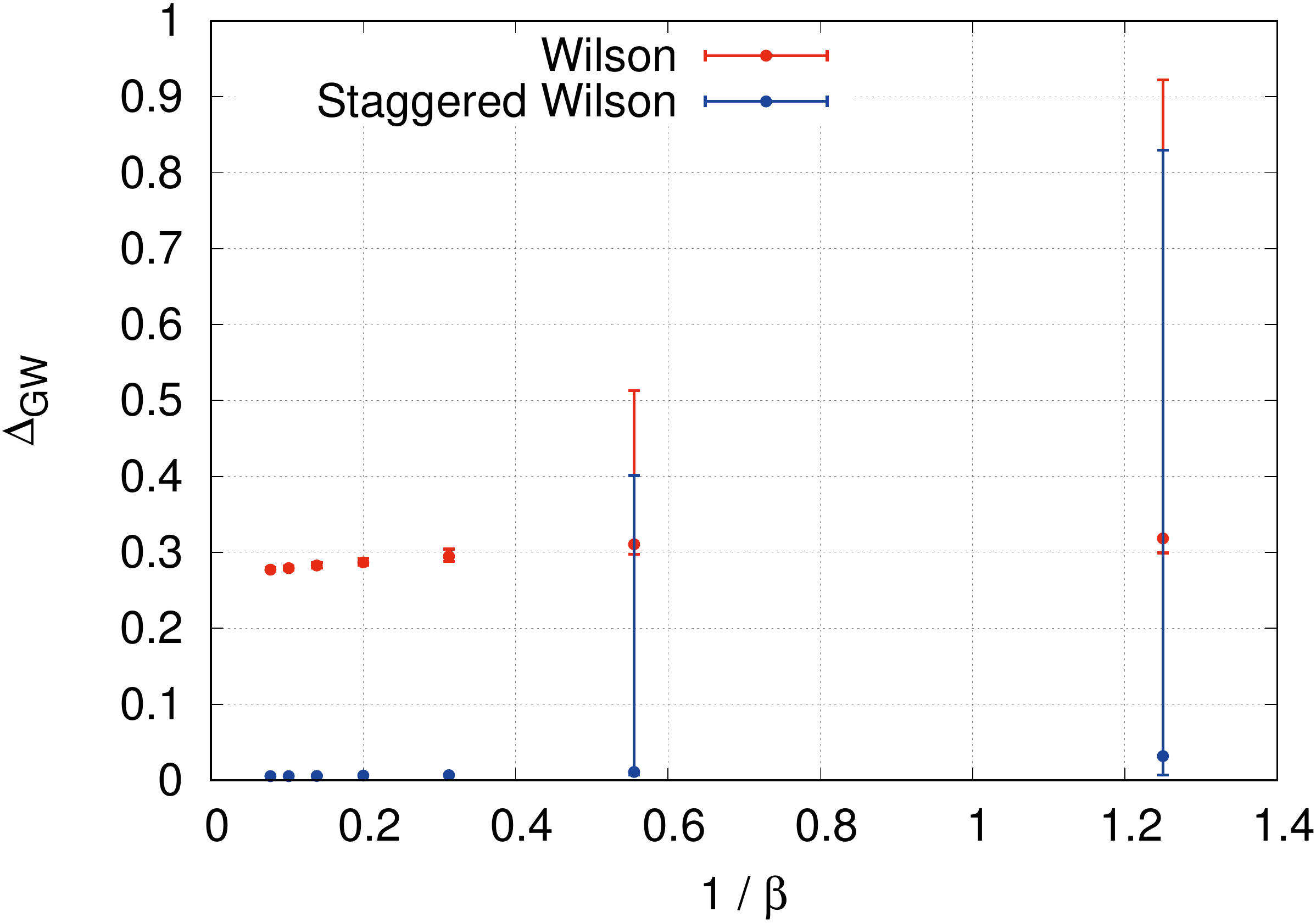}

}\hfill{} 
\par\end{centering}
\caption{$\Delta_{\mathsf{GW}}$ of $\varrho D_{\mathsf{eff}}$ in Boriçi's
construction at $N_{s}=4$ as a function of $1/\beta$. \label{fig:cont-limit-gwr}}
\end{figure*}

In order to judge the performance of the different fermion formulations
when approaching the continuum, we evaluated them on seven different
ensembles with $1000$ configurations each. We kept the physical volume
constant and considered the following lattices: $8^{2}$ at $\beta=0.8$,
$12^{2}$ at $\beta=1.8$, $16^{2}$ at $\beta=3.2$, $20^{2}$ at
$\beta=5.0$, $24^{2}$ at $\beta=7.2$, $28^{2}$ at $\beta=9.8$
and $32^{2}$ at $\beta=12.8$. Together with the smeared version
of each configuration, we consider $N=14\,000$ configurations in
total.

We do not attempt to carry out a strict continuum limit analysis,
but we are interested in the relative performance of the different
formulations when the lattices become finer. An indication for the
performance are the chiral symmetry violations on our finest lattice
at $\beta=12.8$. We quote the median values in Table \ref{tab:cont-limit-values}.
We restrict ourselves to $N_{s}\in\left\{ 2,4,6\right\} $ as for
$N_{s}\geq8$ some values have the same of order of magnitude as the
rounding errors and, thus, we are unable to quote precise numbers.
We observe that for large $\beta$ the chiral symmetry violations
for the standard construction are comparable in the case of a Wilson
and a staggered Wilson kernel. For Boriçi's and the optimal construction,
on the other hand, the violations are much lower for a staggered Wilson
kernel, often by $1$ to $2$ orders of magnitude. Note that for simplicity
we set the parameters $\lambda_{\mathsf{min}}$ and $\lambda_{\mathsf{max}}$
for optimal domain wall fermions on a configuration basis. This is
intended to give an indication of the performance, while in realistic
simulations one would fix suitable values on an ensemble basis after
having projected out a number of low-lying eigenmodes. One then has
to find a compromise between mapping small eigenvalues accurately
and keeping the overall approximation error small.

In general, we also observe that the optimal construction shows a
better performance than Boriçi's, while Boriçi's construction performs
better than the standard construction. A interesting exception is
$m_{\mathsf{eff}}$, where the optimal construction with a Wilson
kernel is not clearly outperforming the other constructions. This
is not unexpected as the optimal $\sign$-function approximation has
a point of maximal deviation at $\lambda_{\mathsf{min}}$ and as a
result low-lying eigenvalues are not mapped accurately on smooth configurations.
However, for larger $N_{s}$ this phenomenon disappears and the optimal
construction shows a superior performance.

A related question is how the lowest eigenvalue $\lambda_{\mathsf{min}}=\min_{\lambda\in\spec H}\left|\lambda\right|$
of the kernel operators $H=H_{\mathsf{w}}$ and $H=H_{\mathsf{sw}}$
is distributed. Especially close to the origin smooth approximations
to the $\sign$-function tend to be inaccurate, so the mappings of
small eigenvalues suffer from large errors. In Fig.~\ref{fig:cdf-lmin},
we can find the numerically determined cumulative distribution functions
(CDFs) for both kernel at three different $\beta$. As expected, for
larger $\beta$ the tail of small values thins out and near-zero $\lambda_{\mathsf{min}}$
become infrequent. The CDFs for the Wilson and staggered Wilson kernel
look very much alike and the probability to encounter configurations
where $\lambda_{\mathsf{min}}\approx0$ is comparable.

Finally, an example of how chiral symmetry violations vary with $\beta$
can be found in Fig.~\ref{fig:cont-limit-gwr}. We show here the
violation of the Ginsparg-Wilson relation $\Delta_{\mathsf{GW}}$
of $\varrho D_{\mathsf{eff}}$ in Boriçi's construction at $N_{s}=4$
as a function of $1/\beta$. We plot the median value together with
the width of the distribution, characterized by the $q$-quantile
and the $\left(1-q\right)$-quantile, where $q=\left[1-\erf\left(1/\sqrt{2}\right)\right]/2$
and $\erf$ denotes the error function. With this choice $\unit[68.3]{\%}$
of the values are within the error bars. One can clearly see how the
effective operator with a staggered Wilson kernel shows superior chiral
properties when $\beta$ is sufficiently large. As expected, we observe
that smearing improves chiral properties in particular on the coarsest
lattices.

\section{Conclusions \label{sec:Conclusions}}

In this work, we gave an explicit construction of staggered domain
wall fermions and investigated some of their basic properties in the
free-field case, on quenched thermalized gauge configurations in the
Schwinger model and on smooth topological configurations. It appears
that staggered domain wall fermions indeed work as advertised.

Moreover we could generalize existing modifications of domain wall
fermions, such as Boriçi's and the optimal construction to the staggered
case. This gives rise to previously not considered truncated staggered
domain wall fermions and optimal staggered domain wall fermions. These
modified staggered domain wall fermions in particular show significantly
smaller chiral symmetry violations than the traditional Wilson based
formulations, at least in our setting and with respect to the criteria
used in this work. These properties make formulations with a staggered
Wilson kernel potentially interesting when studying phenomena, where
chiral symmetry is of importance.

It is not yet clear how our results in $\Uone$ background gauge fields
will translate to QCD in $3+1$ dimensions with a $\SUthree$ gauge
group, but they are encouraging and warrant further investigations.
\begin{acknowledgments}
We thank Ting-Wai Chiu for helping us to validate our implementation
of optimal domain wall fermions and Stephan Dürr for helpful comments
on the manuscript. C.~H.~is supported by DFG grant SFB/TRR-55. C.~Z.~is
supported by the Singapore International Graduate Award (SINGA) and
Nanyang Technological University. Parts of the computations were done
on the computer cluster of the University of Wuppertal.
\end{acknowledgments}

\appendix

\section{Optimal weights \label{sec:Optimal-weights}}

In the following, we give some example values for the weight factors
as defined in Sec.~\ref{subsec:Optimal-construction}. To this end,
let us consider the free-field case in $1+1$ dimensions with the
particular choices $M_{0}=1$ and $N_{s}=8$. For the Wilson kernel
we then have $\lambda_{\mathsf{min}}=1$ and $\lambda_{\mathsf{max}}=3$,
for the staggered Wilson kernel $\lambda_{\mathsf{min}}=1$ and $\lambda_{\mathsf{max}}=\sqrt{2}$.
The corresponding example weights $\left\{ \omega_{s}\right\} $ for
optimal domain wall fermions can be found in Table \ref{tab:Example-weights}.\\

\begin{table}[H]
\begin{centering}
\begin{tabular}{>{\centering}m{0.05\columnwidth}>{\centering}m{0.4\columnwidth}>{\centering}m{0.4\columnwidth}}
\toprule 
$s$  & $\omega_{s}\left(\lambda_{\mathsf{min}}=1,\lambda_{\mathsf{max}}=3\right)$  & $\omega_{s}\left(\lambda_{\mathsf{min}}=1,\lambda_{\mathsf{max}}=\sqrt{2}\right)$\tabularnewline
\midrule
\midrule 
1  & \texttt{0.989011284192743}  & \texttt{0.996659816028010}\tabularnewline
\midrule 
2  & \texttt{0.908522120246430}  & \texttt{0.971110743060917}\tabularnewline
\midrule 
3  & \texttt{0.779520722603956}  & \texttt{0.925723982088869}\tabularnewline
\midrule 
4  & \texttt{0.641124364053574}  & \texttt{0.869740043520870}\tabularnewline
\midrule 
5  & \texttt{0.519919928211430}  & \texttt{0.813009342796322}\tabularnewline
\midrule 
6  & \texttt{0.427613177773962}  & \texttt{0.763841917102527}\tabularnewline
\midrule 
7  & \texttt{0.366896221792507}  & \texttt{0.728142270322088}\tabularnewline
\midrule 
8  & \texttt{0.337036936444470}  & \texttt{0.709476563432225}\tabularnewline
\bottomrule
\end{tabular}
\par\end{centering}
\caption{$\omega_{s}$ for optimal domain wall fermions. \label{tab:Example-weights}}
\end{table}

\bibliography{literature}

\end{document}